\documentclass[review]{elsarticle}
\usepackage{todonotes}

\usepackage[colorlinks,citecolor=blue,linktoc=all,linkcolor=cyan]{hyperref}
\usepackage{graphicx}
\usepackage{bm}
\usepackage[T1]{fontenc}
\usepackage{dsfont}               
\usepackage{mathrsfs}             
\usepackage{slashed}              
\usepackage{amsmath}
\usepackage{amssymb}
\usepackage{amsbsy}
\usepackage{amsfonts}
\usepackage{MnSymbol}
\usepackage{braket}
\usepackage{multirow}
\numberwithin{equation}{section}
\numberwithin{table}{section}
\numberwithin{figure}{section}
\usepackage{subcaption}

\journal{Progress in Particle and Nuclear Physics}

\usepackage{titlesec}
\usepackage{sectsty}
\titleformat{\section}{\normalfont\Large\bfseries}{\thesection}{1em}{}
\titleformat{\subsection}{\normalfont\large\bfseries}{\thesubsection}{1em}{}
\titleformat{\subsubsection}{\normalfont\normalsize\bfseries}{\thesubsubsection}{1em}{}

\newcommand{\params}{\boldsymbol{\theta}}

\biboptions{numbers,sort&compress}

\begin{document}
	
	\begin{frontmatter}
		
        \title{Neural-network quantum states for the nuclear many-body problem}

		\author[argonne_phy,argonne_cps,trento,ific]{Alessandro Lovato \corref{mycorrespondingauthor}}
		\ead{lovato@anl.gov}
        \address[argonne_phy]{Physics Division, Argonne National Laboratory, Argonne, Illinois 60439, USA}
        \address[argonne_cps]{Computational Science Division, Argonne National Laboratory, Argonne, Illinois 60439, USA}
        \address[trento]{INFN-TIFPA Trento Institute of Fundamental Physics and Applications, 38123 Trento, Italy. }
        \address[ific]{Instituto de Física Corpuscular (IFIC), Consejo Superior de Investigaciones Científicas (CSIC) and Universidad de Valencia
        E-46980 Paterna, Valencia, Spain}
		
		\author[epfl]{Giuseppe Carleo}  
        \address[epfl]{Institute of Physics, École Polytechnique Fédérale de Lausanne (EPFL), CH-1015 Lausanne, Switzerland}

        \author[argonne_phy]{Bryce Fore}

        \author[oslo]{Morten Hjorth-Jensen}
        \address[oslo]{Department of Physics and Center for Computing in Science Education, University of Oslo, N-0316 Oslo, Norway}

        \author[argonne_phy,ohio]{Jane Kim}
        \address[ohio]{Institute of Nuclear and Particle Physics and Department of Physics and Astronomy, Ohio University, Athens, OH, 45701, USA}
               
        \author[barcelona-fqa,barcelona-icc]{Arnau Rios}
        \address[barcelona-fqa]{Departament de Física Quàntica i Astrofísica, Universitat de Barcelona (UB), c. Martí i Franquès         1, E08028 Barcelona, Spain}
        \address[barcelona-icc]{Institut de Ciències del Cosmos, Universitat de Barcelona (UB), c. Martí i Franquès 1, E08028 Barcelona, Spain}
        
        \author[fermilab,ific]{Noemi Rocco}
        \address[fermilab]{Theoretical Physics Department Fermi National Accelerator Laboratory P.O. Box 500 Batavia Illinois 60510 USA}
		
        \begin{abstract}
        A long-standing goal of nuclear theory is to explain how the structure and dynamics of atomic nuclei and neutron-star matter emerge from the underlying interactions among protons and neutrons. Achieving this goal requires solving the nuclear quantum many-body problem with high accuracy across a wide range of length scales and density regimes. In this review, we discuss how artificial neural network representations of the nuclear many-body wave function have significantly extended the capabilities of continuum quantum Monte Carlo methods. In particular, neural-network quantum states enable calculations of larger systems than were previously accessible and provide a flexible framework for capturing phenomena that challenge conventional approaches, including the emergence of nuclear clusters and superfluid phases in dense matter. We highlight recent applications to finite nuclei, infinite nuclear and neutron matter, and dynamical processes relevant to lepton–nucleus and nucleus–nucleus scattering. We also discuss conceptual and methodological connections with condensed matter physics, emphasizing developments in neural-network quantum states that bridge strongly correlated systems across disciplines. Together, these developments demonstrate how neural-network methods open new avenues toward unified and accurate descriptions of nuclear structure, matter, and reactions.
        \end{abstract}
        		
        \begin{keyword}
        Neural Network Quantum States\sep Nuclear Quantum Many-Body Problem\sep Atomic Nuclei \sep Neutron Star Matter \sep Nuclear Electroweak Interactions
        \end{keyword}
		
	\end{frontmatter}
	
	\newpage
	
	\thispagestyle{empty}
	\tableofcontents
	


\newpage
\section{Introduction}
\label{sec:intro}

At the energy scales relevant for the dynamics of atomic nuclei, the fundamental theory of strong interactions, quantum chromodynamics (QCD), becomes non-perturbative, and quarks and gluons are confined within hadrons. Following the seminal works of Weinberg~\cite{weinberg_nuclear_1990,weinberg_effective_1991}, effective field theories (EFTs) have emerged as the framework of choice for computing low-energy nuclear observables with quantified uncertainties. Two prominent examples are pionless EFT and chiral EFT, both of which exploit a separation between a ``hard'' momentum scale $\Lambda$ and a ``soft'' scale $Q$ that characterizes the typical momenta in low-energy nuclear systems~\cite{epelbaum_modern_2009,machleidt_chiral_2011}. Pionless EFT is applicable when $Q \ll m_\pi$, and its breakdown scale is set by $\Lambda \sim m_\pi$. In contrast, chiral EFT retains explicit pion degrees of freedom, and its breakdown scale is considerably higher, $\Lambda \sim 600$--$700$ MeV, while the soft scale remains $Q \sim m_\pi$. At low energies, protons and neutrons interact through nonrelativistic potentials systematically organized as an expansion in $Q/\Lambda$~\cite{van_kolck_few_1994,epelbaum_modern_2009,machleidt_chiral_2011}. High-momentum physics is integrated out and encoded in low-energy constants fitted to nucleon--nucleon scattering data and observables in light nuclei. 

Interactions and electroweak currents derived consistently within the EFT framework provide the primary input to \emph{ab initio} methods designed to solve the nuclear many-body Schr\"odinger equation with controlled and systematically improvable approximations~\cite{hergert_guided_2020,ekstrom_what_2023}. Single-particle basis methods such as the no-core shell model (NCSM)~\cite{barrett_ab_2013}, coupled-cluster (CC) theory~\cite{hagen_coupled-cluster_2014}, the in-medium similarity renormalization group (IMSRG)~\cite{hergert_-medium_2016}, and self-consistent Green's function (SCGF) techniques~\cite{Carbone:2013eqa} have achieved impressive success. The NCSM and its resonating-group extension~\cite{navratil_recent_2009} accurately describe clustering and collective dynamics~\cite{quaglioni_ab_2008,roth_similarity-transformed_2011,mccoy_emergent_2020} but are generally restricted to light nuclei, and convergence for long-range observables can be slow~\cite{caprio_robust_2022}. Polynomially scaling approaches such as CC and IMSRG now reach nuclei as heavy as $^{100}$Sn~\cite{hagen_structure_2016,morris_structure_2018,mougeot_mass_2021} and $^{208}$Pb~\cite{hu_ab_2022}, providing access to spectra, response functions, and transition rates~\cite{sobczyk_ab_2021,acharya_16o_2024,bonaiti_electromagnetic_2024}. These developments have deepened our understanding of the long-standing quenching of $g_A$~\cite{gysbers_discrepancy_2019} and enabled uncertainty-quantified predictions of neutrinoless double-$\beta$ decay~\cite{yao_ab_2020,wirth_ab_2021,belley_ab_2021}. Nevertheless, truncation of model spaces limits their ability to fully capture multi-scale phenomena, especially alpha clustering and short-range correlations. Symmetry-adapted shell-model approaches~\cite{launey_symmetry-guided_2016,dytrych_physics_2020} mitigate some of these challenges but they may not  be able to fully capture short-distance dynamics.

A second class of \emph{ab initio} methods relies on stochastic formulations, such as nuclear lattice EFT (NLEFT)~\cite{lee_lattice_2025} and continuum quantum Monte Carlo (QMC) techniques~\cite{carlson_quantum_2015,lynn2019,gandolfi2020}. NLEFT has advanced to medium-mass nuclei~\cite{elhatisari_wavefunction_2024}, revealing compelling signatures of alpha clustering~\cite{shen_emergent_2023} and providing insights into nuclear matter at finite temperature~\cite{Ren:2023ued,ma_structure_2024}, enabled by substantial algorithmic progress~\cite{sarkar_floating_2023}. However, the fermion sign problem and the computational cost of fine lattice spacings limit its ability to resolve short-range dynamics with high accuracy. Continuum QMC methods, such as Green’s function Monte Carlo (GFMC)~\cite{carlson_alpha_1988} and auxiliary-field diffusion Monte Carlo (AFDMC)~\cite{Schmidt:1999lik}, accurately describe nuclear dynamics across long, intermediate, and short ranges. In practice, however, GFMC scales exponentially with the mass number, restricting applications to nuclei with $A \lesssim 13$, while AFDMC encounters a severe sign problem when realistic correlations are introduced, limiting its applicability to nuclei with $A \sim 16$~\cite{novario_trends_2023,martin_auxiliary_2023}.

These challenges have motivated the exploration of neural-network representations of quantum many-body wave functions, a framework that has already achieved notable success in condensed-matter physics and quantum chemistry~\cite{carleo_solving_2017,hermann_deep-neural-network_2020,pfau_ab_2020}. In the nuclear domain, however, neural-network quantum states (NQS) methods face a distinct set of difficulties. Nuclear interactions are highly non-perturbative and exhibit strong spin--isospin dependence, placing them well outside the regime where mean-field approximations provide a reliable starting point~\cite{wiringa_accurate_1995}. As a result, NQS methods are confronted with correlations at all length scales. At the same time, realistic nuclear Hamiltonians require treating both continuous spatial degrees of freedom and discrete spin and isospin degrees of freedom, significantly enlarging the structure of the underlying Hilbert space.

In this review, we discuss the development and application of NQS ans\"atze in nuclear physics and closely related areas. Following the pioneering application of NQS to the deuteron~\cite{keeble_machine_2020}, variational Monte Carlo methods based on NQS~\cite{adams_variational_2021,yang_consistent_2022,lovato_hidden-nucleons_2022,Yang:2023rlw,gnech_distilling_2024} have proven to provide a systematically improvable framework with polynomial computational scaling and have begun to address several long-standing challenges in \emph{ab initio} nuclear theory. Our focus is on first-quantized architectures, which operate directly on the continuous spatial coordinates and discrete spin--isospin degrees of freedom of the nucleons. This representation naturally accommodates short-, intermediate-, and long-range nuclear dynamics, from the repulsive core and tensor correlations to clustering and collective degrees of freedom. Moreover, NQS approaches capture the multiscale behavior characteristic of nuclei and neutron-star matter~\cite{fore2023}, including superfluidity, the emergence of nuclear clusters, and dominant short-range correlations~\cite{kim_neural-network_2024,fore_investigating_2024}.

First, we briefly review existing continuum QMC methods, focusing in particular on GFMC and the computational challenges associated with treating nuclei with more than $A=13$ nucleons. This naturally leads to a discussion on how the Hilbert space is sampled in VMC methods, with emphasis on the nuclear case, where both continuous (spatial coordinates) and discrete (spin--isospin) degrees of freedom must be sampled efficiently. After providing a short recap of the elements of deep learning relevant to NQS approaches, we introduce the main first-quantized NQS architectures proposed to date and discuss how physical symmetries—such as parity, time-reversal, and translation invariance—are incorporated into these models.

In the second part of the review, we summarize the most important recent results for finite nuclei and for infinite nuclear and neutron matter. We then discuss initial applications of NQS wave functions to lepton--nucleus and nucleus--nucleus scattering, highlighting the potential of these methods to deliver unified and accurate \emph{ab initio} descriptions of both nuclear structure and reactions, and to extend the reach of nuclear many-body theory beyond the traditional limits imposed by basis truncations and the sign problem. Before concluding, we provide a selective overview NQS approaches to condensed matter theory, focusing particularly on those that have direct links to nuclear theory.

\newpage
\section{The nuclear quantum many-body problem}
\label{sec:nmbp}
The nuclear quantum many-body problem presents formidable challenges due to the strong spin–isospin dependence and intrinsically non-perturbative character of realistic nuclear interactions. In contrast to quantum chemistry, where the Coulomb interaction is known \emph{a priori} and provides a universal two-body potential, nuclear forces must be derived within the EFT framework. The construction of accurate two- and three-nucleon interactions with quantified uncertainties remains an active research frontier~\cite{epelbaum_modern_2009,machleidt_chiral_2011,ekstrom_what_2023}.

Recent advances in Bayesian parameter estimation and regulator optimization have further improved the consistency and predictive power of modern chiral EFT Hamiltonians.
The complexity of these interactions has motivated the development of a broad suite of \emph{ab initio} many-body approaches~\cite{hergert_guided_2020}. Single-particle basis methods, including the NCSM, CC theory, SCGF and the IMSRG, have achieved high-precision results across wide regions of the nuclear chart. Complementary stochastic formulations such as NLEFT have revealed striking signatures of clustering~\cite{shen_emergent_2023} and have advanced toward medium-mass nuclei, though their finite lattice spacing limits the resolution of short-range dynamics.
Continuum QMC methods~\cite{carlson_quantum_2015,lynn2019,gandolfi2020}, including GFMC~\cite{carlson_alpha_1988} and AFDMC~\cite{Schmidt:1999lik}, accurately capture nuclear dynamics from long to short distances, but their applicability across the nuclear chart remains restricted. GFMC scales exponentially with the mass number, while AFDMC suffers from a worsening sign problem for realistic chiral interactions~\cite{novario_trends_2023,martin_auxiliary_2023}.

Because this review focuses on neural quantum states (NQS) applied to continuum QMC, this section summarizes the local coordinate-space Hamiltonians derived from EFT that serve as inputs to both conventional QMC and NQS-based VMC calculations~\cite{gezerlis_local_2014}. We then discuss the principal scaling bottlenecks encountered by GFMC and AFDMC as the mass number increases, highlighting the aspects most relevant for constructing efficient NQS ans\"{a}tze. These considerations motivate the exploration of first-quantized NQS architectures that act directly on the continuous spatial coordinates and discrete spin–isospin degrees of freedom, offering a promising path toward extending the reach of continuum QMC methods beyond their traditional limits.

\subsection{Nuclear Hamiltonian}
\label{sub:hamiltonian}

To a remarkable extent, the properties of atomic nuclei and neutron-star matter can be described by point-like nucleons whose dynamics are governed by a nonrelativistic Hamiltonian  
\begin{align}
H_{LO} &= \sum_i \frac{\mathbf{p}_i^2}{2m_N} 
+ \sum_{i<j} v_{ij} 
+ \sum_{i<j<k} V_{ijk} \,,
\label{eq:ham}
\end{align}
where $\mathbf{p}_i$ is the three-momentum of the $i$-th nucleon, $m_N$ is its mass, and $v_{ij}$ and $V_{ijk}$ are the potentials describing nucleon-nucleon ($N\!N$) and three-nucleon ($3N$) interactions, respectively.

With the notable exception of Ref.~\cite{keeble_machine_2020}, and more recently Refs.~\cite{wen_neural-network_2025,yang_zemach_2025}, most applications of neural quantum states (NQS) to nuclear physics employ $N\!N$ and $3N$ potentials based on the premise that the momentum scales relevant for modeling the structure of atomic nuclei are much smaller than the pion mass, $m_\pi \simeq 140$ MeV. In this regime, pions can be integrated out, giving rise to pionless EFT~\cite{chen_nucleon-nucleon_1999,bedaque_effective_2002}, in which the charge-independent (CI) nuclear interactions consist solely of contact terms between two or more nucleons.
Pionless EFT provides a controlled and systematically improvable description of nuclear systems in the low-momentum regime, where short-range correlations and universal behavior dominate the structure of light nuclei and dilute nuclear matter. At the same time, its reduced operator complexity offers a natural setting for developing and benchmarking first-quantized NQS architectures, allowing one to isolate representational and optimization challenges before confronting the additional multiscale structure and strong tensor correlations induced by explicit pion exchange.

In this review, we focus on the leading-order (LO) pionless EFT expansion of Ref.~\cite{schiavilla_two-_2021}. The LO $N\!N$ interaction derived in that work is constrained to act only in even partial waves. In coordinate space, it is given by  
\begin{equation}
v_{LO}^{CI}(r_{ij}) = C_{01} C_1(r_{ij}) P_0^\sigma P_1^\tau 
                   + C_{10} C_0(r_{ij}) P_1^\sigma P_0^\tau ,
\label{eq:vNN_LO}
\end{equation}
where $r_{ij} = |{\bf r}_i - {\bf r}_j|$ and $P_{0,1}^\sigma$ ($P_{0,1}^\tau$) are the spin (isospin) projection operators for the nucleon pair $ij$ with total spin $S$ and isospin $T$ equal to 0 or 1:  
\begin{equation}
P_0^\sigma = \frac{1 - \sigma_{ij}}{4},\qquad
P_1^\sigma = \frac{3 + \sigma_{ij}}{4},\qquad
P_0^\tau   = \frac{1 - \tau_{ij}}{4},\qquad
P_1^\tau   = \frac{3 + \tau_{ij}}{4}\, .
\end{equation}
Here, $\sigma_{ij} = {\boldsymbol \sigma}_i \cdot {\boldsymbol \sigma}_j$ and $\tau_{ij} = {\boldsymbol \tau}_i \cdot {\boldsymbol \tau}_j$, with ${\boldsymbol \sigma}_i$ and ${\boldsymbol \tau}_i$ denoting the Pauli spin and isospin operators acting on nucleon $i$, respectively. In this implementation, the contact interactions are regularized using Gaussian cutoff functions,  
\begin{equation}
C_{\alpha}(r) = \frac{1}{\pi^{3/2} R_\alpha^3} \, e^{-(r / R_\alpha)^2} ,
\end{equation}
where $R_\alpha$ controls the range of the regulator.

\begin{figure}[!t]
    \centering
    \includegraphics[width=0.49\linewidth]{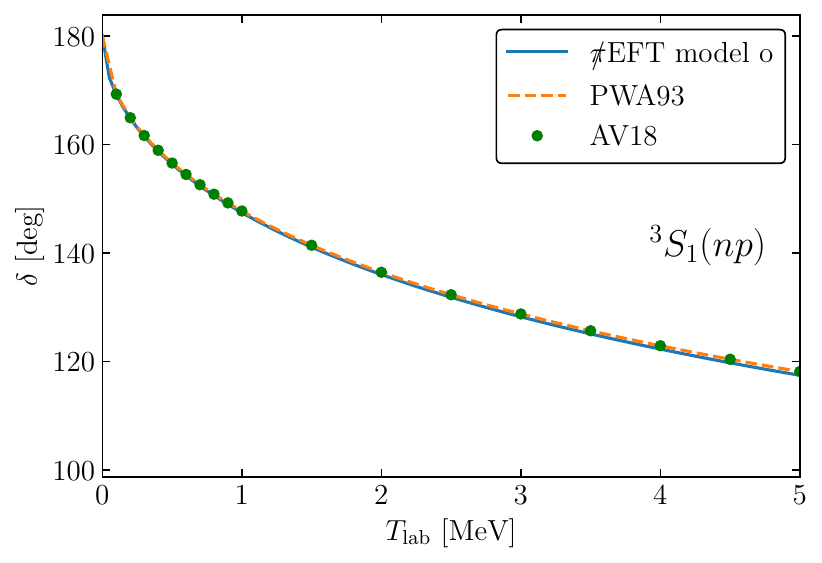}
    \includegraphics[width=0.49\linewidth]{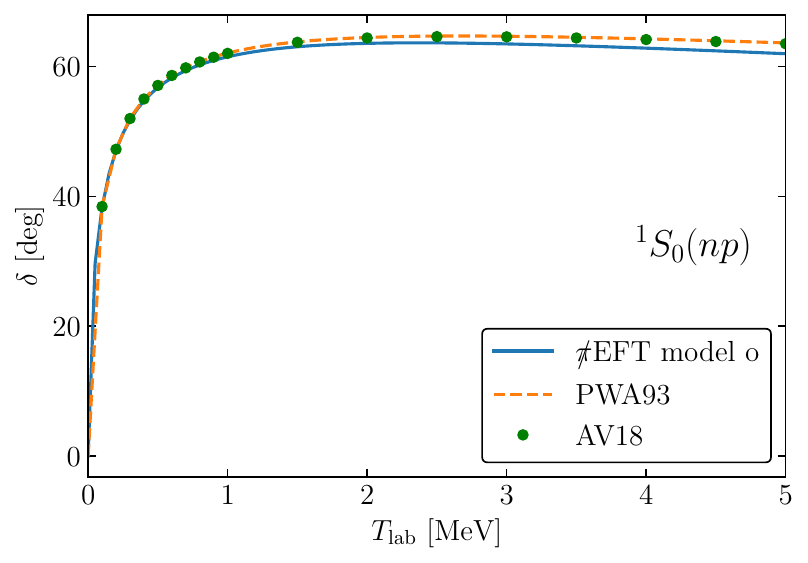}
\caption{Phase shifts in the $^3S_1$ and $^1S_0$ channels for $np$ scattering computed using the LO EFT Hamiltonian ``o'' of Ref.~\cite{schiavilla_two-_2021}, compared to the PWA93 analysis and results from the realistic Argonne v$_{18}$ potential~\cite{wiringa_accurate_1995}.}
 \label{fig:phases_np}
\end{figure}

\begin{figure}[!b]
    \centering
    \includegraphics[width=0.49\linewidth]{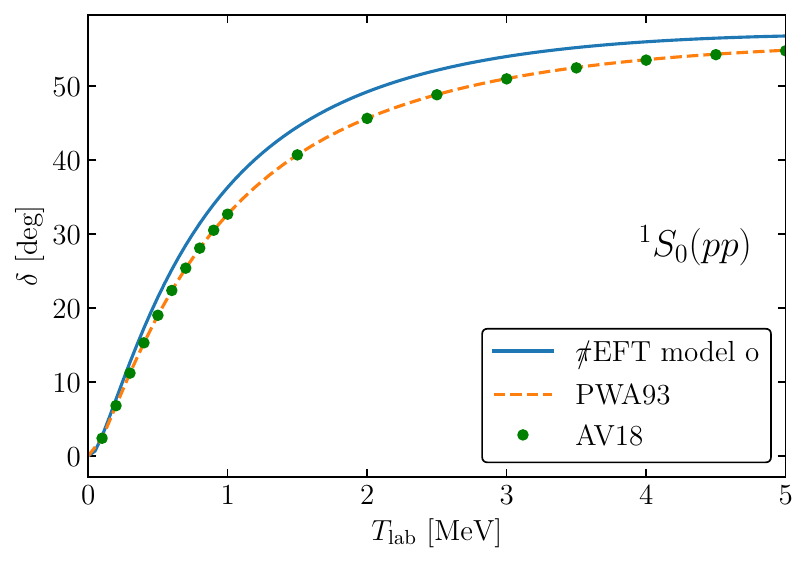}
    \includegraphics[width=0.49\linewidth]{figures/many_body_problem/phase_shifts_S1_T0_MT0.pdf}
\caption{Phase shifts in the $^1S_0$ channels for $pp$ and $nn$ scattering computed using the LO EFT Hamiltonian ``o'' of Ref.~\cite{schiavilla_two-_2021} compared to the PWA93 analysis and results from the realistic Argonne v$_{18}$ potential~\cite{wiringa_accurate_1995}..}
 \label{fig:phases_pp_nn}
\end{figure}

Most of the results presented in this review are obtained using model ``o'' of Ref.~\cite{schiavilla_two-_2021}, in which the cutoff radii $R_0 = 1.5459$ fm and $R_1 = 1.8304$ fm, as well as the low-energy constants $C_{01} = -5.27518671$ fm$^2$ and $C_{10} = -7.04040080$ fm$^2$, were adjusted to reproduce the neutron--proton scattering lengths and effective ranges in the singlet and triplet channels, as well as the deuteron binding energy. To demonstrate the performance of this interaction, Fig.~\ref{fig:phases_np} displays the phase shifts in the $^3S_1$ and $^1S_0$ channels for $np$ scattering. 

On the one hand, we observe excellent agreement with both the PWA93 analysis and those obtained from the highly realistic Argonne v$_{18}$ potential, extending up to $T_{\rm lab} = 5$ MeV.
On the other hand, the phase shifts of the model ``o'' in the $^1S_0$ channels for $pp$ and $nn$ scattering, displayed in Fig.~\ref{fig:phases_pp_nn}, exhibit some discrepancies compared to both the PWA93 analysis and those obtained from the realistic Argonne v$_{18}$ potential~\cite{wiringa_accurate_1995}. These differences can be attributed to the absence of a charge-dependent term in model ``o'', which enters at next-to-leading order (NLO) in the pionless EFT expansion. It has yet to be determined whether including these terms will stabilize $^6$He, $^8$Li, $^8$B, $^9$C, and $^{17}$F against breakup into smaller clusters, or if P-wave terms need to be incorporated into the nucleon-nucleon interaction~\cite{gattobigio_embedding_2019}.

\begin{figure*}[!t]
\centering
  \includegraphics[width=0.6\textwidth]{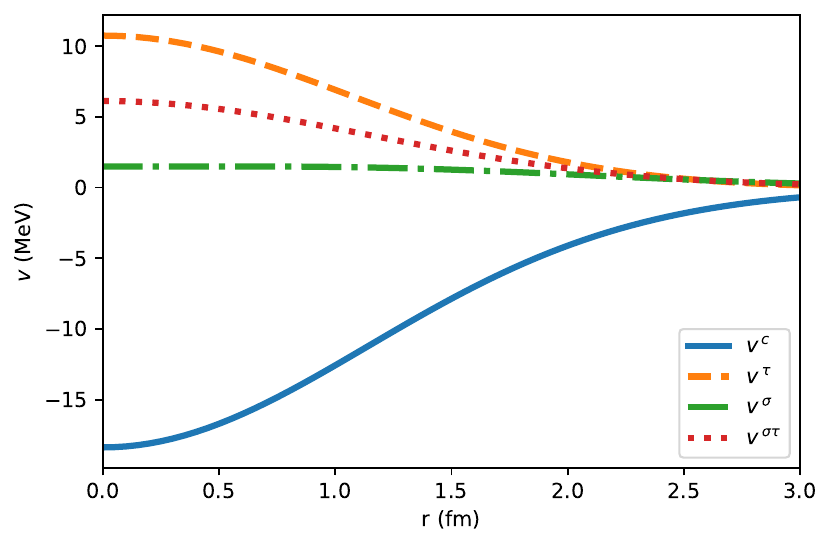}
\caption{Radial functions of the model ``o'' $N\!N$ potential at LO in the pionless EFT expansion, expressed in the spin--isospin basis.}
\label{fig:v_NN}
\end{figure*}

The $N\!N$ potential of Eq.~\eqref{eq:vNN_LO} can be expressed compactly in the spin--isospin operator basis as  
\begin{align}
    v^{\rm CI}_{ij} = \sum_{p=1}^4 v^{p}(r_{ij}) O^p_{ij} \,,
    \label{eq:NN_op}
\end{align}
where $O^{p=1,4}_{ij} = \bigl( 1, \tau_{ij}, \sigma_{ij}, \sigma_{ij} \tau_{ij} \bigr)$, akin to the $v_4^\prime$ potential~\cite{wiringa_evolution_2002}. The explicit forms of the radial functions $v^p(r_{ij})$ are provided in Appendix~A of Ref.~\cite{schiavilla_two-_2021}, and their radial dependence is illustrated in Fig.~\ref{fig:v_NN}.

In addition to the CI component, the two-nucleon interaction includes an electromagnetic (EM) contribution, such that $v = v^{\rm EM} + v^{\rm CI}_{\rm LO}$. The full $v^{\rm EM}$ consists of one- and two-photon Coulomb terms, the Darwin-Foldy term, vacuum polarization, and magnetic moment interactions; see Ref.~\cite{wiringa_accurate_1995} for their explicit expressions. However, most NQS applications retain only the Coulomb repulsion between finite-size (rather than point-like) protons.  

In pionless EFT, solving systems with $A \geq 3$ using purely attractive LO $N\!N$ potentials leads to the so-called Thomas collapse~\cite{yang_we_2020} as the regulator is removed. This pathological behavior is avoided by promoting a contact $3N$ force to LO~\cite{bedaque_renormalization_1999}. In this review, we consider a regularized $3N$ potential of the form  
\begin{align}
    V_{ijk} = \frac{c_E}{f_\pi^4 \Lambda_\chi} \frac{(\hbar c)^6}{\pi^3 R_3^6} 
    \sum_{\rm cyc} e^{-(r_{ij}^2 + r_{jk}^2)/R_3^2} ,
    \label{eq:o_3NF}
\end{align}
where $\Lambda_\chi = 1$ GeV, $f_\pi = 92.4$ MeV is the pion decay constant, and $\sum_{\rm cyc}$ denotes cyclic permutations of the indices $i$, $j$, and $k$. The low-energy constant $c_E$ is fixed to reproduce the $^3$H binding energy, $B(^3$H$) = 8.475$ MeV, for a given value of the cutoff $R_3$.  The analysis of Ref.~\cite{schiavilla_two-_2021} shows that the choice $R_3 = 1.0$ fm and $c_E = 1.0786$ provides a satisfactory description of nuclear binding energies across a broad mass range, up to $^{90}$Zr. However, subsequent VMC–NQS calculations have indicated that this choice leads to overbinding in $^{16}$O and heavier nuclei. Increasing the cutoff to $R_3 = 1.1$ fm with $c_E = 1.2945$ largely resolves this issue~\cite{gnech_distilling_2024}, as the extended range of the $3N$ force introduces additional repulsion in heavier systems.

\begin{figure*}[!t]
\centering
  \includegraphics[width=0.6\textwidth]{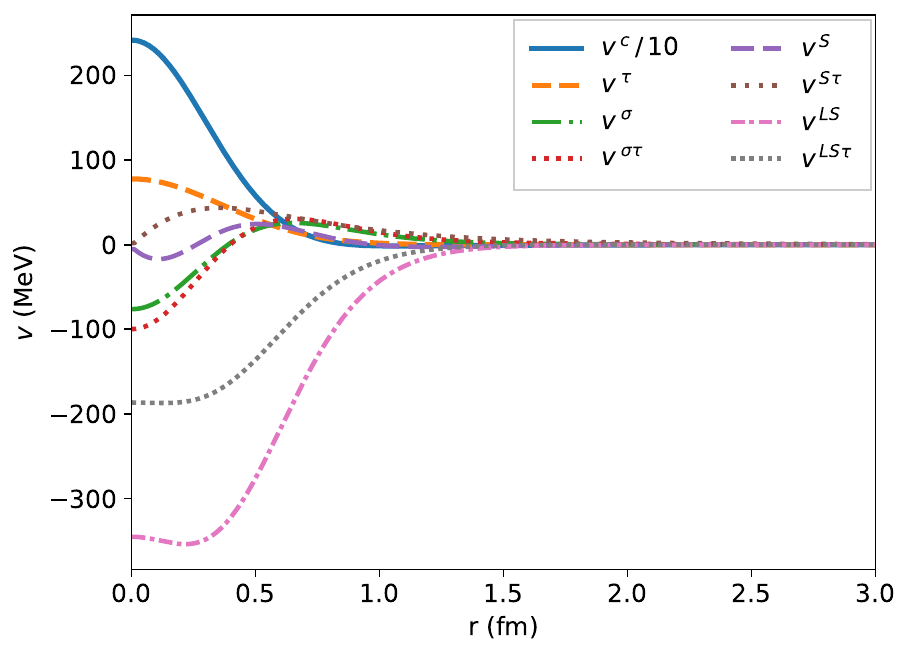}
\caption{Radial functions of the Argonne $v_8^\prime$ potential, expressed in the spin–isospin basis.}
\label{fig:av8p}
\end{figure*}

Very recent applications of nuclear NQS employ higher-resolution interactions, either phenomenological or derived within chiral EFT. In particular, Refs.~\cite{yang_chiral_2025,wen_neural-network_2025,yang_zemach_2025} use consistent $NN$ and $3N$ potentials derived at next-to-next-to-leading order (N$^2$LO) in the chiral EFT expansion that are local in coordinate space~\cite{gezerlis_quantum_2013,gezerlis_local_2014,lynn_chiral_2016}. Notably, the authors of Ref.~\cite{yang_chiral_2025} were also able to employ NQS with the phenomenological Argonne $v_8^\prime$ + Urbana IX Hamiltonian~\cite{pudliner_quantum_1995}, which features a strongly repulsive core at short distances.
The charge-independent (CI) part of both the Argonne $v_8^\prime$ interaction and the local N$^2$LO chiral EFT potentials can be written in operator form as
\begin{equation}
v_{ij} = \sum_{p=1}^{8} v^p(r_{ij}) O^p_{ij}\,,
\label{eq:vij_op_sum}
\end{equation}
where $r_{ij} = |{\bf r}_i - {\bf r}_j|$ is the interparticle distance.
For the local N$^2$LO chiral EFT potential, the sum is restricted to the first seven operators.
Compared to Eq.~\eqref{eq:NN_op}, the additional operators entering this expansion include the tensor and spin-orbit components, for a total of eight operators, where $O^{p=5,8}_{ij} = \bigl( S_{ij}, S_{ij}\, \tau_{ij}, \mathbf{L} \cdot \mathbf{S}, \mathbf{L} \cdot \mathbf{S}\, \tau_{ij} \bigr)$. 
The tensor operator is
\begin{equation}
S_{ij}= \frac{3}{r^2_{ij}}(\boldsymbol{\sigma}_i\cdot {\bf r}_{ij})(\boldsymbol{\sigma}_j \cdot{\bf r}_{ij})- \sigma_{ij}\, ,
\end{equation}
while the spin-orbit contribution is expressed in terms of the relative angular momentum $\mathbf{L}=\frac{1}{2 i} (\mathbf{r}_i -\mathbf{r}_j) \times (\nabla_i - \nabla_j)$ and the total spin $\mathbf{S}=\frac{1}{2}(\boldsymbol{\sigma}_i+\boldsymbol{\sigma}_j)$ of the pair. The radial functions of the Argonne $v_8^\prime$ potential are displayed in Fig.~\ref{fig:av8p}.  
In contrast to the model ``o'' interaction of Fig.~\ref{fig:v_NN}, the Argonne potential exhibits a very strong short-range repulsive core, exceeding 2~GeV (in the figure the central channel is divided by a factor of 5 for visibility). The sizable tensor and spin--orbit components, which extend to relatively large distances, are also clearly visible. This long-range behavior is a consequence of one-pion exchange. While the associated tensor interaction is singular at short distances in its bare form, scaling as $1/r^3$, this divergence is regularized in the Argonne interaction. The interaction nevertheless remains long-ranged in the Yukawa sense at large separations, decaying as $e^{-m_\pi r}/r$ up to polynomial factors, and provides the dominant source of tensor correlations at intermediate distances in nuclei. Note that the Coulomb interaction, which is the longest-ranged component of the nuclear Hamiltonian, decays as $1/r$ but is comparatively weak and acts only between protons. 

The earliest example of a $3N$ force dates back to the Fujita--Miyazawa interaction, whose main contributions arise from the virtual excitation of a $\Delta(1232)$ resonance in processes involving three interacting nucleons~\cite{fujita_pion_1957}. This term is included in both chiral EFT and in the Urbana IX potential~\cite{pudliner_quantum_1995} and reads 
\begin{equation}
V_{ijk}^\Delta = V_{ijk}^{\Delta,a} + V_{ijk}^{\Delta,c}\, .
\end{equation}
The ``anticommutator'' and ``commutator'' contributions are  
\begin{equation}
V_{ijk}^{\Delta,a} = \sum_{\rm cyc} 
A_{2\pi}\,\{X_{ij}^\pi, X_{jk}^\pi\}\,\{\tau_{ij}, \tau_{jk}\} \quad , \quad 
V_{ijk}^{\Delta,c} = \sum_{\rm cyc} 
C_{2\pi}\,[X_{ij}^\pi, X_{jk}^\pi]\,[\tau_{ij}, \tau_{jk}]\,,
\end{equation}
where $\sum_{\rm cyc}$ denotes a sum over the three cyclic permutations of the nucleons $i,j,k$.  
The operator $X_{ij}^\pi$ is defined as 
\begin{equation}
X_{ij}^\pi = T(r_{ij})\, S_{ij} + Y(r_{ij})\,\sigma_{ij}\,,
\end{equation}
where the radial Yukawa and tensor functions, $Y(r)$ and $T(r)$, are defined consistently with those appearing in the corresponding $NN$ interaction~\cite{wiringa_accurate_1995,pudliner_quantum_1997}.

Generally, a three-body potential composed only of $V_{ijk}^\Delta$ does not reproduce the empirical saturation density of isospin-symmetric nucleonic matter~\cite{lagaris_variational_1981}; a repulsive three-body contribution must be included,
\begin{equation}
V^{R}_{ijk} = A_R \sum_{\rm cyc} T^2(r_{ij})\, T^2(r_{jk}) \,.
\end{equation}
The constants $A_{2\pi}$ and $A_R$ entering the Urbana IX model are determined by fitting the binding energy of ${}^3$H and the saturation density of isospin-symmetric nucleonic matter, $\rho_0 = 0.16~\mathrm{fm}^{-3}$~\cite{Akmal:1998cf}.  
Chiral EFT interactions likewise include a short-range three-body operator proportional to the low-energy constant $c_E$. This contribution may be purely scalar or contain isospin dependence, although the latter appears to be incompatible with astrophysical constraints.

In addition to the terms above, chiral EFT three-body forces at N$^2$LO contain a one-pion-exchange (OPE) three-body component proportional to the low-energy constant $c_D$, as well as a second short-range contact proportional to $c_E$. The OPE three-body structure is also included in the phenomenological Illinois-7 force~\cite{pieper_illinois_2008}. For detailed discussions of these terms, we refer the reader to Refs.~\cite{lynn_quantum_2017}, while Ref.~\cite{lovato_comparative_2012} provides a comprehensive comparison between local phenomenological and chiral EFT interactions.

\subsection{Progress and challenges of quantum Monte Carlo methods}
\label{sub:qmc_methods}

The Hamiltonians discussed in the previous section are the main input to \emph{ab initio} many-body methods that, within controlled approximations, solve the nuclear Schr\"odinger equation
\begin{equation}
H |\Psi_n\rangle = E_n |\Psi_n\rangle\,,
\end{equation}
where $|\Psi_n\rangle$ denotes the $n$th eigenstate and $E_n$ its associated eigenvalue.
Several stochastic many-body frameworks have been developed to address this problem. 
Two broad and complementary approaches are lattice formulations based on path-integral methods and continuum quantum Monte Carlo techniques.

Nuclear lattice effective field theory (NLEFT)~\cite{epelbaum_ab_2011,lee_lattice_2025}, aided by notable algorithmic advances~\cite{sarkar_floating_2023}, has expanded its reach to medium-mass nuclei~\cite{elhatisari_wavefunction_2024}, provided compelling evidence for $\alpha$ clustering~\cite{shen_emergent_2023}, and explored nuclear and neutron matter at finite temperature~\cite{Ren:2023ued,ma_structure_2024}. 
However, NLEFT continues to face limitations arising from the fermion sign problem and the rapidly growing computational cost associated with fine lattice spacings, which constrain its ability to resolve short-range dynamics unless simplified interactions are used.

Continuum QMC methods~\cite{carlson_quantum_2015,gandolfi2020}, originally developed in condensed-matter physics and later adapted to nuclear systems, operate directly in coordinate space. 
The variational Monte Carlo (VMC) method optimizes a parametrized trial state to capture the most important correlations and enforce the correct asymptotic behavior, thereby accurately describing long-range dynamics and $\alpha$ clustering. 
Building on VMC, Green’s function Monte Carlo (GFMC)~\cite{carlson_greens_1987,carlson_alpha_1988} projects the variational state in imaginary time toward exact eigenstates, while auxiliary-field diffusion Monte Carlo (AFDMC)~\cite{Schmidt:1999lik} employs Hubbard--Stratonovich auxiliary fields to sample spin--isospin operators and access larger systems.  
These continuum approaches model nuclear dynamics across long, intermediate, and short distances and can accommodate ``stiff'' interactions with high-momentum components.  
Recent applications include ground-state properties~\cite{piarulli_light-nuclei_2018}, inclusive electron- and neutrino-scattering~\cite{lovato_ab_2020,andreoli_quantum_2024}, and a variety of electroweak observables such as electromagnetic moments and form factors, low-energy transitions and $\beta$ decays, and muon-capture processes~\cite{king_ab_2023,chambers-wall_quantum_2024}.

In this review, we focus primarily on the VMC method, first outlining the computational challenges of conventional variational states. VMC relies on the Rayleigh–Ritz variational principle
\begin{equation}
\frac{\langle \Psi_V(\params) \, | \, H \, |\,  \Psi_V(\params)\rangle}{\langle \Psi_V(\params) \, | \, \Psi_V(\params)\rangle} \equiv E_V(\params) \geq E_0
\label{eq:variational}
\end{equation}
to determine the optimal set of parameters $\params$ defining the variational state $\Psi_V$. For the nuclear many-body problem, it is customary to assume that the trial state factorizes into long- and short-range components,
\begin{equation}
\Psi_V(R,S) \equiv \langle  R,S \, |\Psi_V\rangle = \langle R S \, | \left(1 + \sum_{i<j<k} F_{ijk}\right)
\left(\mathcal{S}\prod_{i<j} F_{ij}\right) | \Phi \rangle\,,
\label{eq:variational_product}
\end{equation}
where $F_{ij}$ and $F_{ijk}$ are two- and three-body correlation operators, respectively. The symbol $\mathcal{S}$ denotes a symmetrized product over nucleon pairs, since—in general, owing to their spin--isospin dependence—the $F_{ij}$ need not commute. The symbol $R$ denotes the spatial coordinates of all nucleons, $(\mathbf{r}_1, \ \dots \ , \ \mathbf{r}_A )$, while $S$ stands for the spin--isospin degrees of freedom, to be discussed in detail below. 

The long-range antisymmetric part $\Phi$ is commonly expressed as a linear combination of a few Slater determinants built from single-particle orbitals appropriate to the system of interest. For homogeneous matter, the orbitals are usually plane waves and may include pairing correlations~\cite{gandolfi2009a}. For atomic nuclei, the single-particle orbitals are generally taken in the $ls$- or $jj$-coupling schemes and combined to yield the desired total angular momentum $J$ and parity $P$ of the nucleus~\cite{pudliner_quantum_1997}. Importantly, VMC calculations can explicitly incorporate the strong $\alpha$-cluster structure of light nuclei: the wave function of $p$-shell systems is constructed as a sum of independent-particle components, each with four nucleons in an $\alpha$-like core and the remaining $(A-4)$ nucleons occupying $p$-shell orbitals~\cite{pieper_quantum_2002}.

The spin--isospin structure of the two-body correlation operator mirrors that of the $NN$ potential and is written as
\begin{equation}
F_{ij} = \sum_{p=1}^{6} f^{p}(r_{ij}) O^{p}_{ij} .
\end{equation}
Spin-orbit correlations, corresponding to $p=7$, $8$, may also be included, but are often neglected due to the significant computational cost and the relatively small gain in variational energy~\cite{lonardoni_variational_2017}. The optimal radial functions $f^{p}(r)$ are typically obtained by minimizing the two-body cluster contribution to the ground-state energy, subject to the correct asymptotic behavior~\cite{lomnitz-adler_monte_1981}.
Appropriate three-body correlation form factors have been derived within perturbation theory, as discussed in Ref.~\cite{carlson_quantum_2015},
\begin{equation}
F_{ijk} = \sum_x \epsilon_x V_{ijk}^x(\tilde r_{ij}, \tilde r_{ik}, \tilde r_{jk}) ,
\label{eq:three_body_corr}
\end{equation}
where $\tilde r = y_x r$, with $y_x$ a variational scaling parameter and $\epsilon_x$ a small (negative) strength parameter, both optimized variationally. The superscript $x$ labels the different components of the three-nucleon interaction (e.g., $\Delta$, $R$).

\begin{figure}[!t]
    \centering
    \includegraphics[width=0.8\linewidth]{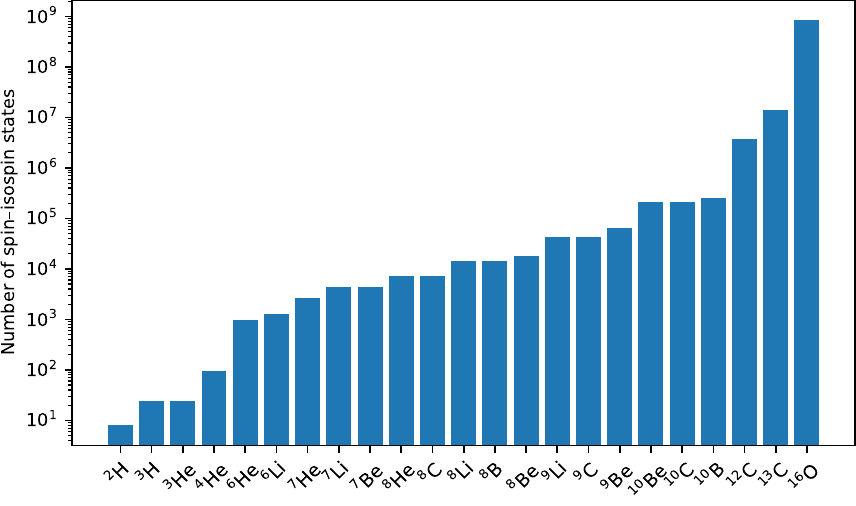}
\caption{Number of many-body spin--isospin states, $2^A \binom{A}{Z}$, relevant to VMC and GFMC calculations for selected light nuclei.\label{fig:gfmc_scaling}}
\end{figure}

Both the VMC and GFMC methods employ an explicit many-body basis for the
spin--isospin sector of the Hilbert space. A generic basis state can be written as $|S\rangle \equiv |\chi_{i_s}\rangle \otimes |\chi_{i_t}\rangle, $
where the indices $i_s$ and $i_t$ enumerate the many-body spin and isospin basis states, respectively. The $2^A$ spin basis states can be written as
\begin{align}
|\chi_1\rangle &= |\downarrow_1, \downarrow_2,\dots,\downarrow_A\rangle \nonumber, \\
|\chi_2\rangle &= |\uparrow_1, \downarrow_2,\dots,\downarrow_A\rangle \nonumber, \\
& \  \vdots \nonumber\\
|\chi_{n_s}\rangle &= |\uparrow_1, \uparrow_2,\dots,\uparrow_A\rangle\,,
\end{align}
with $n_s = 2^A$. The corresponding isospin states $|\chi_{i_t}\rangle$ are
obtained by replacing $\downarrow$ with $n$ and $\uparrow$ with $p$.
For fixed proton number $Z$ (charge conservation), the dimension of the isospin basis is $\binom{A}{Z}$.
Figure~\ref{fig:gfmc_scaling} shows the spin--isospin Hilbert-space dimension for nuclei treated with GFMC to date. Data shown includes $^{13}$C, currently being computed on ALCF’s Aurora supercomputer, and $^{16}$C, anticipated to be within reach of the next generation of leadership-class machines.

The symmetrized product of pair correlation operators is evaluated by successive operations
for each pair, sampling their ordering. As an example, consider the application of the
operators $\sigma_{12} \sigma_{13} \sigma_{23}$ on the three-body spin state $|\uparrow_1, \downarrow_2, \downarrow_3\rangle$. Noting that $\sigma_{ij} = 2P^\sigma_{ij} - 1$, where $2P^\sigma_{ij}$ exchanges the spin of particles $i$ and $j$, we obtain:
\begin{align}
\sigma_{12}\, \sigma_{13}\, \sigma_{23} |\uparrow_1, \uparrow_2, \downarrow_3\rangle & = \sigma_{12}\, \sigma_{13} ( 2 |\uparrow_1, \downarrow_2, \uparrow_3\rangle - |\uparrow_1, \uparrow_2, \downarrow_3\rangle) \nonumber \\
& = \sigma_{12}\, ( 2 |\uparrow_1, \downarrow_2, \uparrow_3\rangle - 2 |\downarrow_1, \uparrow_2, \uparrow_3\rangle + |\uparrow_1, \uparrow_2, \downarrow_3\rangle) \nonumber \\
& = 4 |\downarrow_1, \uparrow_2, \uparrow_3\rangle - 6 |\uparrow_1, \downarrow_2, \uparrow_3\rangle - 2 |\downarrow_1, \uparrow_2, \uparrow_3\rangle + |\uparrow_1, \uparrow_2, \downarrow_3\rangle) \,.
\end{align}
Hence, even starting from a single spin state, the action of the correlation operator generates four of them (even more are generated by the tensor correlations). In general, the the number of operations necessary to calculate the wave function grows exponentially with the number of nucleons, limiting the applicability of the VMC and GFMC methods to $A\leq 13$ nuclei. Sampling the spin--isospin state and evaluating variational wave functions for that sampled state still requires a number of operations  that is exponential in the particle number, bringing little savings in terms of computing time~\cite{carlson_quantum_2015,gandolfi2020}. For this reason, both VMC and GFMC are limited to relatively small mass numbers, up to $A \simeq 13$. 

To circumvent this difficulty, the authors of Ref.~\cite{gandolfi2014} developed a linearized version of the correlator in Eq.~\eqref{eq:variational_product}, in which the spin--isospin-independent correlations are retained in full, while the spin--isospin-dependent ones are kept only to first order:
\begin{equation}
\left(\mathcal{S}\prod_{i<j}\sum_{p=1}^{6} f^{p}(r_{ij})O^{p}_{ij}\right)
\longrightarrow
\left(\prod_{i<j} f^{c}(r_{ij})\right)
\left(1+\sum_{i<j}\sum_{p=1}^{6} u^{p}(r_{ij}),O^{p}_{ij}\right)\,,
\label{eq:variational_linear}
\end{equation}
where $u^{p}(r_{ij}) \equiv f^{p}(r_{ij})/f^{c}(r_{ij})$. Because only linear terms in the spin--isospin-dependent two-body correlations are kept, this wave function is significantly cheaper to evaluate for a single spin--isospin amplitude than the full product of spin--isospin-dependent two-body correlations. This approach scalie as $A^{5}$ rather than exponentially. However, since only pairs of nucleons are correlated at a time, the cluster property is violated. Here, the \emph{cluster property} means that, for two widely separated groups of nucleons the many-body wave function factorizes, $\Psi(A\cup B)\to\Psi(A)\Psi(B)$, so that connected cross-cluster correlations (and mixed contributions to extensive observables) should vanish.

Nevertheless, the use of these linearized spin-dependent correlations has enabled AFDMC calculations of properties of nuclei up to $A\sim 20$ with local chiral EFT interactions. However, the decreasing accuracy with system size prevents the application of AFDMC to medium-mass nuclei, as the method relies on accurate variational wave functions to control the fermion sign problem~\cite{piarulli_benchmark_2020}. To address this issue, the AFDMC variational wave function has been improved by incorporating quadratic pair correlations~\cite{lonardoni_auxiliary_2018}, at a substantially higher computational cost that scales as $A^{7}$. These increased costs, together with residual violations of the cluster property, substantially limit the current reach of AFDMC, with current applications around $A \sim 16$~\cite{novario_trends_2023,martin_auxiliary_2023}.

Extending VMC to larger nuclei calls for variational ans\"atze that capture the key physics, including short- and long-range correlations, clustering, correct asymptotics, and symmetries, while keeping the computational cost polynomial in $A$. This motivates the development of nuclear \emph{neural-network quantum states}, which provide compact, flexible parameterizations of many-body wave functions and can be optimized efficiently within VMC.

\newpage
\section{Neural-network quantum states for nuclear physics}
\label{sec:nqs}

\begin{figure}[!hb]
    \centering
    \includegraphics[width=0.7\linewidth]{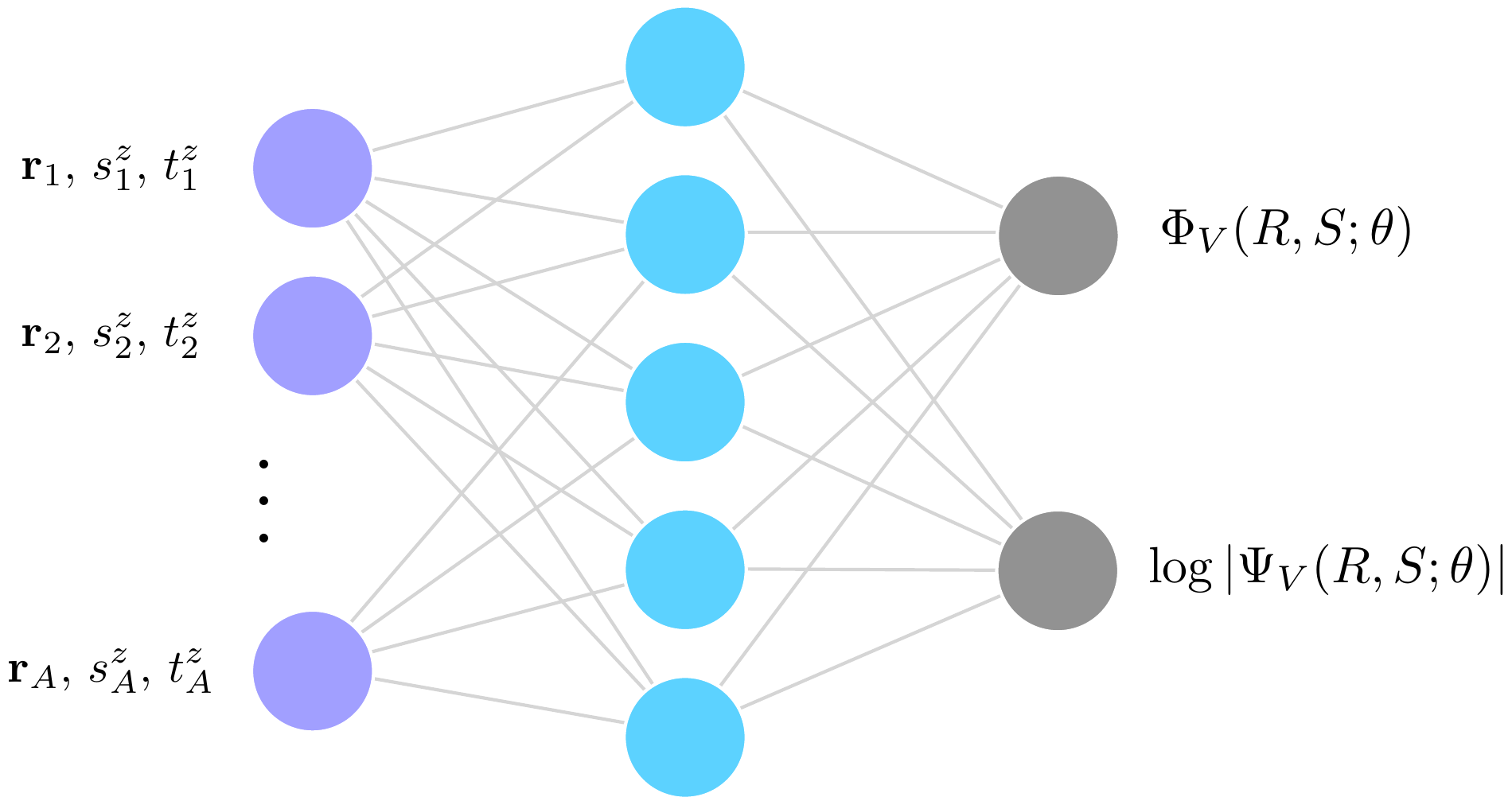}
\caption{Schematic representation of a NQS to solve the nuclear quantum many-body problem. An anti-symmetric artificial neural network takes as input the spatial and spin-isospin coordinates of the  $A$ nucleons, $\{\mathbf{r}_i, s^z_i, t^z_i\}$, and outputs the logarithmic amplitude and phase of the quantum many-body wave function, $\log(\Psi_V(R\, S;\theta)) = \log|\Psi_V(R\, S;\theta)| + i\Phi_V(R\, S;\theta)$
\label{fig:nqs_schematic}}

\end{figure}

In recent years, VMC methods leveraging NQS have been developed to model the wave functions of complex nuclear systems with high accuracy and favorable polynomial scaling with the number of nucleons. Similar to conventional continuum QMC methods, working in coordinate space enables NQS-based approaches to efficiently capture short-range nuclear dynamics. As illustrated schematically in Fig.~\ref{fig:nqs_schematic}, NQS architectures typically take as input the spatial and spin-isospin coordinates of the nucleons and output the amplitude and phase of the quantum many-body wave function. A generic neural-network architecture with this construction will not guarantee fermion anti-symmetry, so it must be explicitly imposed. To this aim, in parallel with condensed-matter and quantum-chemistry applications, different ans\"atze, discussed in detail in Section~\ref{sec:wavefunction}, have been developed. 

In contrast, in second-quantized approaches, fermion antisymmetry is encoded directly by the basis and the operator algebra: states are expanded in the ordered occupation basis, and fermionic signs follow from the canonical anticommutation relations upon reordering operators. Consequently, in second quantization no additional permutation constraint is imposed on the variational wave function. The fermionic signs are determined by the ordered Fock basis and the anticommutation relations, not by any symmetry of the coefficients~\cite{choo_fermionic_2020}. 

\subsection{Sampling the Hilbert space }
Evaluating the variational energy in Eq.~\eqref{eq:variational} requires
carrying out a $3A$-dimensional integral over the continuous Cartesian
coordinates of the nucleons,
$R = (\mathbf{r}_1, \ldots, \mathbf{r}_A)$,
and a sum over the $2^A \binom{A}{Z}$ discrete spin--isospin configurations,
$S=\big((s_1^z, t_1^z), \ldots, (s_A^z, t_A^z)\big),$
with $s_i^z$ and $t_i^z$ denoting the spin and isospin projections of the $i$th nucleon:
\begin{equation}
E_V =
\frac{\langle \Psi_V | H | \Psi_V \rangle}{\langle \Psi_V | \Psi_V\rangle}
= \frac{\sum_{S} \int dR\, \langle \Psi_V | R S\rangle \langle R S | H | \Psi_V\rangle}
{\sum_{S} \int dR\, \langle \Psi_V | R S\rangle \langle R S | \Psi_V\rangle},
\label{eq:variational_2}
\end{equation}
where, for brevity, we have dropped the explicit dependence of the variational state on a set of parameters $\params$. Since a deterministic evaluation is computationally prohibitive, these expressions are estimated stochastically using Monte Carlo sampling. To this end, it is convenient to rewrite the variational energy as
\begin{equation}
E_V
= \frac{\sum_{S} \int dR\, |\Psi_V(R,S)|^2 \dfrac{\langle R S | H | \Psi_V \rangle}
{\langle R S | \Psi_V \rangle}}
{\sum_{S} \int dR\, |\Psi_V(R,S)|^2}
= \sum_{S} \int dR\, \pi_V(R,S)\, E_L(R,S)\,,
\label{eq:en_exp}
\end{equation}
where we introduced the probability distribution and the local energy,
\begin{equation}
\pi_V(R,S) =
\frac{|\Psi_V(R,S)|^2}{\sum_{S} \int dR\, |\Psi_V(R,S)|^2}\,,
\qquad
E_L(R,S) =
\frac{\langle R S | H | \Psi_V \rangle}{\langle R S | \Psi_V \rangle}\,.
\end{equation}
This procedure is unbiased, since configurations with
$\Psi_V(R,S)=0$ do not contribute to Eq.~\eqref{eq:variational_2}, and the
reweighting by $|\Psi_V(R,S)|^2$ is therefore well defined.

In most NQS-based applications of the VMC method to the nuclear many-body
problem~\cite{adams_variational_2021,Gnech2021,yang_consistent_2022,Yang:2023rlw,
gnech_distilling_2024}, with the notable exception of
Ref.~\cite{wen_neural-network_2025}, the Metropolis--Hastings
algorithm~\cite{metropolis_equation_1953,hastings_monte_1970} is used to
generate Markov chain samples of spatial and spin--isospin configurations
distributed according to $\pi_V(R,S)$. The resulting $N_{\rm conf}$
configurations are then used to estimate ground-state expectation values via
\begin{equation}
E_V \simeq \frac{1}{N_{\rm conf}} \sum_{\{R,S\}\sim \pi_V} E_L(R,S)\, .
\end{equation}

The initial $3A$ spatial coordinates of the nucleons are typically sampled from a Gaussian distribution centered at the origin, with per–coordinate variance chosen to reproduce the target rms point radius. For example, taking $R_{\rm pt}\approx 1.2\,A^{1/3}\,\mathrm{fm}$, one may set $\sigma^2 = R_{\rm pt}^2/3$. The $z$-projections of the spin–isospin degrees of freedom are randomly initialized so that the total charge and total spin of the nucleus of interest are reproduced. Each Monte Carlo step consists of a Gaussian move of the $3A$ spatial coordinates. Since charge is conserved, an ergodic isospin move within fixed $Z$ is to swap the isospin labels of a randomly selected pair of nucleons. The same swap is a legitimate move for the spin labels if tensor and spin–orbit contributions in the NN potential are neglected, so that the total spin projection $S_z=\sum_i s_i^z$ is conserved. However, when tensor or spin–orbit interactions are present, one also needs moves that do not conserve $S_z$, such as single or double spin flips, to ensure ergodicity.

The newly proposed spatial configuration $R'$ and spin–isospin projection $S'$ are accepted separately (in two successive Metropolis updates) with probabilities
\begin{equation}
A(R\!\to\! R')=\min\!\left[1,\frac{|\Psi_V(R',S)|^2}{|\Psi_V(R,S)|^2}\right],\qquad
A(S\!\to\! S')=\min\!\left[1,\frac{|\Psi_V(R,S')|^2}{|\Psi_V(R,S)|^2}\right],
\end{equation}
where we have assumed symmetric proposal kernels for the Gaussian spatial moves. The width of the Gaussian ``kick'' is tuned to keep the spatial acceptance rate near $0.6$.

After $N_{\rm therm}$ thermalization steps, which are necessary to equilibrate the Markov chain, the Metropolis–Hastings algorithm produces configurations $\{R,S\}$ distributed according to the target probability $\pi_V(R,S)$. Because successive samples are correlated, $N_{\rm void}$ ``void'' (non-measurement) steps are typically inserted between measurements, and chosen such that the residual autocorrelation falls below a prescribed threshold. After equilibration, a total of $N_{\rm meas}$ measurements are performed. To improve efficiency, the embarrassingly parallel nature of the Metropolis–Hastings algorithm is exploited by propagating $N_{\rm chains}$ independent Markov chains in parallel, so that the total number of configurations used to estimate observables is
$N_{\rm conf}=N_{\rm chains}\times N_{\rm meas}$.

\subsection{Optimization strategies}
\label{subsec:opt}
The energy estimate in Eq.~\eqref{eq:variational} provides an upper bound to the ground-state energy. One must still minimize this quantity with respect to the parameters $\params$, which turns VMC into a nonlinear optimization problem:
\begin{equation}
E^{*}=\min_{\params}\, E_V(\params)\,, \qquad
\params^{*}=\arg\min_{\params}\, E_V(\params)\, .
\end{equation}
A standard strategy is to solve this problem iteratively. A succinct account of such iterative approaches in the NQS setting is given below. For comprehensive treatments in machine learning and optimization (including non-iterative methods), see Refs.~\cite{nocedal_numerical_2006,amari_information_2016}.

In an iterative scheme, one starts from an initial parameter vector $\params$ and updates it as
$\params \leftarrow \params + \boldsymbol{\delta}$, where the increment $\boldsymbol{\delta}$ is chosen to reduce the energy objective. In the NQS setting, a common choice is
\begin{equation}
\boldsymbol{\delta} \;=\; -\,\eta\, S^{-1}\, \mathbf{g} \,,
\label{eq:update_pars}
\end{equation}
which corresponds to minimizing the local quadratic model
\begin{equation}
M(\boldsymbol{\delta}) \;=\; E_V(\params) \;+\; \boldsymbol{\delta}^{\top} \mathbf{g} \;+\; \tfrac{1}{2}\, \boldsymbol{\delta}^{\top} S\, \boldsymbol{\delta}\,.
\label{eq:quadratic_sr}
\end{equation}
The scalar $\eta$ is the learning rate that controls the overall step size in parameter space. The gradient of the variational energy is
\begin{equation}
g_i
= 2 \left(
\frac{\braket{\Psi_V | O_i^\dagger H | \Psi_V}}{\braket{\Psi_V | \Psi_V}}
- \frac{\braket{\Psi_V | O_i^\dagger | \Psi_V}}{\braket{\Psi_V | \Psi_V}}
  \frac{\braket{\Psi_V | H | \Psi_V}}{\braket{\Psi_V | \Psi_V}}
\right)\!,
\label{eq:gradient}
\end{equation}
where $O_i|\Psi_V(\params)\rangle = \partial_{\theta_i} |\Psi_V(\params)\rangle$ denotes a derivative of the wave function with respect to $i-$th parameter. The matrix $S$ serves as a preconditioner, and different optimization schemes prescribe different choices. In the information-geometry paradigm, $S$ accounts for the curvature and geometry of parameter space, beyond the Euclidean (flat-space) gradient descent corresponding to $S=I$~\cite{amari_information_2016}. In quadratic optimization, there is also a close relationship between $S$ and the Hessian of the loss function~\cite{nocedal_numerical_2006}.

An extension of the information geometry approach in many-body quantum mechanics is the \emph{stochastic reconfiguration} (SR) method~\cite{stokes_quantum_2020}. Originally and independently introduced in the QMC literature at the end of the last century~\cite{sorella1998,sorella_wave_2005}, SR translates naturally to the NQS setting. It is commonly justified by minimizing the Fubini–Study distance between an updated NQS state and a small imaginary-time step for energy minimization~\cite{sorella1998,stokes_quantum_2020,park_geometry_2020}. This leads to identifying $S$ with the real part of the quantum geometric tensor (QGT),
\begin{align}
S_{ij}
&= \frac{\braket{\Psi_V| O_i^\dagger O_j | \Psi_V}}{\braket{\Psi_V|\Psi_V}}
 - \frac{\braket{\Psi_V|O_i^\dagger|\Psi_V}}{\braket{\Psi_V|\Psi_V}}
   \frac{\braket{\Psi_V|O_j|\Psi_V}}{\braket{\Psi_V|\Psi_V}}\, .
\end{align}
The QGT acts as a metric under the Fubini–Study distance~\cite{stokes_quantum_2020,park_geometry_2020} and thus encodes the curvature of the manifold of normalized wave functions~\cite{dash_efficiency_2025}. Moreover, for real, non-negative ans\"atze, one can write $\Psi_V=\sqrt{p}$ with $p$ the Born probability density. In this case, the QGT reduces to one quarter of the classical Fisher information matrix of $p$, a central object in information geometry~\cite{amari_information_2016}.

Whatever optimization setting is employed, both formal and numerical issues can affect the update in Eq.~\eqref{eq:update_pars} and thereby hamper convergence. On the formal side, one wants the inverse to be well behaved so as to avoid, for instance, excessively large steps. In many schemes, \(S\) is designed to be symmetric positive semidefinite, but it can be ill conditioned or rank deficient. A common remedy is to add a small diagonal damping term, \(S \to S + \epsilon I\) with \(\epsilon>0\), which makes the system positive definite. This regularization augments the quadratic model in Eq.~\eqref{eq:quadratic_sr} with the term \(\tfrac{\epsilon}{2}\,\|\boldsymbol{\delta}\|_2^{2}\), where \(\|\boldsymbol{\delta}\|_2\) is the Euclidean norm of the parameter update, thereby penalizing large displacements and favoring small-norm updates. However, with this simple regularization, all diagonal elements of the $S$ matrix are shifted by the same amount, neglecting potential order-of-magnitude differences in their typical changes~\cite{sorella_weak_2007}. 

To address this shortcoming, the authors of Ref.~\cite{lovato_hidden-nucleons_2022} introduce an RMSProp-inspired regularization by accumulating an exponentially decaying average of the squared gradients,
\begin{equation}
\mathbf{v}_t = \beta \mathbf{v}_{t-1} + (1 - \beta)\,\mathbf{g}_t^2 ,
\label{eq:sr_rmsprop}
\end{equation}
and modifying the QGT as $S \to S + \epsilon\,\mathrm{diag}\!\left(\sqrt{\mathbf{v}_t}+10^{-8}\right)$. In the original formulation of the SR algorithm, increasing $\epsilon$ suppresses the magnitude of the parameter update and rotates it toward the stochastic-gradient direction. In the RMSProp variant, the regularization instead biases the update toward the RMSProp direction, which typically leads to faster and more stable convergence than simple stochastic gradient descent.

Since the Fisher (QGT) matrix is of size $N_\text{par}\!\times\! N_\text{par}$, its storage scales as $\mathcal{O}(N_\text{par}^2)$ in memory, which quickly becomes infeasible. One way to overcome this limitation is to use iterative solvers, such as conjugate gradient~\cite{neuscamman_optimizing_2012} or MINRES~\cite{drissi_second-order_2024}, to solve the linear system associated with the SR update. Alternatively, Ref.~\cite{chen_empowering_2024} introduces an accelerated scheme, \emph{minSR}, which leverages neural tangent kernel ideas to provide highly efficient inversions. Traditional ML preconditioners like K-FAC have also been employed to approximate $S$ and its inverse~\cite{pfau_ab_2020}, although they were not originally devised for unsupervised settings such as standard NQS simulations~\cite{drissi_second-order_2024}.

SR is the most common approach for NQS optimization and, as noted above, aligns directly with the information geometry paradigm. However, alternative strategies based on different geometric principles can also be employed. Ref.~\cite{drissi_second-order_2024} introduces the use of \emph{decision geometry} within the NQS setting. Decision geometry~\cite{dawid_geometry_2007} reduces to information geometry for specific choices of scoring rules and can therefore be viewed as a game-theoretic generalization of standard probabilistic optimization schemes. An appealing feature of this framework is the flexibility to choose among different (proper) scoring rules, while still obtaining well-behaved updates with positive semidefinite preconditioning metrics~\cite{parry_proper_2012}. A scoring rule tailored to one-dimensional continuous fermion systems showed promising performance in Ref.~\cite{drissi_second-order_2024}, and further explorations for other systems are underway.

Finally, it is worth noting that iterative ML optimizers often exploit the history of the search via a \emph{momentum} term: the update $\delta$ from iteration $k\!-\!1$ is stored so that the update at iteration $k$ depends on both the current gradient and the previous step. Momentum-based schemes used in NQS include variants of RMSProp~\cite{park_geometry_2020,keeble_machine_2020}. By contrast, standard choices widely used in other neural-network contexts (e.g., stochastic gradient descent or Adam~\cite{kingma_adam_2014}) can be insufficient to achieve the percent-level accuracy typically required for ground-state energies in NQS.
A recent example that incorporates step history in a natural-gradient setting is the \emph{Subsampled Projected-Increment Natural Gradient} (SPRING) method~\cite{goldshlager_kaczmarz-inspired_2024}. In this approach, the quadratic model of Eq.~\eqref{eq:quadratic_sr} is augmented with a term $\tfrac{\epsilon}{2}\,\|\boldsymbol{\delta} - \mu\,\boldsymbol{\delta}_{\mathrm{prev}}\|_2^{2}$, where $\boldsymbol{\delta}_{\mathrm{prev}}$ denotes the previous parameter update and $\mu<1$ is a damping factor. This proximal term is centered at $\mu\,\boldsymbol{\delta}_{\mathrm{prev}}$, pulling the new step toward the previous one rather than merely penalizing its magnitude.

\subsection{Basics of deep learning}
\label{sec:deep_learning}
In nuclear physics, the simultaneous treatment of continuous spatial degrees of freedom and discrete spin and isospin degrees of freedom typically requires deep neural architectures composed of multiple multilayer perceptrons (MLPs). These individual MLPs are interconnected in structured ways to enforce the physical symmetries of the quantum many-body system. MLPs are widely recognized as one of the most prominent forms of artificial neural networks (ANN), primarily due to their training simplicity and scalability. However, they can be prone to overfitting in situations where a large data set cannot be made available. VMC-NQS is less prone to this issue, since the number of Monte Carlo samples used to estimate the energy can be increased to improve the effective training set.

A multilayer perceptron is a type of deep feedforward neural network defined by alternating compositions of affine transformations and simple, nonlinear transformations. It consists of an input layer, at least one hidden layer, and an output layer, all of which are densely connected to the adjacent layers only. As the name implies, information flows only in one direction, starting from the input layer and ending at the output layer. For an MLP with $L$ layers (including the hidden and output layers, but excluding the input layer), we define the layers as 
\begin{align}
    \boldsymbol{h}^{(0)} &= \boldsymbol{v} \;\in \mathbb{R}^{d_0},\\
    \boldsymbol{h}^{(\ell)} &= f_\ell \!\left( W^{(\ell)} \boldsymbol{h}^{(\ell-1)} + \boldsymbol{b}^{(\ell)} \right) \; \in \mathbb{R}^{d_\ell}, 
    \hspace{5mm}
    \mathrm{for} \ \ell = 1, \dots, L,
\label{eq:forward-prop}
\end{align}
where $\boldsymbol{v}$ denotes the vector of visible nodes for the input layer $\boldsymbol{h}^{(0)}$; 
$\boldsymbol{h}^{(\ell)}$ are the hidden layers for $\ell = 1, \dots, L-1$; 
$\boldsymbol{h}^{(L)}$ is the output layer; and $f_\ell$ represents the nonlinear activation function in the $\ell$-th layer.
Each hidden layer has dimension $d_\ell$, meaning $W^{(\ell)}$ is a $d_\ell \times d_{\ell-1}$ weight matrix and $\boldsymbol{b}^{(\ell)}$ is a $d_\ell$-dimensional bias vector. 
The activation functions $f_\ell$ introduce nonlinearity to the network, enabling it to learn and approximate complex functions. 
Without at least one nonlinear activation function, the neural network reduces to a purely linear model. 
In NQS applications for continuous-space systems, the activation functions in the hidden layers should be at least twice continuously differentiable to enable the stable computation of the local kinetic energy. 
If the target space is unbounded, the activation function for the output layer $f_L$ is usually chosen to be the identity.

Gradients of MLPs with respect to the weights $W^{(\ell)}$ and biases $\boldsymbol{b}^{(\ell)}$ are efficiently computed through backpropagation, which applies the chain rule recursively:
\begin{align}
    \frac{\partial \boldsymbol{h}^{(L)}}{\partial W^{(\ell)}} 
    = \left( \prod_{k=\ell+1}^L \frac{\partial \boldsymbol{h}^{(k)}}{\partial \boldsymbol{h}^{(k-1)}} \right) \frac{\partial \boldsymbol{h}^{(\ell)}}{\partial W^{(\ell)}}, \ \ \ \ \
    \frac{\partial \boldsymbol{h}^{(L)}}{\partial \boldsymbol{b}^{(\ell)}} 
    = \left( \prod_{k=\ell+1}^L \frac{\partial \boldsymbol{h}^{(k)}}{\partial \boldsymbol{h}^{(k-1)}} \right) \frac{\partial \boldsymbol{h}^{(\ell)}}{\partial \boldsymbol{b}^{(\ell)}}.
\end{align}
In deep architectures, the repeated multiplication of layer Jacobians can cause gradients to vanish or explode, as the eigenvalues of these matrices can be much smaller or larger than unity. Modern architectures mitigate this effect through skip connections, which facilitate gradient flow by providing alternative paths for information propagation. Two common forms are the residual skip connection,
\begin{equation}
    \boldsymbol{h}^{(\ell+1)} = \boldsymbol{h}^{(\ell)} + F_\ell \!\left(\boldsymbol{h}^{(\ell)}\right),
\end{equation}
and the concatenative skip connection,
\begin{equation}
    \boldsymbol{h}^{(\ell+1)} = \mathrm{concat} \!\left(\boldsymbol{h}^{(\ell)}, \  F_\ell \!\left(\boldsymbol{h}^{(\ell)}\right)\right),
\end{equation}
where $F_\ell$ typically denotes a feedforward transformation (e.g.\ an MLP block). The additive, or residual, form preserves dimensionality and is widely used for stabilizing deep networks, while the concatenative form aggregates intermediate features, allowing later layers to access information from all previous representations.

To represent complex wave functions using deep neural networks with real parameters, one possible strategy is to predict the real and imaginary parts of the wave function separately, for instance using two output nodes in the final layer or two distinct networks. However, optimization is typically more stable when the network instead predicts the log-amplitude $\log |\Psi_V|$ and phase $\Phi_V = \arg(\Psi_V)$. In this parameterization,
$\Psi_V = e^{\log |\Psi_V| + i \Phi_V},$
the log-derivative reduces to
$\nabla \log \Psi_V = \nabla \log|\Psi_V| + i\,\nabla \Phi_V$, so that the quantities entering the VMC estimators are directly given by smooth network outputs. This leads to numerically well-behaved gradients and reduces the risk of vanishing or exploding updates during training.

A major advantage of deep neural networks is their flexibility to incorporate physical symmetries directly into the architecture. For systems of identical particles, the wave function must be symmetric under particle exchange for bosons and antisymmetric for fermions. In both cases, it is useful to construct permutation-invariant components so that the network naturally treats particles as indistinguishable, without requiring explicit handling of all possible particle permutations, whose number grows factorially with system size.
The Deep Sets~\cite{zaheer_deep_2017} construction provides a simple way to build such invariant representations,
\begin{equation}
    f(\{ \mathbf{x}_i \}_{i=1}^A) = \rho \!\left( \mathrm{pool}\!\left( \{ \phi(\mathbf{x}_i)\}_{i=1}^A \right) \right),
\end{equation}
where $\mathbf{x}_i=(\mathbf{r}_i, s_i^z, t_i^z)$ denotes the single-particle degrees of freedom of the $i$th nucleon; $\phi$ and $\rho$ are learnable maps; and $\mathrm{pool}$ is a permutation-invariant aggregation operation (e.g., sum, mean, or logsumexp). 
Here, $\phi$ embeds each element of the set into a higher-dimensional latent space, $\mathrm{pool}$ combines all elements into a single latent representation of the set, and $\rho$ maps this representation to the desired output space. 
This construction can be applied similarly to the set of all pairwise degrees of freedom.

\begin{figure}[t]
    \centering
    \includegraphics[width=0.75\textwidth,trim=2.5in 4in 2in 3.7in,clip]{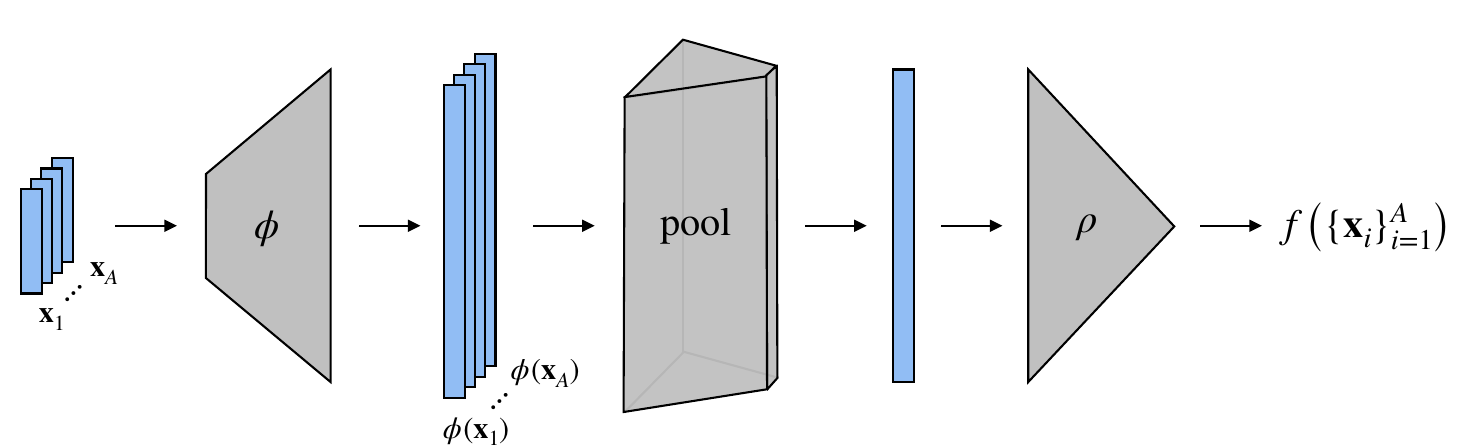}
    \caption{Cartoon of a Deep Set architecture (adapted from Ref.~\cite{fuchs_deepsets_2019})}
    \label{fig:deepset}
\end{figure}

A wide range of architectural components can be incorporated to further enhance the flexibility and expressive power of neural networks. One prominent example is the use of attention mechanisms~\cite{vaswani_attention_2017}, which dynamically weigh the importance of different elements in an input sequence when making predictions.
Instead of processing all input features equally, attention allows the network to focus selectively on the most relevant parts, effectively learning context-dependent weighting. They form the foundation of modern large language models, enabling them to capture long-range dependencies and nuanced contextual relationships, and have recently been incorporated into neural quantum state architectures to enhance the representation of complex correlations in many-body wavefunctions~\cite{von_glehn_self-attention_2022,pescia_message-passing_2024,geier_self-attention_2025}.

The standard formulation of the self-attention mechanism~\cite{vaswani_attention_2017}
computes queries, keys and values for each element of the input through linear
transformations,
\begin{equation}
    Q = X W_Q, \hspace{1cm} 
    K = X W_K, \hspace{1cm} 
    V = X W_V .
\label{eq:attn-weights}
\end{equation}
As a minimal example, consider the input matrix $X \in \mathbb{R}^{A \times 5}$, chosen to represent the row-wise collection of all single-nucleon degrees of freedom, corresponding to the three spatial coordinates together with the spin and isospin labels. One could alternatively construct $X$ to represent pair degrees of freedom or, more generally, the output of a preceding neural network layer.
Choosing the query and key weight matrices $W_Q$ and $W_K$ to have dimension
$5 \times d_k$, the value weight matrix $W_V$ will have dimension
$5 \times d_v$, yielding $Q, K \in \mathbb{R}^{A \times d_k}$ and
$V \in \mathbb{R}^{A \times d_v}$.
The output of the attention layer is obtained by comparing the query of each particle with the keys of all others, producing a set of normalized weights that are then applied to the values to generate a context-aware representation,
\begin{equation}
    \mathrm{attn}(Q, K, V)
    = \mathrm{softmax}_i \!\left( \frac{Q K^T}{\sqrt{d_k}} \right) V .
\end{equation}
The intermediate matrix $QK^T \in \mathbb{R}^{A \times A}$ encodes all pairwise
similarities between particles and is effectively bilinear in $X$. The row-wise
$\mathrm{softmax}_i$ activation function converts each row into a normalized probability
distribution, and the factor of $1/\sqrt{d_k}$ aids numerical stability.

\subsection{Wave function ansätze}
\label{sec:wavefunction}
Here we review the principal wave function ansätze that have been used in ``first-quantized'' NQS approaches to nuclear physics. These architectures act directly on the continuous spatial coordinates and spin–isospin degrees of freedom of the nucleons.

\subsubsection{Slater--Jastrow}
\label{subsub:slater_jastrow}
Initial applications of the NQS framework to atomic nuclei~\cite{adams_variational_2021}
extended the Slater--Jastrow (SJ) architecture, originally developed for quantum
chemistry~\cite{hermann_deep-neural-network_2020,pfau_ab_2020}, to handle the
complex spin--isospin structure of nuclear interactions. Such an ansatz can be
schematically written as
\begin{equation}
\Psi_{\rm SJ}(X) = F(X)\,\Phi(X)\,,
\label{eq:SJ}
\end{equation}
where $F(X)$ is a permutation-invariant Jastrow factor, and antisymmetry is
enforced by the mean-field-like component $\Phi(X)$. 
For compactness, we use $X=(x_1, \ldots, x_A)$ to denote the collection of all single-particle degrees of freedom $x_i = (\mathbf{r}_i, s_i^z, t_i^z)$, namely the three-dimensional spatial coordinates $\mathbf{r}_i$ and the spin and isospin projections along the $z$-axis.

The mean-field term $\Phi(X)$ is typically expressed as a sum of Slater determinants of single-particle orbitals~\cite{adams_variational_2021,gnech_calculation_2020,yang_consistent_2022}. For atomic nuclei, it takes the form
\begin{equation}
\Phi(X)
= \Bigg[\sum_\mu C_\mu \,
\mathcal{A}\!\left[\phi_{\alpha_1^\mu}(x_1)\,\phi_{\alpha_2^\mu}(x_2)\cdots 
\phi_{\alpha_A^\mu}(x_A)\right]\Bigg]_{J^\pi T}\!,
\label{eq:mf}
\end{equation}
where $\mathcal{A}$ is the antisymmetrization operator acting on the particle labels $x_i$, and $\{C_\mu\}$ are configuration amplitudes that incorporate the Clebsch--Gordan couplings required to form a many-body state $\mu$ with the desired total angular momentum, parity, and isospin $J^\pi T$ for the nucleus of interest. The single-particle orbitals are evaluated as
\begin{equation}
\phi_{\alpha}(x_i) 
= \mathcal{R}_{n l}(r_i)\, Y_{l\,l^z}(\hat{r}_i)\,
\langle s_i^z | s^z \rangle \,
\langle t_i^z | t^z \rangle \, ,
\label{eq:single_particle_nuclei}
\end{equation}
where $\alpha = (n, l, l^z, s^z, t^z)$ are the quantum numbers, 
$\langle s_i^z | s^z \rangle = \delta_{s_i^z,s^z}$ and 
$\langle t_i^z | t^z \rangle = \delta_{t_i^z,t^z}$ project onto the fixed spin and isospin, $\mathcal{R}_{nl}(r_i)$ are the 
radial functions parametrized by MLPs, and $Y_{l\,l^z}(\hat{r}_i)$ are the spherical harmonics.

To automatically remove the spurious center-of-mass contribution from the kinetic energy, the spatial coordinates are replaced by $\mathbf{r}_i \to \mathbf{r}_i - \mathbf{R}_{\rm CM}$, with
$\mathbf{R}_{\rm CM} = \frac{1}{A}\sum_{i=1}^A \mathbf{r}_i$
being the center-of-mass coordinate~\cite{massella_exact_2020}.
Since the network parameters are randomly initialized, in the early stages of
training the Metropolis--Hastings walk can let the nucleons drift away from
$\mathbf{R}_{\mathrm{CM}}$. To control this behavior, a Gaussian factor is
multiplied by the single-particle radial functions to confine the nucleons
within a finite volume,
\begin{equation}
\mathcal{R}_{nl}(r_i) \;\to\; \mathcal{R}_{nl}(r_i)\, e^{-\alpha r_i^2}\,.
\end{equation}
Typical values for the confining constant are in the range
$\alpha = 0.02$--$0.05~\text{fm}^{-2}$.

In Refs.~\cite{adams_variational_2021,gnech_calculation_2020,yang_consistent_2022},
the real-valued Jastrow correlator was taken to be of the form
\begin{align}
F(X) = e^{\mathcal{U}(X)}\,\tanh\!\big[\mathcal{V}(X)\big]\,.
\label{eq:jastrow_tanh}
\end{align}
Here, the positive-definite exponential is modulated by a hyperbolic tangent, which acts as a smooth surrogate for the sign. The functions $\mathcal{U}(X)$ and $\mathcal{V}(X)$ are implemented as permutation-invariant neural networks. Several architectures have been developed to represent such functions efficiently, including point-cloud~\cite{charles_pointnet_2017} and attention-based~\cite{vinyals_pointer_2015,vaswani_attention_2017,lee_set_2018}
models. Among these, the authors of Ref.~\cite{adams_variational_2021} found the Deep Sets architecture~\cite{zaheer_deep_2017,wagstaff_limitations_2019} to be a practical and sufficiently accurate choice, discussed further in Section~\ref{sec:deep_learning}.

In particular, each single-particle input $x_i$ is mapped independently to a latent representation, and a sum aggregation is applied to enforce permutation invariance:
\begin{equation}
\mathcal{U}(X)
= \rho_{\mathcal U}\!\left(\sum_{i=1}^A \phi_{\mathcal U}(x_i)\right),
\qquad
\mathcal{V}(X)
= \rho_{\mathcal V}\!\left(\sum_{i=1}^A \phi_{\mathcal V}(x_i)\right).
\label{eq:deep_sets_single}
\end{equation}
Both $\phi_{\mathcal U}$ and $\rho_{\mathcal U}$ (and analogously $\phi_{\mathcal V}$ and $\rho_{\mathcal V}$) are typically implemented as MLPs. Computing the kinetic energy (Laplacian) requires activation functions that are differentiable (ideally twice). Standard choices include hyperbolic tangent, softplus~\cite{dugas_incorporating_2001}, and GELU~\cite{hendrycks_gaussian_2023}.

Instead of single-particle inputs as in Ref.~\cite{adams_sensitivity_2021}, Ref.~\cite{gnech_calculation_2020} maps the coordinates of each particle pair directly to a latent-space representation, in close analogy with earlier condensed-matter applications~\cite{pfau_ab_2020,hermann_deep-neural-network_2020}. As before, a sum aggregation removes any dependence on particle ordering and therefore enforces permutation invariance:
\begin{equation}
\mathcal{U}(X)
= \rho_{\mathcal U}\!\left(\sum_{i \neq j}
\phi_{\mathcal U}(x_i, x_j)\right),
\qquad
\mathcal{V}(X)
= \rho_{\mathcal V}\!\left(\sum_{i \neq j}
\phi_{\mathcal V}(x_i, x_j)\right).
\label{eq:deep_sets_pair}
\end{equation}
When single-particle coordinates are used as input, correlations are generated
in the latent feature space by the aggregation network $\rho_{\mathcal F}$, where we use $\mathcal{F}\in\{\mathcal{U},\mathcal{V}\}$ to
denote either of the two correlator networks. By
contrast, using pair coordinates introduces correlations already at the level of
the real-space inputs. The conventional two-body Jastrow ansatz is recovered
when the latent dimension is one and $\rho_{\mathcal F}$ is taken to be the
identity, $\rho_{\mathcal F}(x)=x$.
In a similar spirit, Ref.~\cite{yang_consistent_2022} feeds only rotationally
invariant one- and two-body features to $\phi_{\mathcal U}$ and
$\phi_{\mathcal V}$, namely $ \|\mathbf r_i \|$, $ \|\mathbf r_j \|$, and
$\| \mathbf r_i-\mathbf r_j \|$, while neglecting spin--isospin dependence. This
choice significantly reduces the number of variational parameters without
compromising the accuracy of the resulting energies.

\subsubsection{Operator-dependent Slater--Jastrow}
\label{sec:chiral_ansatz}

Very recently, VMC methods based on NQS~\cite{yang_chiral_2025,yang_zemach_2025,wen_neural-network_2025} have been applied to the nuclear many-body problem using high-resolution phenomenological and chiral-EFT Hamiltonians that are local in coordinate space. To address the strongly non-perturbative character of these interactions, in particular their tensor component, the NQS includes explicit spin–isospin operator dependence, as in Eq.~\eqref{eq:variational_product}. To keep a polynomial computational cost  in the number of nucleons, the ansatz introduced in Ref.~\cite{yang_chiral_2025} is expressed in the same linearized form used in AFDMC, reported in Eq.~\eqref{eq:variational_linear}, namely
\begin{equation}
\Psi_V(R,S) = e^\mathcal{U}(R) \tanh[\mathcal{V}(R)]
\left( 1 + \sum_{i<j}\sum_{p=1}^6 W^p_{ij}(R) O^p_{ij} \right).
\label{eq:SJ_chiral}
\end{equation}
As a reminder, the first six spin-isospin operators are $O^{p=1,6}_{ij} = \bigl( 1, \tau_{ij}, \sigma_{ij}, \sigma_{ij} \tau_{ij}, S_{ij}, S_{ij} \tau_{ij}\bigr)$. The central correlation functions $\mathcal{U}(R)$ and $\mathcal{V}(R)$ are represented by the spin–isospin independent permutation-invariant neural network of Eq.~\eqref{eq:deep_sets_pair}, introduced in Ref.~\cite{yang_consistent_2022}. Importantly, this construction ensures both translational and rotational invariance. 

Contrary to conventional GFMC and AFDMC trial wave functions, in which the pair
correlation functions $W^p_{ij}$ depend only on the relative coordinate of
particles $i$ and $j$, the present ansatz makes $W^p_{ij}$ a functional of the
full configuration $R$. This is achieved using the Deep Sets architecture, discussed above and in Section~\ref{sec:deep_learning}, which incorporates information from all particles through
\begin{equation}
W^p_{ij}(R) =
\rho_{\mathcal{W}}^p\!\left(
\sum_{k=1}^A \phi_{\mathcal{W}}(r_{ik}, r_{jk}, r_{ij})
\right),
\label{eq:SJ_chiral_2}
\end{equation}
where $\phi_{\mathcal{W}}$ and $\rho_{\mathcal{W}}$ are feedforward neural networks, each with a single fully connected hidden layer. This global dependence allows the pair correlations to adapt to the surrounding many-body environment. As a result, these wave functions provide an excellent starting point for GFMC calculations and were used in Ref.~\cite{yang_chiral_2025} to compute peripheral neutron--$\alpha$ scattering with high accuracy. A more detailed discussion of these results is given in Section~\ref{sec:applications}.

A similar ansatz, but incorporating the \emph{symmetrized product of two-body correlations}---see Eq.~\eqref{eq:variational_product} for its conventional counterpart---has been proposed by the authors of Ref.~\cite{wen_neural-network_2025}:
\begin{equation}
\Psi_V(R,S) =
\left( 1 + \sum_{i<j<k} F_{ijk} \right)
\prod_{i<j} f^c_{ij}(R)
\left( 1 + \sum_{p=2}^6 u_{ij}^p(R)\, O_{ij}^p \right).
\label{eq:wen_ansatz}
\end{equation}
The three-body correlation functions $F_{ijk}$ are defined analogously to
Eq.~\eqref{eq:three_body_corr}. As in the ansatz of Ref.~\cite{wen_neural-network_2025},
\emph{all} radial functions entering the two- and three-body correlations depend on the full
configuration $R$. A representative example is the operator-dependent correlation,
\begin{equation}
u_{ij}^p(R) =
\rho_u^p\!\left(
r_{ij},\,
\sum_{k\neq i,j}
\left[\, \phi_u(r_{ij}) + \phi_u(r_{ik}) \,\right]
\right),
\label{eq:wen_u_corr}
\end{equation}
where both $\rho_u^p$ and $\phi_u$ are feed-forward neural networks composed of multiple one-dimensional transformation layers connected by residual blocks and employing the RiLU activation function.

An important difference from previous NQS implementations is that this approach leverages normalizing flows to produce uncorrelated samples to solve the nuclear Schr\"odinger equation, as opposed to Metropolis–Hastings sampling. As a consequence, it avoids the autocorrelation issues inherent in Metropolis–Hastings and reduces the overall computational cost.
Normalizing flows~\cite{albergo_flow-based_2019,brady_normalizing_2021,wen_application_2024} generate samples by drawing an initial batch from a simple uniform distribution, implemented here using a Sobol quasi-random sequence, and passing it through a series of invertible change-of-variable transformations with tractable derivatives and Jacobians. By optimizing the parameters of the flow, the resulting distribution can be made to closely approximate the target distribution.
Results for the $A=3$ nuclei obtained with this ansatz and the local chiral-EFT interactions of Ref.~\cite{gezerlis_local_2014} will also be summarized in Section~\ref{sec:applications}.

\subsubsection{Hidden Nucleons}
\label{subsub:hidden_nucleons}
A systematically improvable family of variational wave functions for strongly correlated fermionic systems was introduced in Ref.~\cite{moreno_fermionic_2022}. These wave functions are built from Slater determinants in an augmented Hilbert space that includes additional, ``hidden'' fermionic degrees of freedom. The authors demonstrated that this ansatz is universal on the lattice and applied it to the ground-state properties of the Hubbard model on the square lattice, achieving accuracies competitive with state-of-the-art variational methods.
This framework was subsequently extended to nuclear systems, whose Hilbert spaces involve both continuous and discrete degrees of freedom~\cite{lovato_hidden-nucleons_2022}. In this formulation, the Hilbert space includes fictitious coordinates for $A_h$ Hidden Nucleons (HN), defined as functions of the visible coordinates $X_h = f(X)$. The amplitudes of the HN wave function in the $X$ basis can be written schematically as
\begin{equation} 
\Psi_{\rm HN}(X)\equiv \det 
\begin{bmatrix} \Phi_v(X) & \Phi_v(X_h)\\ \chi_h(X) & \chi_h(X_h)\\ \end{bmatrix} ,
\label{eq:HN} 
\end{equation}
where $\Phi_v(X)$ denotes the $A\times A$ matrix of visible single-particle orbitals evaluated at the visible coordinates, equivalent to the mean-field term $\Phi(X)$ of Eq.~\eqref{eq:mf}. This would be the only component of the wave function in a Hartree–Fock description. In contrast to the original implementation of Ref.~\cite{moreno_fermionic_2022}, the columns of the matrix in Eq.~\eqref{eq:HN} correspond to different particles, while the rows correspond to different single-particle states. The $A_h\times A_h$ matrix $\chi_h(X_h)$ contains the amplitudes of hidden orbitals evaluated at hidden coordinates, whereas $\chi_h(X)$ and $\Phi_v(X_h)$ provide the amplitudes of hidden orbitals at visible coordinates and visible orbitals at hidden coordinates, respectively.
The SJ ansatz of Eq.~\eqref{eq:SJ} is recovered in the limit $A_h=1$, $\chi_h(X)=0$, and $\Phi_v(X_h)=0$, allowing one to interpret it as a special case of the HN formulation. If the function $f$ is permutation invariant, $\Psi_{\rm HN}(X)$ is automatically antisymmetric under particle exchange. The expressivity of this construction for discrete degrees of freedom was formally proven in Ref.~\cite{moreno_fermionic_2022}, provided that the functions $\chi_h$ and $f$ are sufficiently general. To avoid the combinatorial complexity of $f$, the $i$-th columns of $\Phi_v(X_h)$ and $\chi_h(X_h)$ are parametrized by independent, permutation-invariant neural networks as
\begin{equation}
\Phi_v^i(X_h) = e^{\mathcal{U}^i_\Phi(X)}\tanh[\mathcal{V}^i_\Phi(X)], \qquad
\chi_h^i(X_h) = e^{\mathcal{U}^i_\chi(X)}\tanh[\mathcal{V}^i_\chi(X)].
\end{equation}
A more recent implementations of the HN ansatz~\cite{gnech_distilling_2024} employs complex-valued matrices of the form
\begin{equation}
\Phi_v^i(X_h) = e^{\mathcal{U}^i_\Phi(X) + i \mathcal{V}^i_\Phi(X)}, \qquad
\chi_h^i(X_h) = e^{\mathcal{U}^i_\chi(X) + i \mathcal{V}^i_\chi(X)}.
\end{equation}
In both formulations, permutation invariance is enforced using Deep Sets architectures~\cite{zaheer_deep_2017,wagstaff_limitations_2019} to express the functions $\mathcal{U}^i_\Phi$, $\mathcal{V}^i_\Phi$, $\mathcal{U}^i_\chi$, and $\mathcal{V}^i_\chi$. For applications to finite nuclei~\cite{lovato_hidden-nucleons_2022} and dilute neutron matter~\cite{fore2023}, a $\mathrm{logsumexp}$ pooling operation replaces the simple sums in Eqs.~\eqref{eq:deep_sets_single} and~\eqref{eq:deep_sets_pair},
\begin{equation}
\mathcal{F}(X)= \rho_\mathcal{F}\left[\log\Big(\sum_i e^{\phi_\mathcal{F}(x_i)}\Big)\right]\,.
\end{equation}
In these works, $\phi_\mathcal{F}$ and $\rho_\mathcal{F}$ are implemented as MLPs with two hidden layers of 16 nodes each, and a 16-dimensional latent space connecting them. The output layers of $\rho_\mathcal{F}$ contain $A$ nodes for $\mathcal{F}=\mathcal{U}^i_\Phi,\mathcal{V}^i_\Phi$ and $A_h$ nodes for $\mathcal{F}=\mathcal{U}^i_\chi,\mathcal{V}^i_\chi$. An ablation study in Ref.~\cite{lovato_hidden-nucleons_2022} examined the convergence of the $^4$He ground-state energy with network size and found that increasing the number of hidden layers or nodes beyond this configuration provided no substantial improvement, while smaller networks slightly degraded the accuracy (by about 0.2 MeV).
The single-particle orbitals defining $\Phi_v(X)$ and $\chi_h(X)$ are also represented by MLPs that take as input the single-particle coordinates $x_i$. In Refs.~\cite{lovato_hidden-nucleons_2022,fore2023}, these networks had two hidden layers with ten nodes each, $\tanh$ activation functions, and one-dimensional linear outputs. Other differentiable activations such as $\mathrm{softplus}$ and $\mathrm{GELU}$ were tested without appreciable differences. Differentiability is required to compute the kinetic energy, which involves second derivatives of the wave function.
Unlike in the Slater–Jastrow case, confinement is not imposed at the single-particle level. Instead, the full wave function in Eq.~\eqref{eq:HN} is multiplied by a global Gaussian factor,
\begin{equation}
\Psi_{\rm HN} (X) \to \Psi_{\rm HN} (X), e^{-\alpha \sum_{i=1}^A r_i^2}\,,
\end{equation}
which ensures spatial localization of the system.

When introducing the HN ansatz, the authors of Ref.~\cite{lovato_hidden-nucleons_2022} found it beneficial
to enforce point symmetries, such as parity and time reversal. For positive- and negative-parity states, one can impose
\begin{align}
\Psi_{\rm HN}^{P^+}(X) &= \Psi_{\rm HN}(R,S) + \Psi_{\rm HN}(-R,S)\,, \nonumber\\
\Psi_{\rm HN}^{P^-}(X) &= \Psi_{\rm HN}(R,S) - \Psi_{\rm HN}(-R,S)\,.
\label{eq:psi_P}
\end{align}
For even-even nuclei, such as $^4$He and $^{16}$O, it is also possible to enforce time-reversal symmetry,
\begin{align}
\Psi_{\rm HN}^{T}(X) \equiv \Psi_{\rm HN}(R,S) + \Psi_{\rm HN}^{*}(R,\theta S)\,.
\label{eq:psi_PT}
\end{align}
where $\theta S$ is obtained by applying the operator $-i\sigma_y$ to all single-particle spinors~\cite{bohr_nuclear_1998}.
Note that, unlike the expression reported in Ref.~\cite{lovato_hidden-nucleons_2022}, we explicitly take the complex conjugate, so that the construction applies to both real- and complex-valued NQS. Importantly, time reversal and parity have been combined, to form a simultaneous eigenstate of parity and time reversal, dubbed $\Psi_{\rm HN}^{PT}(X)$. 

\begin{figure}[!tb]
\begin{center}
    \raisebox{0.4mm}{\includegraphics[width=0.49\textwidth, height=0.325\textwidth]{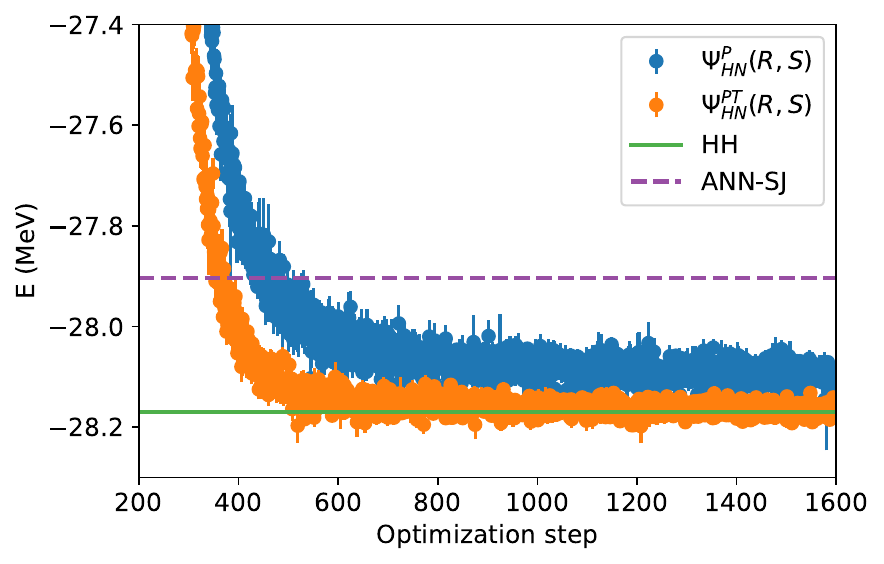}} 
    \includegraphics[width=0.49\textwidth]{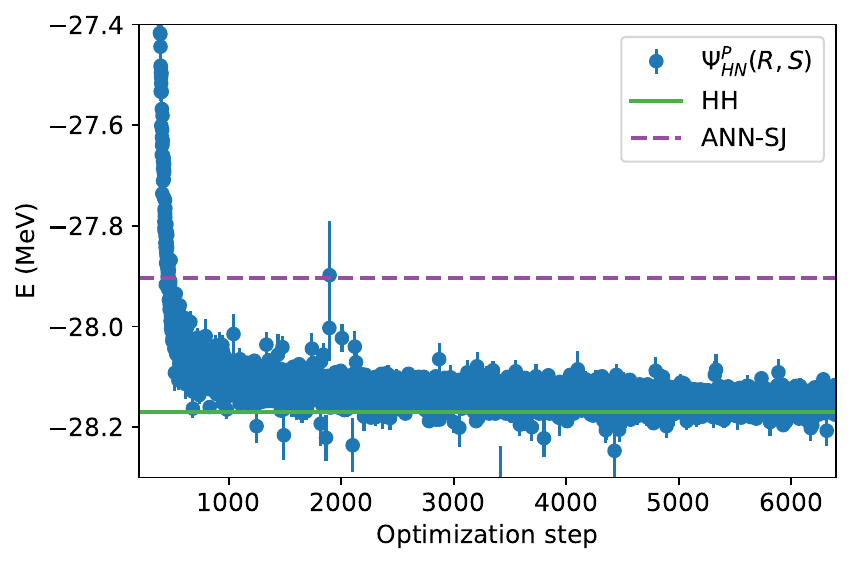} 
    \caption{From Ref.~\cite{lovato_hidden-nucleons_2022} with permission from the Authors. Left panel: Convergence of the $^4$He ground-state energy obtained with the parity-projected HN ansatz (blue solid circles) and with the ansatz that enforces both parity and time-reversal symmetry (orange solid circles). The SJ and the hyper-spherical harmonics ground-state energies from Ref.~\cite{gnech_calculation_2020} are shown by the purple dashed and solid green lines, respectively.
    Right panel: Convergence of the parity-projected ansatz when employing a wider neural-network architecture than in the left panel.}
    \label{fig:4He}
\end{center}
\end{figure}

The convergence of the $^4$He ground-state energy computed with $A_h = 4$ HN is displayed in Fig.~\ref{fig:4He}. The parity-conserving wave function $\Psi_{HN}^{P}(R,S)$ is outperformed by $\Psi_{HN}^{PT}(R,S)$, which additionally preserves time-reversal symmetry. Both provide significantly better energies than the SJ results of Ref.~\cite{Gnech2021}, as they can improve the nodal surface of the single-particle Slater determinant. In fact, $\Psi_{\rm HN}^{PT}(R,S)$ yields a variational energy consistent with the numerically exact HH estimate of Ref.~\cite{Gnech2021}. 

It is worth noting that $\Psi_{HN}^{P}(R,S)$ should, in principle, converge to the exact energy, but this may require wider (or deeper) architectures.
To illustrate this point, the right panel of Fig.~\ref{fig:4He} displays the training of $\Psi_{HN}^{P}(R,S)$ with $A_h=4$, in which the number of nodes in the hidden layers of $\phi_\mathcal{F}$ and $\rho_\mathcal{F}$ has been increased from $16$ to $24$. After about $4800$ optimization steps, the parity-conserving ansatz yields energies that are consistent with the HH method. 
Importantly, though,
enforcing time-reversal symmetry in the ansatz is effective in reducing the training time and appears to enhance the expressivity of the HN ANN architecture.

\subsubsection{Pfaffian--Jastrow}
\label{subsub:pfaffian_jastrow}

Pfaffian--Jastrow neural architectures were originally introduced to describe the unitary Fermi gas~\cite{kim_neural-network_2024}, where strong pairing correlations dominate. The same pairing-based formulation was later shown to be advantageous for nuclei and dilute neutron-star matter~\cite{fore_investigating_2024}. These neural architectures build on a long history of pairing wave functions in quantum Monte Carlo studies of ultracold Fermi gases, in which the many-body state is written as an antisymmetrized product of spin-singlet pairs~\cite{gezerlis_strongly_2008,carlson_superfluid_2003,chang_quantum_2004,gezerlis_heavy-light_2009}. These can be thought of as Bardeen--Cooper--Schrieffer (BCS) pairs within the nuclear context.

In the literature, this class of pairing states appears under several closely related names, including geminal wave functions~\cite{casula_geminal_2003,casula_correlated_2004}, number-projected BCS~\cite{galea_diffusion_2016}, and singlet-pairing wave functions~\cite{bajdich_pfaffian_2008}. These ansätze already provide a substantial improvement over single-determinant Slater states. However, their effectiveness is reduced in partially spin-polarized systems, since only the spin-singlet pairing channel is included~\cite{casula_correlated_2004}. This limitation motivated the introduction of the singlet–triplet–unpaired (STU) Pfaffian wave function~\cite{bajdich_pfaffian_2006,bajdich_pfaffian_2008}, which generalizes the geminal form by explicitly incorporating both singlet and triplet pairing channels, together with a separate sector for unpaired particles. In this formulation, the many-body wave function is written as the Pfaffian of a block matrix collecting singlet, triplet, and unpaired contributions. The original geminal ansatz is recovered as a special case when the triplet blocks are set to zero.

A key limitation of both geminal and STU wave functions in their original formulations is that they assume a fixed spin ordering of the fermions. This is perfectly adequate for model Hamiltonians with conserved spin species, but it is not well suited to nuclear interactions, which contain explicit spin-exchange terms and couple different spin–isospin channels~\cite{piarulli_local_2020}. In neutron-matter applications, this issue is commonly circumvented by expressing the Pfaffian pairing orbital as a plane-wave expansion weighted by BCS amplitudes and multiplied by a spin-singlet component~\cite{Gandolfi2009}:
\begin{equation}
\phi(x_i, x_j)
= \sum_{\alpha} c_{\alpha} \,
e^{i \mathbf{k}_{\alpha} \cdot (\mathbf{r}_i - \mathbf{r}_j)}\,\left(
\frac{\eta_\uparrow(s_i^z)\eta_\downarrow(s_j^z)
      - \eta_\downarrow(s_i^z)\eta_\uparrow(s_j^z)}{\sqrt{2}}\right) \, ,
\label{eq:pf_pairing_orbital_nm}
\end{equation}
where $x_i = (\mathbf{r}_i, s_i^z)$,
$\mathbf{k}_{\alpha} = \tfrac{2\pi}{L}(n_x \hat{\mathbf{x}} + n_y \hat{\mathbf{y}} + n_z \hat{\mathbf{z}})$
denotes the momenta compatible with the periodic boundary conditions, and the complex-valued $c_{\alpha}$ are variational parameters. While in neutron matter the spin-triplet channel is typically neglected, it can be incorporated within the same framework without requiring fixed spin ordering.

By introducing neural network representations of the pairing orbital, the Pfaffian can be employed in its most general form, capturing arbitrary spin–isospin–dependent correlations without relying on fixed assumptions about the pairing channel.
For an even number of nucleons $A$, the Pfaffian--Jastrow wave function is written as
\begin{equation}
\Psi_{PJ}^{\mathrm{even}}(X) = e^{J(X)}\, \mathrm{pf}[P(X)]\,,
\label{eq:pf_even}
\end{equation}
where $J(X)$ is a Jastrow factor and $P(X)$ is an $A \times A$ skew-symmetric matrix whose elements define the pairing orbital between any two nucleons.
The antisymmetric structure of the wave function is thus fully encoded in the pairing matrix $P(X)$, given by~\cite{kim_neural-network_2024,fore_investigating_2024}
\begin{equation}
P(X) =
\begin{bmatrix}
0 & \phi(x_1,x_2) & \phi(x_1,x_3) & \cdots & \phi(x_1,x_A) \\
-\phi(x_1,x_2) & 0 & \phi(x_2,x_3) & \cdots & \phi(x_2,x_A) \\
-\phi(x_1,x_3) & -\phi(x_2,x_3) & 0 & \cdots & \phi(x_3,x_A) \\
\vdots & \vdots & \vdots & \ddots & \vdots \\
-\phi(x_1,x_A) & -\phi(x_2,x_A) & -\phi(x_3,x_A) & \cdots & 0
\end{bmatrix} \, .
\end{equation}
Similarly to the most recent implementations of the HN ansatz, the Jastrow factor is complex-valued~\cite{fore_investigating_2024} and can be written as
\begin{align}
F(X) = e^{\left[\mathcal{U}(X) + i\mathcal{V}(X)\right]}\,,
\label{eq:jastrow_complex}
\end{align}
where $\mathcal{U}$ and $\mathcal{V}$ are real functions that encode two-body correlations in the modulus and phase of the wave function, respectively.
As in the real-valued case, the pairwise Deep-Sets construction of Eq.~\eqref{eq:deep_sets_pair} is typically employed, so that two-body correlations are built directly in coordinate space rather than in the latent feature space.

Alternatively, the pairing orbital can be taken to be a sum over products of single-particle orbitals,
\begin{equation}
\phi(x_i, x_j)
= \sum_{\alpha, \beta} c_{\alpha\beta} \, \phi_{\alpha}^*(x_i)\, \phi_{\beta}(x_j)\,,
\label{eq:pf_pairing_orbital_closed}
\end{equation}
where the single-particle orbitals $\phi_{\alpha}(x)$ are those defined in Eq.~\eqref{eq:single_particle_nuclei}. This strategy has also been adopted in recent condensed-matter applications of the Pfaffian family of wave functions~\cite{gao_neural_2024}, where the elements of the antisymmetric matrix $c_{\alpha\beta}$ are treated as variational parameters. For a closed-shell system, the Slater–Jastrow ansatz is recovered by restricting the sum to the occupied orbitals and choosing $c_{\alpha\beta}$ to be antisymmetric and nonzero only between time-reversed partners $(\alpha, \bar{\alpha})$, e.g. $c_{\alpha \bar{\alpha}} = 1$,  $c_{\bar{\alpha} \alpha} = -1$, with all other $c_{\alpha\beta}=0$. This choice forms antisymmetric pairs of conjugate single-particle states (the usual time-reversed pairs of a closed shell), and in this limit the Pfaffian reduces to the determinant built from the same one-body orbitals, thus recovering the SJ form as a special case.

In contrast to these orbital-product expansions, and to both the SJ and HN ansätze, one can require only a single trainable pair orbital $\phi$, so the number of variational parameters does not grow with the number of particles $A$. Skew-symmetry is enforced at the level of the orbital itself by parameterizing $\phi$ through a neural network $\nu$,
\begin{equation}
\phi(x_i, x_j)
= \nu(x_i, x_j) - \nu(x_j, x_i)\,,
\end{equation}
so that $\phi(x_i, x_j) = - \phi(x_j, x_i)$ holds identically. In practice~\cite{kim_neural-network_2024,fore_investigating_2024}, $\nu$ is implemented as an MLP acting on the concatenated features of the two nucleons, which allows the orbital to depend simultaneously on relative position and on spin–isospin quantum numbers, thereby capturing correlations beyond pure pairing.

For systems with an odd number of nucleons, the ansatz is extended by adding a single unpaired orbital $\psi$, parameterized by a separate MLP. In this case, the Pfaffian is taken over an enlarged $(A+1)\times(A+1)$ skew-symmetric matrix:
\begin{equation}
\Psi_{PJ}^{\mathrm{odd}}(X)
= e^{J(X)}
\, \mathrm{pf}
\begin{bmatrix}
P(X) & \mathbf{u}(X) \\
-\mathbf{u}(X)^{T} & 0
\end{bmatrix},
\label{eq:pf_odd}
\end{equation}
where
\begin{equation}
\mathbf{u}(X) =
\begin{bmatrix}
\psi(x_1) \ \psi(x_2) \ \cdots \ \psi(x_A)
\end{bmatrix}^T
\end{equation}
collects the values of the unpaired orbital on all single-particle degrees of freedom. This construction preserves antisymmetry while allowing Pfaffian-based NQS to describe odd-mass nuclei or systems with a blocked quasiparticle.

Although the Pfaffian ansatz for odd $A$ introduces an additional, unpaired orbital relative to the even-$A$ case, the overall number of trainable parameters in both formulations remains independent of the particle number. Moreover, because the even- and odd-$A$ constructions share the same underlying pairing orbital, transfer-learning strategies can be straightforwardly employed between neighboring systems. This favorable scaling has enabled simulations of dilute, isospin-asymmetric neutron-star matter with up to $A=42$ nucleons under periodic boundary conditions~\cite{fore_investigating_2024}. Owing to this efficiency, the Pfaffian ansatz is emerging as a particularly effective tool for larger systems, while benchmark studies continue to compare its performance with the HN ansatz, which is known to provide a universal approximation to antisymmetric wave functions~\cite{moreno_fermionic_2022}.

In neutron matter, pairing is an essential ingredient, whereas in finite nuclei it may or may not be dominant, depending on shell structure and interaction strength~\cite{dean_pairing_2003}. Nevertheless, the Pfaffian formulation remains attractive because (i) for an appropriate choice of pairing orbital it collapses to a single Slater determinant; (ii) the orbital depends simultaneously on spatial and spin–isospin coordinates, allowing it to capture independent-pair correlations beyond pure pairing; (iii) it yields a compact description of open-shell systems without resorting to linear combinations of many Slater determinants; and (iv) its cost scales favorably with system size, since only a single pairing orbital is required regardless of $A$.

\subsubsection{Backflow correlations}
\label{subsub:backflow_correlations}

Backflow correlations encode how the state of a single particle is influenced by the coordinates of all the others, effectively redefining its coordinates according to the many-body configuration,
$x_i \mapsto g_i(x_i; \{ x_j\}_{j \neq i})$.
This idea dates back to Ref.~\cite{feynman_energy_1956} and has since become a widely adopted strategy for enhancing the expressivity of antisymmetric NQS~\cite{luo_backflow_2019,hermann_deep-neural-network_2020,pfau_ab_2020}.
A key requirement is permutation equivariance: permuting the particles in the input must produce the same permutation in the output, $g_i \mapsto g_j$ if $x_i \mapsto x_j$, so as to preserve the antisymmetry of the fermionic wave function.

A relatively simple backflow transformation has been recently incorporated into the HN framework and applied to compute properties of nuclei such as $^{20}$Ne in Ref.~\cite{gnech_distilling_2024}.
The equivariant backflow transformation reads
\begin{equation}\label{eq:mpnn}
g_i = \Big(x_i, \sum_j m(x_i, x_j) \Big)\, ,
\end{equation}
with $m$ an MLP made of two fully-connected layers with $32$ nodes each~\cite{gnech_distilling_2024}.

The calculations in Ref.~\cite{Yang:2023rlw} for $^4$He, $^6$Li, and $^{16}$O employ a \emph{multi-determinant} SJ architecture with neural backflow, dubbed \emph{FeynmanNet}.
The backflow transformation generalizes the Fermionic Neural Network of Ref.~\cite{pfau_ab_2020} to encompass both continuous spatial and discrete spin--isospin degrees of freedom.
As shown in Sec.~\ref{sec:applications}, this architecture yields highly accurate ground-state energies for input Hamiltonians from pionless EFT at leading and next-to-leading order, which include a short-range tensor component.

Many nuclear-physics calculations based on NQS make use of transformations built from message-passing neural networks (MPNNs) to improve the flexibility and scalability of the variational ansatz.
MPNN-based backflow was first integrated into the SJ ansatz to preprocess the input coordinates.
In its first application to the homogeneous electron gas, this strategy enabled simulations with up to $N=128$ electrons at extremely low densities~\cite{pescia_message-passing_2024}.
It was later extended to ultracold Fermi gases~\cite{kim_neural-network_2024}, where strong pairing correlations are particularly challenging.
Most recently, MPNNs have been used to compute properties of dilute neutron-star matter, substantially enhancing the expressivity of the PJ ansatz and allowing a first-principles description of cluster formation in the neutron-star crust~\cite{fore_investigating_2024}.

MPNNs are a class of graph neural networks designed to model correlations in graph-structured data. In quantum systems of identical particles, this graph is fully connected, with each particle interacting with every other.
Importantly, to satisfy the Pauli principle, MPNNs can process inputs in a permutation-equivariant manner.
There are many possible ways to implement an MPNN, but the core idea is to iteratively update the information encoded on the nodes and edges of a graph.
The input graph typically encodes single-particle features on the nodes and pairwise features on the edges.
At each iteration, messages are passed along the edges based on these features, allowing each node to incorporate information from its neighbors.
As a result, the node and edge representations become increasingly enriched with nonlocal, correlated information from the entire system.
The output is then another graph with the same connectivity but with node and edge features that encode many-body correlations in a permutation-equivariant way.
These features can then be fed into other parts of the NQS in place of the original node and edge features.

Here, we outline the structure of the MPNN following Ref.~\cite{fore_investigating_2024} based on earlier developments in Refs.~\cite{pescia_message-passing_2024,kim_neural-network_2024}, with modifications appropriate for finite nuclei.
We take the input single-particle ``visible'' features as \( \mathbf{v}_i = [\bar{\mathbf{r}}_i, s_i^z, t_i^z ] \). Note that—unlike in periodic systems, where only the spin and isospin of particle \( i \) are used as inputs—here the Cartesian coordinates of the nucleons are also included. To automatically remove spurious center-of-mass contributions from all observables~\cite{massella_exact_2020}, we define the intrinsic spatial coordinates as \( \bar{\mathbf{r}}_i = \mathbf{r}_i - \mathbf{R}_{\rm CM} \), where \( \mathbf{R}_{\rm CM} \) denotes the center of mass of the nucleus. The ``visible'' pairwise features encode spatial as well as spin–isospin coordinates and are defined by  
\[
\mathbf{v}_{ij} = [ \,\mathbf{r}_i-\mathbf{r}_j, \| \mathbf{r}_i - \mathbf{r}_j \|, s_i^z, s_j^z, t_i^z, t_j^z] \, .
\]

The initial hidden features for the nodes and the edges are obtained by concatenating the original and transformed single-particle and two-particle features, respectively, as
\begin{equation*}
\mathbf{h}_i^{(0)} = [\mathbf{v}_i, f_A(\mathbf{v}_i)] \quad , \quad
\mathbf{h}_{ij}^{(0)} = [\mathbf{v}_{ij}, f_B(\mathbf{v}_{ij})] \, .
\end{equation*}
Here, \( f_A \) and \( f_B \) are MLPs that preprocess the input coordinates and map them into a common latent space, ensuring that the dimensions of the hidden features \( \mathbf{h}_i^{(t)} \) and \( \mathbf{h}_{ij}^{(t)} \) remain independent of the iteration index \( t \).

The MPNN update is performed iteratively for \( t = 1, \dots, T \). At each step, information between the node and edge features is exchanged through the message
\begin{equation}
\mathbf{m}_{ij}^{(t)} = f_M^{(t)}\!\left( \mathbf{h}_i^{(t-1)}, \mathbf{h}_{ij}^{(t-1)}, \mathbf{h}_j^{(t-1)} \right) ,
\label{eq:mpnn-message}
\end{equation}
where \( f_M^{(t)} \) is an MLP. For each particle \( i \), the relevant messages are collected and pooled to eliminate any ordering with respect to the other particles \( j \neq i \). As in Ref.~\cite{kim_neural-network_2024}, we use logsumexp pooling, a smooth alternative to max pooling,
\begin{align*}
\mathbf{m}_{i}^{(t)} = \log\!\left(\sum_{j \neq i} \exp\!\left(\mathbf{m}_{ij}^{(t)}\right)\right) .
\end{align*}
The hidden node and edge features are then updated as
\begin{align*}
\mathbf{h}_i^{(t)} &= \left[ \mathbf{v}_i, f_F^{(t)}\!\left(\mathbf{h}_i^{(t-1)}, \mathbf{m}_{i}^{(t)}\right) \right] , \\
\mathbf{h}_{ij}^{(t)} &= \left[ \mathbf{v}_{ij}, f_G^{(t)}\!\left(\mathbf{h}_{ij}^{(t-1)}, \mathbf{m}_{ij}^{(t)}\right) \right] .
\end{align*}
The functions \( f_M^{(t)} \), \( f_F^{(t)} \), and \( f_G^{(t)} \) are distinct MLPs whose output dimensions match those of \( f_A \) and \( f_B \). Including concatenated skip connections to the visible features ensures that the signal from the raw input remains accessible even as the MPNN depth \( T \) increases.

Finally, after the \( T \)-th iteration, we aggregate the hidden node and edge features into single-particle and pairwise feature vectors:
\begin{align*}
\mathbf{g}_{ij} &= \left[ \mathbf{h}_i^{(T)}, \mathbf{h}_j^{(T)}, \mathbf{h}_{ij}^{(T)} \right] \,, \\ 
\mathbf{g}_i &=  \log\left(\sum_{j \neq i} \exp\left(\mathbf{g}_{ij}\right) \right)\,.
\end{align*}
By construction, \( \mathbf{g}_i \) and \( \mathbf{g}_{ij} \) define a permutation-equivariant
many-body feature embedding of the original single-particle and pairwise degrees of freedom. Rather than representing a fixed coordinate transformation, the MPNN learns a nonlinear, nonlocal reparameterization of the input graph that can be used as a drop-in replacement for the original node and edge features in a wide class of NQS.

In recent works~\cite{kim_neural-network_2024,fore_investigating_2024}, the pairwise features \( \mathbf{g}_{ij} \) are used to parameterize
the Pfaffian pairing orbital and the Jastrow factor, replacing the bare inputs
\( (x_i, x_j) \) with their correlated counterparts. However, this choice is not intrinsic
to the construction: the same embedding can be coupled to SJ~\cite{pescia_message-passing_2024}, HN~\cite{gnech_distilling_2024},
or other operator-based ansätze, and the single-particle features \( \mathbf{g}_i \) may
equally be used to define backflow-modified one-body orbitals. In this sense, the
MPNN provides a general, architecture-agnostic mechanism for incorporating
many-body correlations into neural wave functions.

\begin{figure}[!tb]
\centering
\includegraphics[width=0.95\columnwidth]{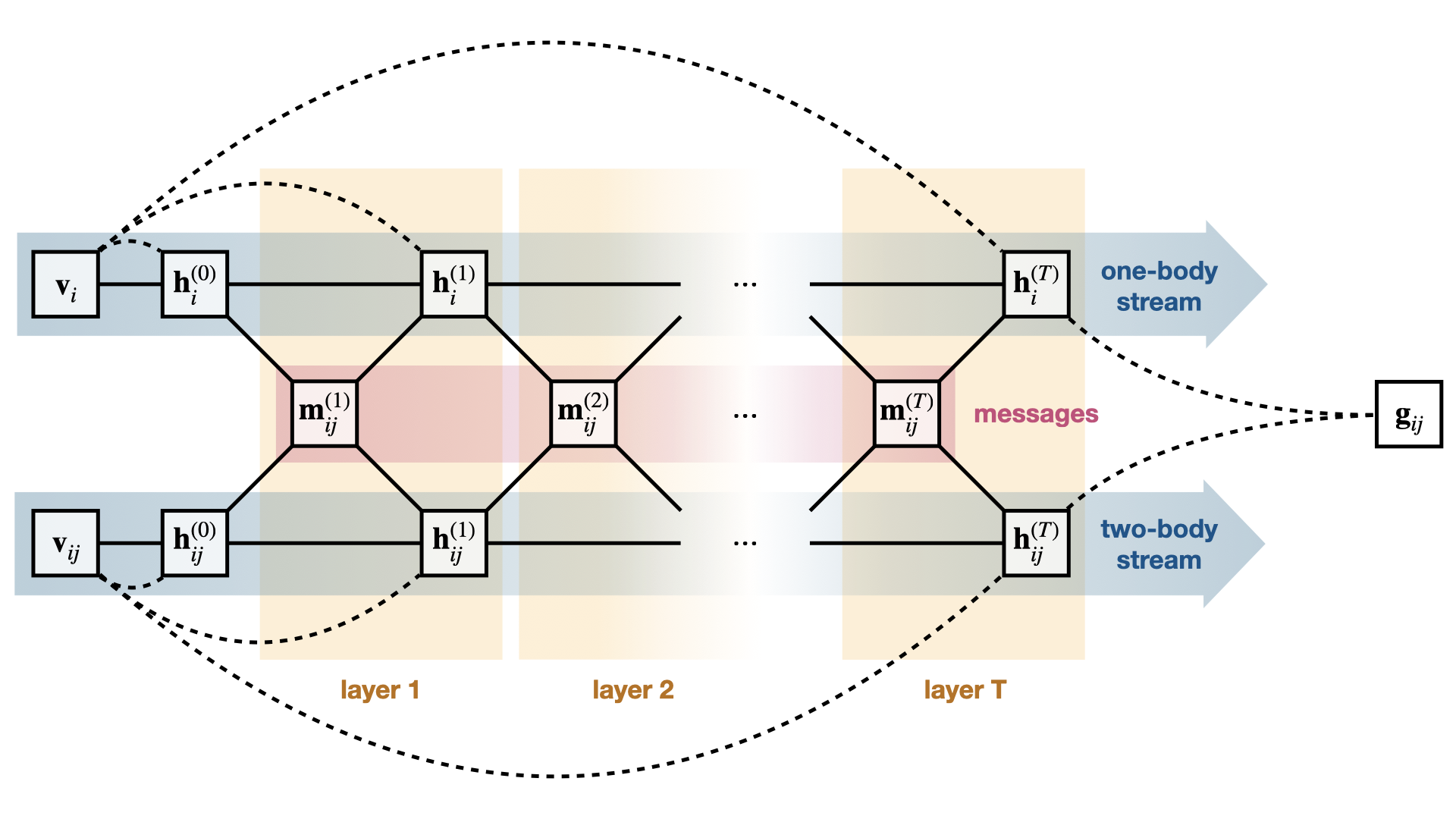}
\caption{Schematic representation of the message-passing neural network. Each of the $T$ total iterations of the network is represented by a yellow box. 
  Dashed lines represent the concatenation operations, while solid lines represent the parameterized transformations (linear transformations and nonlinear feedforward neural networks). Messages, highlighted in pink, mediate the exchange of information between the one- and two-body streams, in blue.}
\label{fig:mpnn}
\end{figure}
   
\newpage
\section{Nuclear Physics Applications}
\subsection{The deuteron}
\label{sub:deuteron}

While the first applications of NQS to many-body problems date to 2017--18~\cite{carleo_solving_2017,Saito2017,Saito2018}, the first application to a nuclear system arrived in 2020 with Ref.~\cite{keeble_machine_2020}. That work presented a proof-of-principle calculation of the deuteron, the only bound state of a neutron and a proton. The setup was deliberately minimal, allowing the authors to focus on the capabilities of the NQS framework rather than on details of the ansatz architecture.

\begin{figure}[!b]
\centering
\includegraphics[width=0.7\linewidth]{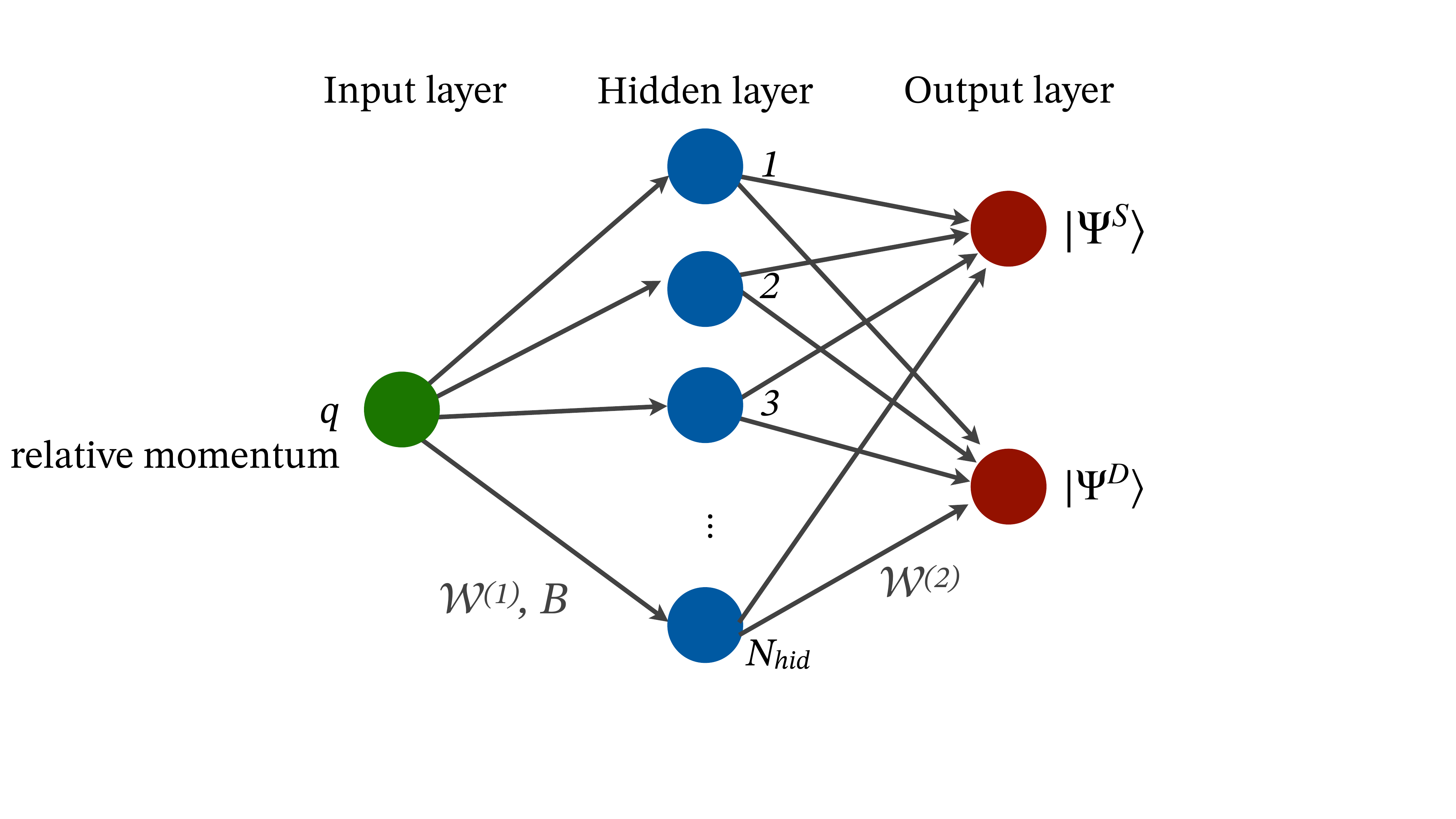}
\caption{\label{fig:NN_deuteron} 
Feed-forward neural network architecture used in Ref.~\cite{keeble_machine_2020}. The input is a single value of relative momentum, $q$, and the wave functions are modeled in terms of a minimal single-layer network with $N_\text{hid}$ nodes. The ansatz has two outputs, one for the $S$ and one for the $D$ state.
}  
\end{figure}

Because nucleon--nucleon interactions are naturally formulated in momentum space~\cite{Epelbaum:2009rkz}, and many existing numerical routines operate in this representation, the initial application was carried out in momentum space. In this basis, the two-body problem factorizes into a center-of-mass component and a relative component, with the former being irrelevant for the bound-state problem. The relative component is characterized by the relative momentum $q$ between the neutron and the proton. An additional practical advantage of working in momentum space is that no derivatives associated with the kinetic energy operator are required. In this formulation, the bound-state wave function can be decomposed into partial waves of definite orbital angular momentum.
The deuteron exhibits a known quadrupole deformation~\cite{Epelbaum:2009rkz}, implying that its wave function contains both a relative orbital angular momentum $L=0$ component, denoted $\ket{\Psi^S}$, and an $L=2$ component, denoted $\ket{\Psi^D}$.

With these nuclear properties in mind, a minimal ansatz for the deuteron wave function was chosen, as shown in Fig.~\ref{fig:NN_deuteron}. The ansatz consists of a simple MLP with one input (the relative momentum $q$) and two outputs (the two components $\ket{\Psi^{S,D}}$). The first layer included a set of weights ${\bf \mathcal{W}}^{(1)}$ and biases $\mathbf{b}$, whereas the second layer did not incorporate any bias. The network contained $N_\text{hid}$ hidden nodes and was assumed to be fully connected, as indicated in Fig.~\ref{fig:NN_deuteron}. The wave-function ansatz has the form
\begin{align}
    \psi_{\textrm{ANN}}^L (q) =  \sum_{i=1}^{N_\text{hid}} \mathcal{W}^{(2)}_{i,L} \, \sigma \left( \mathcal{W}^{(1)}_{i} q + b_{i} \right) \, ,
    \label{eq:wfANNdeuteron}
\end{align}
where $\sigma(x)$ denotes the chosen activation function. Since the deuteron wave function is a smooth, continuous function of $q$, it is natural to employ continuous activation functions in this case. The initial exploration of Ref.~\cite{keeble_machine_2020} considered both sigmoid and softplus functions.

The one-dimensional nature of the problem in relative momentum coordinates allows the total energy to be evaluated straightforwardly by quadrature. For example, the kinetic energy expectation value is approximated as
\begin{align}
    \braket{K} = \sum_L 
    \int_0^\infty dq \, q^2 
    \frac{\hbar^2 q^2}{\mu} | \Psi^L(q) |^2
    \approx 
    \sum_L \sum_{i=1}^{N_q} \omega_i \, q_i^2 
    \frac{\hbar^2 q_i^2}{\mu} | \Psi^L(q_i) |^2 \, ,
    \label{eq:kinetic}
\end{align}
where $\mu = m_n m_p /(m_n + m_p)$ is the reduced mass of the $np$ system. 
A set of $N_q = 64$ Gauss--Legendre quadrature points, tangentially transformed to extend up to $k_\textrm{max} = 500$ fm$^{-1}$, was sufficient to obtain accurate results. The same quadrature scheme can be used to evaluate the potential energy.
This initial study employed the N3LO Entem--Machleidt interaction~\cite{Entem2003}, although in principle any other interaction could be used. The quadrature-discretized Hamiltonian can of course be diagonalized directly, providing an exact benchmark on the same footing as the NQS simulation.

With this quadrature set up, the calculation of the energy is rather straightforward. The original set-up employed a pre-training step to pre-optimize the shape of the ansatz in Eq.~\eqref{eq:wfANNdeuteron} to physically sound values. This step is not necessary, but helps accelerate the optimization procedure~\cite{rozalen_sarmiento_machine_2024}. 
After the pre-training, the energy minimization was performed over $250,000$ steps employing RMSprop~\cite{hinton_lecture_2012} as the optimizer of choice. 

\begin{figure}
\centering
\includegraphics[width=0.7\linewidth]{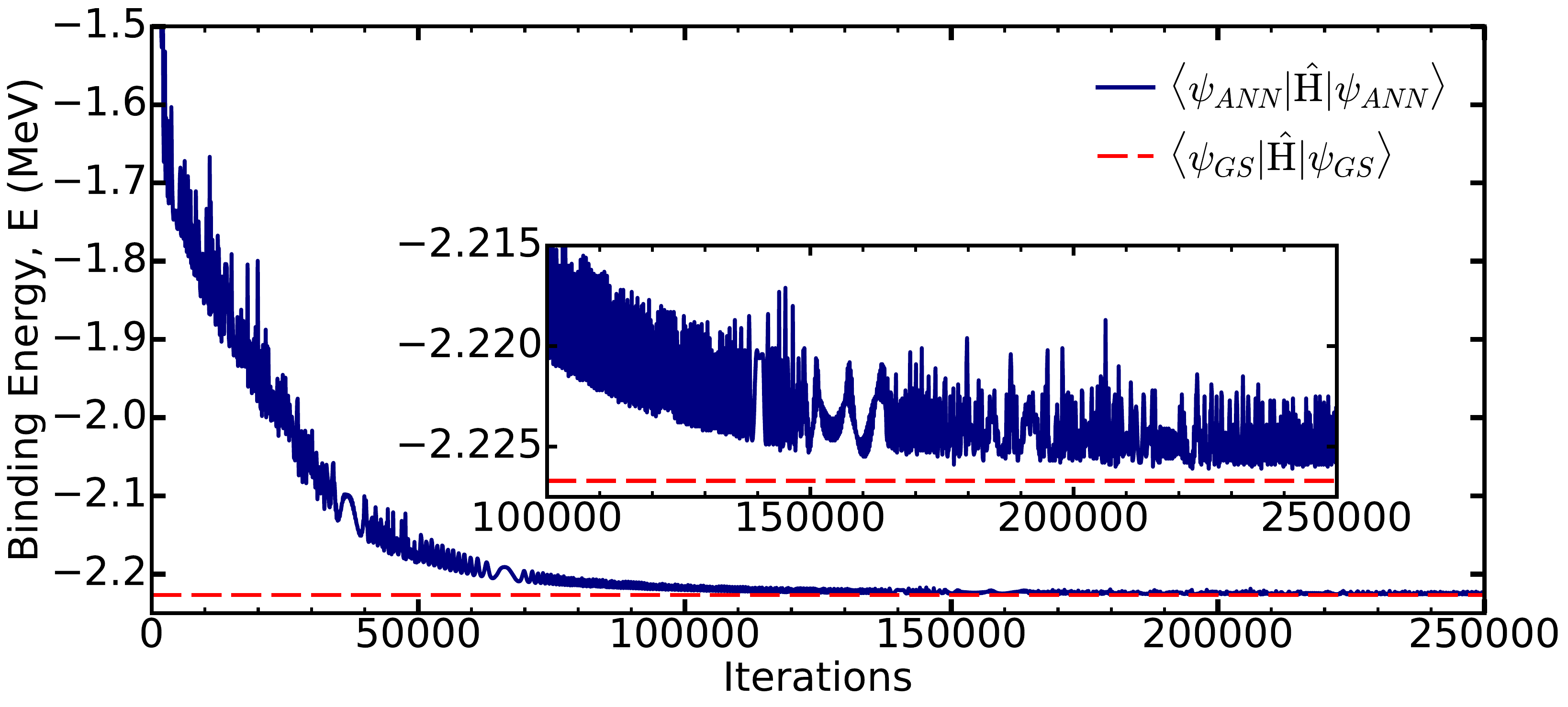}
\caption{\label{fig:energy_deuteron_minimisation} 
Deuteron binding energy as a function of iteration number for a network with 
$N_\text{hid} = 10$ nodes and a softplus activation function~\cite{keeble_machine_2020}.}  
\end{figure}

A typical minimization curve for the case with $N_\text{hid}=10$ is shown in Fig.~\ref{fig:energy_deuteron_minimisation}. For this very simple architecture, the energy converged to $1 \%$ within a few tens of thousands of iterations. 
After around $150,000$ iterations, the energy is very close to the minimum, but oscillates above the actual benchmark minimum within $3-4$ eV of the exact result. 
This minimal set-up already provides an excellent reproduction of the exact wave function and physical properties of the deuteron~\cite{keeble_machine_2020}.

\begin{figure}
\centering
\includegraphics[width=0.8\linewidth]{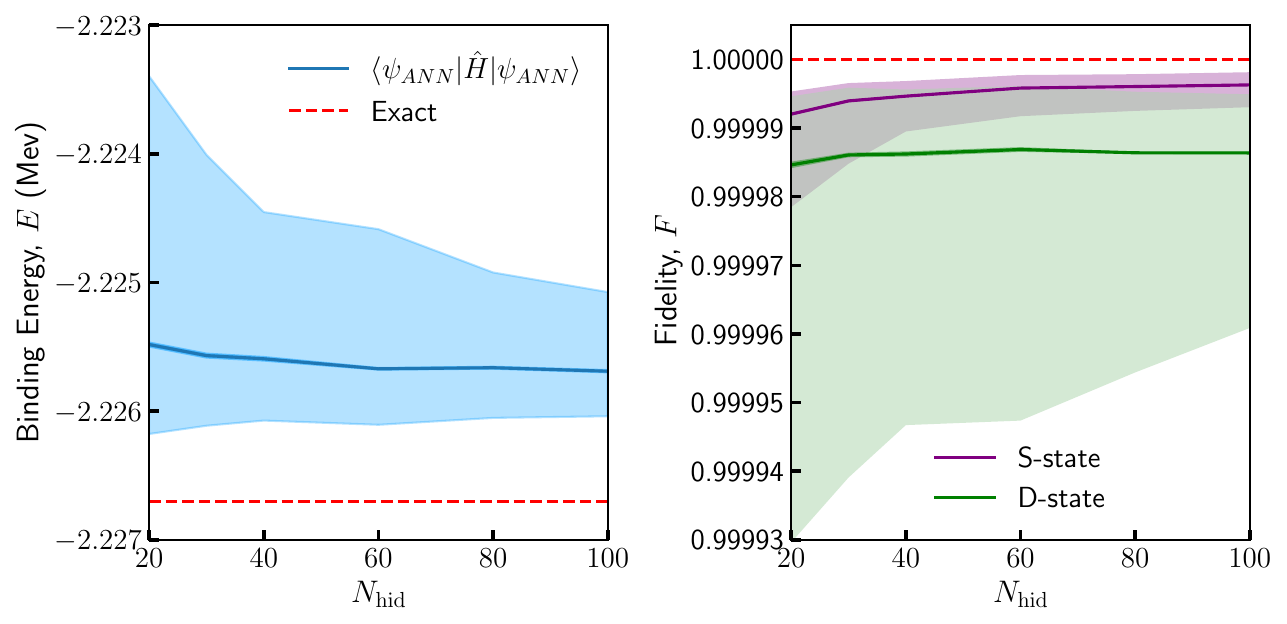}
\caption{\label{fig:deuteron_energies} 
Binding energy of the deuteron (left panel) and fidelity $\mathcal{F}$ (right panel) as a function of the number of hidden layer nodes, $N_\text{hid}$ from Ref.~\cite{rozalen_sarmiento_machine_2024}. See main text for an explanation of the error bands. Horizontal (dashed) lines show the benchmark result. 
}  
\end{figure}

To address systematic errors in the NQS method as opposed to specific minimisations, the analysis of Refs.~\cite{keeble_machine_2020} and \cite{rozalen_sarmiento_machine_2024} focused on the characterization of uncertainties. Simulations of the deuteron were run for different numbers of hidden nodes, $N_\text{hid}$. Results of these numerical experiments for the simple architecture of Eq.~\eqref{eq:wfANNdeuteron} are presented in Fig.~\ref{fig:deuteron_energies}~\cite{rozalen_sarmiento_machine_2024}. 
Two different sets of uncertainties were identified to characterize the NQS minimization generally (as opposed to a minimization-by-minimization basis). First, different initializations of the network lead to different final results after $250,000$ iterations. This uncertainty can be assessed by running the minimization process $20$ times and computing the associated standard deviation. This out-of-sample uncertainty is represented in dark bands in the top panel of Fig.~\ref{fig:deuteron_energies}. We stress that this uncertainty amounts to only a fraction of a keV in energy and that it is rather independent of the number of nodes. Moreover, the fidelity, $F$, between the ansatz at the end of minimization and the corresponding exact benchmark is given in the bottom panel of Fig.~\ref{fig:deuteron_energies}. We represent the fidelities of both the $S-$ and the $D-$states. Again, out-of-sample uncertainties, represented by dark bands, are very small for fidelities~\cite{keeble_machine_2020}.

As illustrated in Fig.~\ref{fig:energy_deuteron_minimisation},  even at the end of a full minimization, the results still show some residual oscillations. 
These oscillations are typically within a few keV of the total energy and are hence larger than the out-of-sample uncertainty. In this context, Ref.~\cite{rozalen_sarmiento_machine_2024} looked into these post-training oscillation amplitudes as an additional source of uncertainty that is associated not so much to minimization instances, but rather to the minimization process itself. To characterize this uncertainty, the model was minimized initially and then evolved for $300$ final iterations. The number of these post-evolution epochs is small, to guarantee that mean energy values do not improve, but also large enough to observe periodicity in the oscillations. Upper and lower values of these oscillations were recorded for $20$ different minimization instances, and their average values are shown as the light bands in Fig.~\ref{fig:deuteron_energies}. 

These results clearly indicate that  oscillation errors at the end of the minimization dominate over any out-of-sample uncertainties. Moreover, these uncertainties show a decreasing trend with $N_\text{hid}$. Whereas for $N_\text{hid}=2$ the post-evolution uncertainty in the energy is of the order of $2.5$ keV, at $N_\text{hid}=100$ this falls just below $1$ keV. A similar decrease in post-evolution uncertainty is observed in the fidelities of the bottom panel. In this case, the fidelity of the $D-$state presents a much larger uncertainty. 

\begin{figure}
\centering
\includegraphics[width=0.75\linewidth]{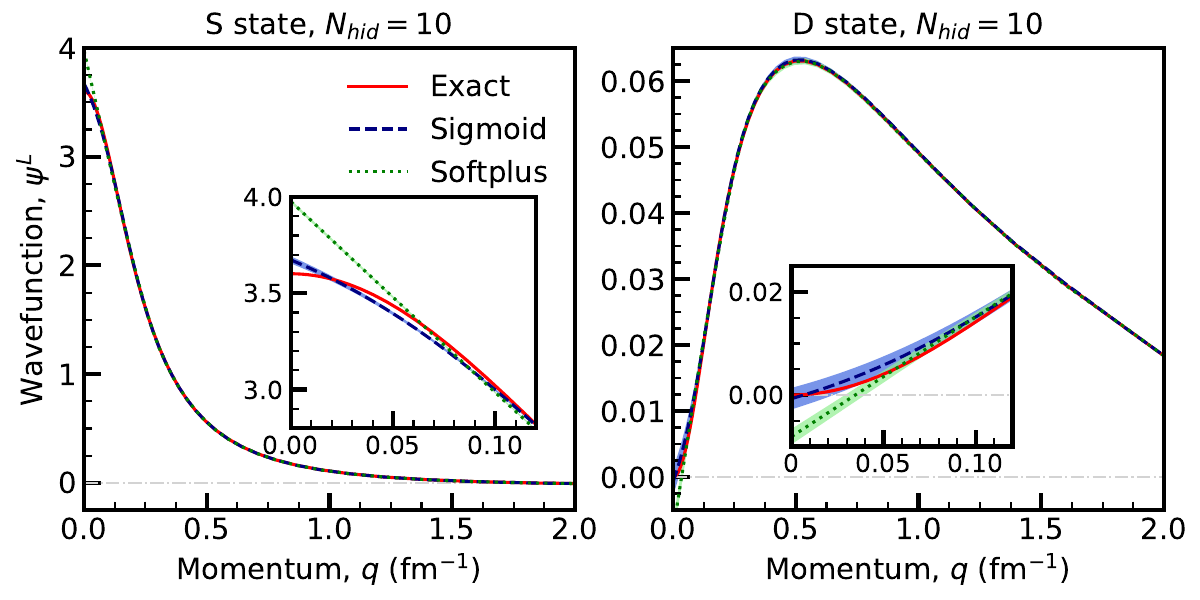}
\caption{\label{fig:deuteron_wave functions} 
Left (right) panel: the $S$ ($D$) state wave function as a function of momentum, taken from Ref.~\cite{keeble_machine_2020}. The benchmark exact solutions (solid lines) are compared to the feed-forward ansatz with $N_\text{hid}=10$ using sigmoid (dashed) and softplus (dotted) activation functions. The shaded bands indicate the standard deviation over 50 different initialization runs.
}  
\end{figure}

The high fidelity shown in Fig.~\ref{fig:deuteron_energies} indicates a strong level of agreement between the exact wave function and the NQS ansatz. This agreement is illustrated more explicitly in Fig.~\ref{fig:deuteron_wave functions}, which displays the $S$- (left panel) and $D$-wave (right panel) components as functions of momentum for minimizations with $N_\text{hid}=10$~\cite{keeble_machine_2020}. The solid (red) lines correspond to the benchmark exact solution, while the dashed (blue) and dotted (green) lines show the results of averaging 50 independent minimizations using sigmoid and softplus activation functions, respectively.

The agreement is excellent across nearly all momenta, with small deviations appearing near the origin, $q \approx 0$ fm$^{-1}$. Because the wave functions are spherically symmetric, the energy functional includes a $q^2$ factor in the integration measure. As a result, the region near $q=0$ contributes very little to the cost function, leaving the wave function weakly constrained there. A simple network with $N_\text{hid}=10$ has limited flexibility to adjust its behavior in this region. As $N_\text{hid}$ increases, however, additional flexibility tends to manifest primarily near the origin, where the weak energetic penalty may allow for large variations in shape, including non-physical behaviors~\cite{keeble_machine_2020}.

While our discussion has focused on the simple architecture of Eq.~\eqref{eq:wfANNdeuteron}, Ref.~\cite{rozalen_sarmiento_machine_2024} presents an extensive analysis of more elaborate architectural extensions. These include separate, independent branches for the $S$ and $D$ channels, as well as deeper, multi-layer networks.
Adopting these more complex architectures leads to two main consequences. First, the fraction of models that converge to an acceptable energy range, $E \in (-2.220, -2.227)$ MeV, is significantly reduced. Whereas the simple one-layer MLP achieves a $100\%$ success rate for nearly all values of $N_\text{hid}$, the acceptance rates of more complex architectures depend strongly on $N_\text{hid}$ and range from $1\%$ to $90\%$~\cite{rozalen_sarmiento_machine_2024}. Second, more elaborate architectures exhibit substantially larger post-training uncertainties,  with a pronounced dependence on $N_\text{hid}$. In some cases, indications of overfitting are also observed. Although understanding the internal behavior of NQS models is not easy, the systematic error analysis and architectural comparisons performed for the deuteron provide a useful starting point for probing these issues in nuclear systems.

\subsection{Atomic nuclei}
\label{sub:nuclei}

\subsubsection[Nuclei with up to A=6 nucleons]{Nuclei with up to A=6 nucleons}
The first application of VMC methods based on NQS to the nuclear many-body problem~\cite{adams_variational_2021} computed the ground-state energies and single-particle densities of $^2$H, $^3$H, and $^4$He. It is worth noting that the LO pionless-EFT Hamiltonian used in that work differs slightly from the one discussed in Sec.~\ref{sec:nmbp}. Specifically, the NN interaction is
\begin{equation}
v_{\mathrm{LO}}^{\mathrm{CI}}(r_{ij}) = \bigl(C_1 + C_2 \sigma_{ij}\bigr)
e^{-r_{ij}^2 \Lambda^2 / 4},
\label{eq:v_nn_contessi}
\end{equation}
where, following Ref.~\cite{kirscher_spectra_2015}, the low-energy constants $C_1$ and $C_2$ are fitted to the deuteron binding energy and the neutron–neutron scattering length. The 3N force is taken as
\begin{equation}
V_{ijk} = D_0 \sum_{\mathrm{cyc}}
e^{-(r_{ik}^2 + r_{ij}^2)\Lambda^2/4},
\end{equation}
with $D_0$ fixed by the $^3$H ground-state energy.

\begin{table}[!b]
\begin{center}
\begin{tabular}{ c | c | c c c c }
\hline
\hline
& $\Lambda$ & SJ  &  Spline & GFMC\\
\hline
\multirow{2}{*}{$^2$H } & $4$ fm$^{-1}$ & $-2.224(1)$  &   $-2.223(1)$  & $-2.224(1)$ \\
& $6$ fm$^{-1}$ & $-2.224(4)$  &   $-2.220(1)$  & $-2.225(1)$ \\
\hline
\multirow{2}{*}{$^3$H } & $4$ fm$^{-1}$ & $-8.26(1)$  &   $-7.80(1)$  & $-8.38(2)$ \\
& $6$ fm$^{-1}$ & $-8.27(1)$  &   $-7.74(1)$  & $-8.38(2)$ \\
\hline
\multirow{2}{*}{$^4$He } & $4$ fm$^{-1}$ & $-23.30(2)$  &   $-22.54(1)$  & $-23.62(3)$ \\
& $6$ fm$^{-1}$ & $-24.47(3)$  &   $-23.44(2)$  & $-25.06(3)$ \\
\hline
\end{tabular}
\caption{(From Ref.~\cite{adams_variational_2021} with permission from the Authors). Ground-state energies in MeV of the $^2$H, $^3$H, and $^4$He for the LO pionless-EFT Hamiltonian for $\Lambda = 4$ fm$^{-1}$ and $\Lambda = 6$ fm$^{-1}$. Numbers in parentheses indicate the statistical errors on the last digit.}
\label{tab:2h}
\end{center}
\end{table}

Reference~\cite{adams_variational_2021} employed the SJ wave-function ansatz of Eq.~\eqref{eq:SJ}, together with the real-valued Jastrow factor of Eq.~\eqref{eq:jastrow_tanh}. In this construction, the Deep Sets architecture (see Section~\ref{sec:deep_learning}) uses the single-particle sum aggregation of Eq.~\eqref{eq:deep_sets_single}. Table~\ref{tab:2h}, adapted from Ref.~\cite{adams_variational_2021}, reports the ground-state energies of $^2$H, $^3$H, and $^4$He. The SJ results are benchmarked against VMC calculations based on spline parameterizations of the two- and three-body spin–isospin–independent Jastrow functions~\cite{contessi_ground-state_2017}, as well as against virtually exact GFMC results.
All three methods yield statistically consistent energies for $^2$H, demonstrating that the SJ ansatz is flexible enough to accurately represent the deuteron ground state. This agrees with the findings of Ref.~\cite{keeble_machine_2020}, discussed in Section~\ref{sub:deuteron}. Since the LO pionless-EFT Hamiltonian lacks tensor and spin–orbit components, the SJ ansatz without backflow is in fact exact for this system.
For $^3$H, the neural SJ ansatz provides an improvement of about $0.5$~MeV over conventional VMC for both $\Lambda = 4$~fm$^{-1}$ and $6$~fm$^{-1}$. The GFMC energies are roughly $0.1$~MeV more bound than the SJ ones. As noted in Ref.~\cite{adams_variational_2021}, this residual difference originates from spin-dependent correlations that are automatically generated in GFMC imaginary-time evolution but are only partially encoded in the Jastrow factor of Eq.~\eqref{eq:jastrow_tanh}. These correlations modify the nodal structure of the wave function, which the SJ architecture cannot recover without incorporating backflow transformations.
A similar trend is observed for $^4$He: the SJ wave functions outperform the conventional VMC ones, improving the energies by about $0.8$~MeV and $1.0$~MeV for $\Lambda = 4$~fm$^{-1}$ and $6$~fm$^{-1}$, respectively. The remaining discrepancies with GFMC again reflect missing spin–isospin–dependent correlations in the SJ ansatz.

\begin{figure}[!htb]
\centering
\includegraphics[width=0.75\columnwidth]{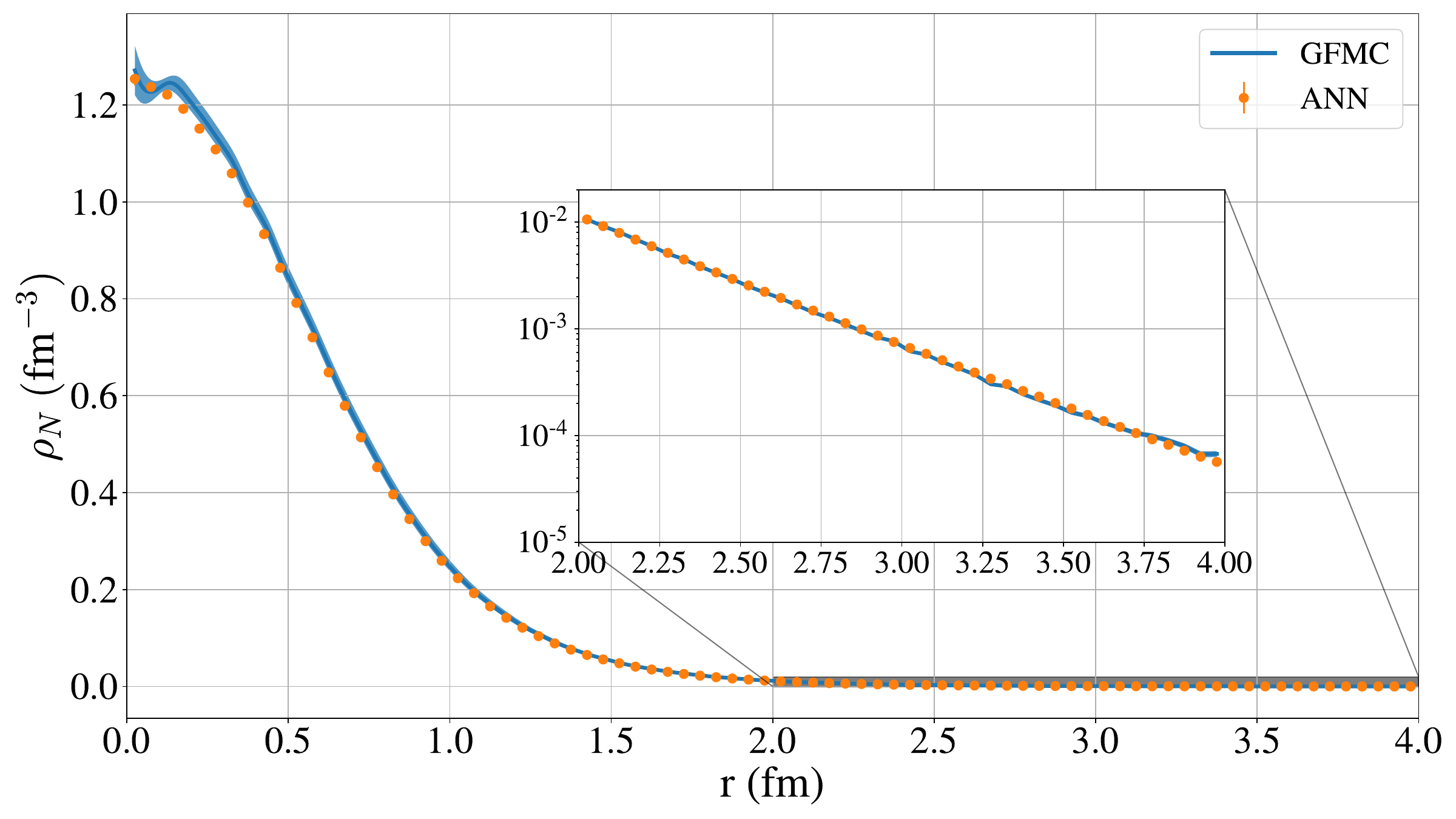}
\caption{(From Ref.~\cite{adams_variational_2021}, with permission from the Authors) Point-nucleon densities of $^4$He (lower panel) for the LO pionless-EFT Hamiltonian with $\Lambda =4$ fm$^{-1}$. The solid points and the shaded area represent the SJ and GFMC results, respectively.}
\label{fig:rho_4he}
\end{figure}

To further elucidate the quality of the ANN wave function, we consider the point-nucleon density
\begin{align}
	\rho_{N}(r) &=\frac{1}{4\pi r^2}\big\langle\Psi_V \big|\sum_i \mathcal \delta(r-|\mathbf{r}_i^{\rm int}|)\big|\Psi_V \big\rangle\,, 
	\label{eq:rho_N}
\end{align}
which is of interest in a variety of experimental settings~\cite{de_vries_nuclear_1987,angeli_table_2013}. Fig.~\ref{fig:rho_4he} displays $\rho_{N}(r)$ of $^4$He as obtained from the SJ ansatz compared with GFMC calculations, both using as input the LO pionless-EFT Hamiltonian with $\Lambda =4$ fm$^{-1}$. There is an excellent agreement between the two methods, which further corroborates the representative power of the SJ ansatz for the wave functions of $A\leq 4$ nuclei. The SJ and GFMC densities overlap both at short distances and in the slowly-decaying asymptotic exponential tails, highlighted in the insets of Fig.~\ref{fig:rho_4he}. As already noted in Ref.~\cite{adams_variational_2021}, the NQS learns how to compensate for the original Gaussian confining function and reproduce the correct exponential falls off of the nuclear wave function, which is notoriously delicate to obtain within nuclear methods that rely on harmonic-oscillator basis expansions~\cite{caprio_robust_2022,sharaf_comparing_2019}.

Ref.~\cite{Gnech2021} extended the SJ approach to the $^6$Li and $^6$He nuclei, employing the model “o’’ Hamiltonian of Eqs.~\eqref{eq:vNN_LO}-\eqref{eq:o_3NF} with the three-nucleon regulator set to $R_3 = 1.0$~fm. A key technical advance relative to Ref.~\cite{adams_variational_2021} is the use of the linear combination of Slater determinants of Eq.~\eqref{eq:mf} to represent the mean-field component of the open-shell systems $^6$Li and $^6$He. In addition, pairwise inputs were incorporated into the Deep Sets architecture used for the Jastrow factor, as in Eq.~\eqref{eq:deep_sets_pair}. In that work, the SJ energies and charge radii were benchmarked against the highly accurate hyperspherical harmonics (HH) method~\cite{kievsky_high-precision_2008}.

The binding energies and charge radii of $^2$H, $^3$H, $^3$He, $^4$He, $^6$He, and $^6$Li obtained with the VMC method based on the SJ ansatz and with the HH method are listed in Table~\ref{tab:res_NN}. For nuclei with $A \geq 3$, the table separately shows the results computed using the $NN$ interaction alone and those obtained with the full model ``o'' Hamiltonian, which also includes the $3N$ force.

The expectation value of the charge radius is derived from the point-proton radius using the relation:
\begin{align}
	\left\langle r_{\rm ch}^2\right\rangle=
	\left\langle r_{\rm pt}^2\right\rangle+
	\left\langle R_p^2\right\rangle+
	\frac{A-Z}{Z}\left\langle R_n^2\right\rangle+
	\frac{3}{4m_p^2},
	\label{eq:rch}
\end{align}
where $\langle r_{\rm pt}^2\rangle$ is the calculated point-proton radius,
$\langle R_p^2 \rangle = 0.770(9),\mathrm{fm}^2$ is the proton mean-square charge radius,
$\langle R_n^2 \rangle = -0.116(2),\mathrm{fm}^2$ is the neutron mean-square charge radius. Here, consistent with Ref.~\cite{Gnech2021}, we list the 2012 values for these quantities~\cite{beringer_review_2012}. Finally, $(3)/(4m_p^2) \approx 0.033,\mathrm{fm}^2$ is the Darwin–Foldy correction~\cite{friar_nuclear_1997}.
The point-proton radius can be computed as the ground-state expectation value
\begin{align}
	\left\langle r_{\rm pt}^2\right\rangle=\frac{1}{Z}\big\langle\Psi\big|\sum_i P_{p} |\mathbf{r}_i-\mathbf{R}_{\rm cm}|^2\big|\Psi\big\rangle,
\end{align}
where $Z$ is the number of protons and
$P_p = (1 + \tau_{z_i})/2$ projects onto proton states.

\begin{table*}[htb]
\setlength{\tabcolsep}{3.9pt}
\centering
\caption{Ground-state energies and charge radii for selected $A\leq 6$ nuclei obtained from VMC calculations based on the SJ ansatz and HH methods using as input model ``o'' Hamiltonian of Ref.~\cite{schiavilla_two-_2021} with and without the $3N$ force. We report also the experimental binding-energies from
Ref.~\cite{national_nuclear_data_center_nudat_nodate} and the charge radius taken from Refs.~\cite{tiesinga_codata_2021,amroun_3h_1994,morton_nuclear_2006,krauth_measuring_2021,wang_laser_2004,puchalski_ground_2013}.}
\label{tab:res_NN}  
\begin{tabular}{c c c c c c c c}
\hline\hline
\multirow{2}{*}{Nucleus} & \multirow{2}{*}{Potential} & \multicolumn{2}{c}{SJ}               & \multicolumn{2}{c}{HH} & \multicolumn{2}{c}{Exp.}                \\
\noalign{\smallskip}
        &           & $E\,(\rm MeV)$ & $r_{\rm ch}\,(\rm fm)$ & $E\,(\rm MeV)$ & $r_{\rm ch}\,(\rm fm)$ & $E\,(\rm MeV)$ & $r_{\rm ch}\,(\rm fm)$ \\     
\noalign{\smallskip}\hline\noalign{\smallskip}
$^2$H  & $N\!N$ & $-2.242(1)$   & $2.120(5)$ & $-2.242$  & $2.110(2)$ & $-2.225$ & $2.128$\\
\noalign{\smallskip}\hline\noalign{\smallskip}                                                               
\multirow{2}{*}{$^3$H} & $N\!N$& $-9.511(1)$   & $1.658(4)$ & $-9.744$  & $1.656(4)$ & \multirow{2}{*}{$-8.475$} & \multirow{2}{*}{$1.755(86)$}  \\
        & $3N$  & $-8.232(1)$   & $1.750(3)$ & $-8.475$  & $1.747(6)$ & &  \\
\noalign{\smallskip}\hline\noalign{\smallskip}                                                  
\multirow{2}{*}{$^3$He} & $N\!N$& $-8.800(1)$   & $1.845(3)$ & $-9.035$  & $1.848(6)$ & 
\multirow{2}{*}{$-7.718$} & \multirow{2}{*}{$1.964(1)$}  \\
        & $3N$  & $-7.564(1)$   & $1.961(3)$ & $-7.811$  & $1.969(8)$ & & \\
\noalign{\smallskip}\hline\noalign{\smallskip}
\multirow{2}{*}{$^4$He} & $N\!N$& $-36.841(1)$   & $1.484(3)$ & $-37.06$  & $1.485(4)$ &
\multirow{2}{*}{$-28.30$} & \multirow{2}{*}{$1.678$}  \\
        & $3N$  & $-27.903(1)$   & $1.643(2)$ & $-28.17$  & $1.646(4)$ & & \\
\noalign{\smallskip}\hline\noalign{\smallskip}
\multirow{2}{*}{$^6$He} & $N\!N$& $-37.25(4)$  & $1.895(2)$ & $-37.96(8)$  & $1.71(1)$ &
\multirow{2}{*}{$-29.27$} & \multirow{2}{*}{$2.05(1)$}  \\
        & $3N$  & $-27.46(2)$   & $>4.89(1)$ & $-27.41(8)$  & $>2.73$ & &\\
\noalign{\smallskip}\hline\noalign{\smallskip}
\multirow{2}{*}{$^6$Li} & $N\!N$& $-42.04(1)$   & $2.248(3)$ & $-42.51(5)$  & $2.09(2)$ &
\multirow{2}{*}{$-31.99$} & \multirow{2}{*}{$2.54(3)$}  \\
        & $3N$  & $-30.82(3)$   & $3.049(2)$ & $-31.00(8)$  & $>2.74$ & & \\
\noalign{\smallskip}\hline                                                        
\hline
\end{tabular}
\label{tab:afdmc-gfmc}
\end{table*}

The SJ ansatz reproduces the HH binding energy of $^2$H, with both calculations yielding values slightly more bound than experiment owing to the missing charge-dependent and charge-symmetry-breaking terms—aside from the Coulomb interaction—in the $NN$ potential. The charge radii obtained with the VMC–SJ and HH methods are compatible within uncertainties and only marginally smaller than the experimental value.

Moving to the $A=3$ systems, VMC–SJ underbinds $^3$H and $^3$He by about $0.25$~MeV relative to HH for both the $NN$ and $NN+3N$ Hamiltonians. As discussed in Ref.~\cite{adams_variational_2021}, these small differences stem from the inability of the Jastrow correlator to compensate for zeros in the mean-field component $\langle RS|\Phi\rangle$. Despite this limitation, the VMC–SJ and HH charge radii remain very similar and, once the $3N$ force is included, agree well with experiment. It is also worth noting that the HH binding energy of $^3$H matches the experimental value by construction, whereas the $^3$He energy differs from experiment by about $0.1$~MeV—a discrepancy likely explained by the neutron–proton mass difference and by the missing charge-dependent and charge-symmetry-breaking components of the $NN$ interaction.

A similar pattern is observed in $^4$He: the VMC–ANN ground-state energy is approximately $0.2$~MeV above the HH result, independent of whether the $3N$ force is included. The repulsive character of the $3N$ interaction is essential for improving agreement with experiment for both the binding energy and the charge radius. Including the $3N$ force pushes nucleons to larger distances from the center of mass, thereby increasing the charge radius.

The $A=6$ nuclei provide a more stringent test. With the $NN$ interaction alone, the SJ ansatz yields wave functions that are stable against breakup—$^6$He into $^4$He plus two neutrons, and $^6$Li into $^4$He plus a deuteron. This is a nontrivial result, as conventional VMC calculations with standard two- and three-body Jastrow correlations often fail to place $^6$Li below the $^4$He + d threshold. In this $NN$-only case, the SJ energies are about $0.5$~MeV less bound than the HH results, corresponding to a per-nucleon difference comparable to that observed for the $A=3$ nuclei.
The SJ charge radii of $^6$He and $^6$Li exceed the HH values, although they remain below the experimental ones. Part of this discrepancy is attributable to the slow convergence of HH calculations for $A=6$ systems~\cite{gnech_calculation_2020}: the limited hyperradial truncation tends to underestimate the radii and prevents a reliable extrapolation. As in lighter systems, the inclusion of the $3N$ force significantly increases the charge radii, in some cases pushing them above experimental values. With the $3N$ interaction, $^6$Li becomes only marginally bound against breakup into $^4$He + d, while $^6$He becomes unbound with respect to $^4$He. Its wave function therefore extends to increasingly large distances from the center of mass, producing a very large charge radius. This behavior mirrors that observed in Ref.~\cite{contessi_ground-state_2017} for the unbound $^{16}$O system and is likely to be corrected once the $p$-wave contributions of the $NN$ potential are included~\cite{gattobigio_embedding_2019}.

\subsubsection[Relativistic effects]{Relativistic effects}
Within the standard quantum many-body paradigm, discussed in Section~\ref{sec:nmbp}, nuclei are treated as systems of point-like nucleons interacting through instantaneous potentials. While this description successfully accounts for a broad range of nuclear properties, it is, strictly speaking, incompatible with causality and can lead to unphysical predictions --- such as superluminal sound speeds in dense matter~\cite{Akmal:1998cf,sabatucci_relativistic_2024}. Relativistic extensions of this framework have therefore been explored for decades~\cite{coester_relativistic_1975}, starting from analyses of nuclear matter~\cite{brockmann_relativistic_1990} and few-body systems~\cite{glockle_relativistic_1986} to more recent Quantum Monte Carlo studies employing Poincar\`e-invariant Hamiltonians constructed from phenomenological NN and 3N forces~\cite{forest_quantum_1999}. In this approach, relativity is incorporated by adopting relativistic kinetic energies and by supplementing two- and three-body potentials with the corresponding Lorentz–boost corrections, which encode the dependence of the interaction on the total momentum of the nucleon pair. Calculations of $A=3$ and $A=4$ nuclei have shown that these boost corrections produce repulsive contributions that represent a substantial fraction of the repulsion usually attributed to irreducible three-nucleon forces. 

Building on this framework, the authors of Ref.~\cite{yang_consistent_2022}
derive, for the first time, a microscopic relativistic Hamiltonian at
leading order in covariant pionless EFT that contains consistent
relativistic and $3N$ potentials, and solve nuclei with $A \leq 4$ using a
VMC method based on a SJ ansatz. More specifically, they obtain a
relativistically corrected expression of the LO pionless-EFT $NN$
interaction of Eq.~\eqref{eq:v_nn_contessi},
\begin{equation}
v_{\mathrm{LO / REL}}^{\mathrm{CI}}(r_{ij}) = - \sum_{i<j}^A
\left(C_1 + C_2\,\sigma_{ij} \right)
\left[ 1 + V_b(\mathbf r_{ij}) + V_t(\mathbf r_{ij}) \right]
\, e^{-\frac{\Lambda^2}{4} r_{ij}^2} \, .
\label{eq:V_rel}
\end{equation}
The boost and transfer interactions in coordinate space read
\begin{align}
V_b(\mathbf r_{ij})
&= - \frac{\hat{\mathbf P}_{ij}^2}{8 m_N^2}
   - \frac{\Lambda^2}{16 m_N^2}\,
     \bigl(\hat{\mathbf P}_{ij} \cdot \mathbf r_{ij}\bigr)^2 ,
\label{eq:Vb}\\[2mm]
V_t(\mathbf r_{ij})
&= - \frac{\Lambda^2}{4 m_N^2}
\left[
  3 - \frac{\Lambda^2}{2} r_{ij}^2
  + 2 i\,\mathbf r_{ij} \cdot \hat{\mathbf p}_{ij}
  + 4\,\frac{\hat{\mathbf p}_{ij}^2}{\Lambda^2}
\right] ,
\label{eq:Vt}
\end{align}
where the total and relative momentum operators of the $i$th and $j$th
nucleons are
\begin{equation}
\hat{\mathbf P}_{ij} = -i \bigl(\boldsymbol{\nabla}_i + \boldsymbol{\nabla}_j\bigr),
\qquad
\hat{\mathbf p}_{ij} = -\frac{i}{2}
\bigl(\boldsymbol{\nabla}_i - \boldsymbol{\nabla}_j\bigr) \, .
\label{eq:Pmom}
\end{equation}
Corrections to the kinetic energy are also included, while those to the
$3N$ force are neglected in that work.

In Ref.~\cite{yang_consistent_2022}, the nuclear Schr\"odinger equation for the above relativistically corrected $NN$ interaction is solved using a VMC method based on an SJ ansatz that respects rotational symmetry. As discussed in Section~\ref{sec:nqs}, spin--isospin dependent correlations are omitted, though their impact is expected to be relatively small.

As noted in Section~\ref{sec:nmbp}, the renormalization behaviour of
few-nucleon systems provides a clear illustration of the role of
relativity in pionless EFT. In the nonrelativistic formulation,
calculations of $^3\mathrm{H}$ and $^4\mathrm{He}$ with only two-nucleon
interactions exhibit the well-known Thomas collapse: as the cutoff is
increased, the ground-state energies diverge, reflecting the lack of
renormalizability of the leading-order theory. The standard remedy is to
promote a repulsive three-nucleon force to leading order in order to
stabilize the spectrum.

\begin{figure}[t]
\centering
\includegraphics[width=0.55\linewidth]{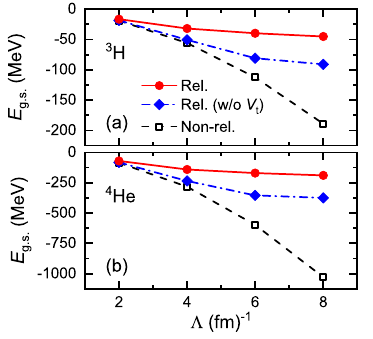}
\caption{\label{fig:relativistic_nuclei} 
(From Ref.~\cite{yang_consistent_2022} with permission from the Authors)
Ground-state energies of $^3$H (panel a) and $^4$He (panel b) obtained
with the nonrelativistic LO Hamiltonian and the relativistic ones
without and with ``transfer'' interactions, as functions of the cutoff
$\Lambda$.}
\end{figure}

As shown in Fig.~\ref{fig:relativistic_nuclei}, taken from
Ref.~\cite{yang_consistent_2022}, in the relativistic framework the
situation changes qualitatively. When the $NN$ potential includes the
relativistic boost and transfer terms, the few-body energies converge
with increasing cutoff, indicating that relativistic dynamics alone can
remove the Thomas collapse without introducing a leading-order
three-nucleon interaction. This stabilizing mechanism is analogous to
that observed in relativistic treatments of three-boson systems. The
boost term generates effective short-range repulsion similar in
character to a three-body force, while the transfer interaction contains
a short-range repulsive core whose strength grows with the cutoff,
preventing the nucleons from approaching arbitrarily close at large
cutoffs.

Reproducing the experimental binding energies nevertheless requires a
$3N$ interaction. In this relativistic framework, the interplay between
relativistic corrections and the $3N$ force leads to a significant
suppression of three-body contributions to the binding energy, largely
independent of the strength of the $3N$ interaction. These findings,
obtained using VMC calculations based on SJ wave functions, provide the
first unified and internally consistent treatment of relativistic
dynamics and many-body forces in light nuclei, and open new avenues for
improving \emph{ab initio} calculations through a more complete
understanding of relativistic effects.

\subsubsection
{Reaching $^{16}$O with systematically improvable ansätze}

As discussed above, the SJ ansatz is not universal, since the Jastrow factor cannot remove the nodes generated by the mean-field component of the wave function. To address this limitation, the authors of Ref.~\cite{lovato_hidden-nucleons_2022} generalized the hidden-fermion family of NQS introduced in Ref.~\cite{moreno_fermionic_2022} to encompass both continuous and discrete degrees of freedom—see Sec.~\ref{sec:nqs} for a detailed discussion of the HN architecture. Using this framework, they solved the nuclear many-body Schr\"odinger equation in a systematically improvable manner. In particular, they showed that augmenting the original Hilbert space with HNs substantially enhances the expressivity of the neural-network ansatz compared to the SJ wave function.

\begin{figure}[t]
\begin{center}
	\includegraphics[width=0.6\textwidth]{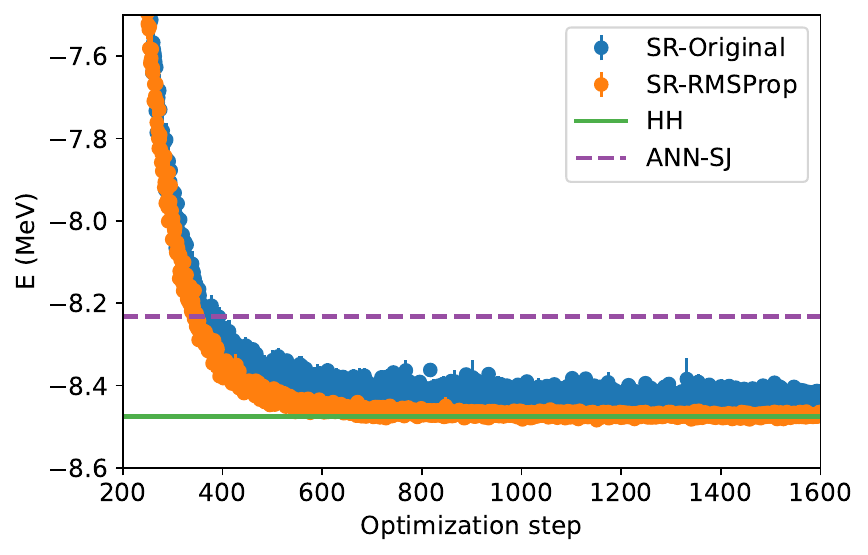} 
	\caption{Convergence of the SR algorithm for $^3$H with the original (blue solid circles) and RMSProp-like (orange solid circles) diagonal shifts. The SJ and the HH energies of Ref.~\cite{Gnech2021} are displayed by the purple dashed and solid green lines, respectively.}
	\label{fig:sr_rmsprop}
\end{center}
\end{figure}

The work in Ref.~\cite{lovato_hidden-nucleons_2022} also demonstrated that explicitly encoding parity and time-reversal symmetries in the wave function, as in Eqs.~\eqref{eq:psi_P} and~\eqref{eq:psi_PT}, significantly accelerates the training. In addition, the convergence of the SR algorithm is substantially improved by introducing the RMSProp-inspired regularization of Eq.~\eqref{eq:sr_rmsprop}. Figure~\ref{fig:sr_rmsprop}, taken from Ref.~\cite{lovato_hidden-nucleons_2022}, shows the convergence of the ground-state energy of $^3$H obtained with $A_h=3$ HNs and the positive-parity ansatz of Eq.~\eqref{eq:psi_P}. The orange solid circles, corresponding to energies obtained using the RMSProp-regularized SR scheme, are systematically closer to the numerically exact hyperspherical-harmonics result of Ref.~\cite{gnech_calculation_2020} than those obtained with the original SR algorithm, shown by solid blue circles. Moreover, the reduced scatter of the SR–RMSProp estimates compared to the SR results indicates improved stability of the optimization. Most notably, regardless of the specific regularization employed, both SR and SR–RMSProp yield energies that are appreciably lower than the SJ value reported in Ref.~\cite{Gnech2021}.

Reference~\cite{lovato_hidden-nucleons_2022} significantly extended the applicability of variational Monte Carlo calculations based on neural-network quantum states to nuclei as large as $^{16}$O, whereas earlier applications~\cite{adams_variational_2021,Gnech2021} were limited to systems with $A\leq 6$. The ground-state energy of $^{16}$O obtained with the HN ansatz was compared with results from the AFDMC method. In particular, comparisons were carried out both with variational energies obtained using the linearized ansatz of Eq.~\eqref{eq:variational_linear} and with full diffusion Monte Carlo results that include imaginary-time projection.

\begin{figure}[!htb]
\centering
\includegraphics[width=0.6\textwidth]{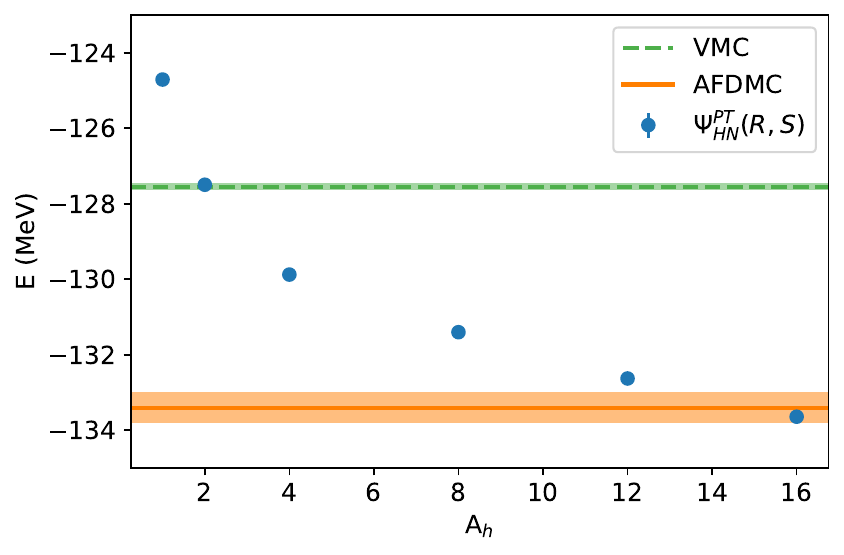} 
\caption{Ground-state energy of $^{16}$O as a function of the number of hidden nucleons $A_h$ (solid blue points). The VMC and AFDMC energies—the latter taken from Ref.~\cite{schiavilla_two-_2021}—are shown by the dashed green and solid orange lines, respectively. The shaded areas represent the corresponding Monte Carlo statistical uncertainties.}
\label{fig:16O_energy}
\end{figure}

Figure~\ref{fig:16O_energy} displays the ground-state energy of $^{16}$O as a function of the number of HNs $A_h$ for the parity- and time-reversal-conserving ansatz of Eq.~\eqref{eq:psi_PT}. For reference, the VMC energy obtained with the correlation operator of Eq.~\eqref{eq:variational_linear} is indicated by the dashed green line, with the shaded band representing the corresponding Monte Carlo statistical uncertainty. The solid horizontal line and the associated shaded region denote the constrained-path AFDMC energy and its statistical uncertainty reported in Ref.~\cite{schiavilla_two-_2021}. Already for $A_h=2$, the HN wave function reproduces the VMC result. Upon further increasing $A_h$, the variational energy is progressively lowered and becomes consistent with the AFDMC value within uncertainties, demonstrating the accuracy of the HN ansatz for $p$-shell nuclei.

A comparable level of accuracy has been achieved in Ref.~\cite{Yang:2023rlw} by augmenting the SJ ansatz with a backflow transformation referred to as \emph{FeynmanNet}. In that work, the authors show that FeynmanNet yields very accurate ground-state energies and wave functions for $^4$He and $^6$Li, and extends to systems as large as $^{16}$O, using leading-order and next-to-leading-order Hamiltonians of pionless effective field theory. Notably, the latter include tensor components, which render the ground-state wave functions complex valued. More specifically, FeynmanNet uses a linear combination of Slater determinants together with a backflow transformation constructed from expressive deep neural networks that act on both the continuous spatial coordinates and the discrete spin–isospin degrees of freedom of the nucleons. To capture many-body correlations induced by tensor and spin–orbit interactions, the networks are designed to represent complex-valued nuclear wave functions. In addition, key features of low-energy nuclear structure, such as the major shell structure and relevant point symmetries, are explicitly encoded in the architecture. As a result, the FeynmanNet ansatz achieves high accuracy while remaining robust and efficient during the training process.

\begin{figure}[!htb]
\centering
\includegraphics[width=0.8\textwidth]{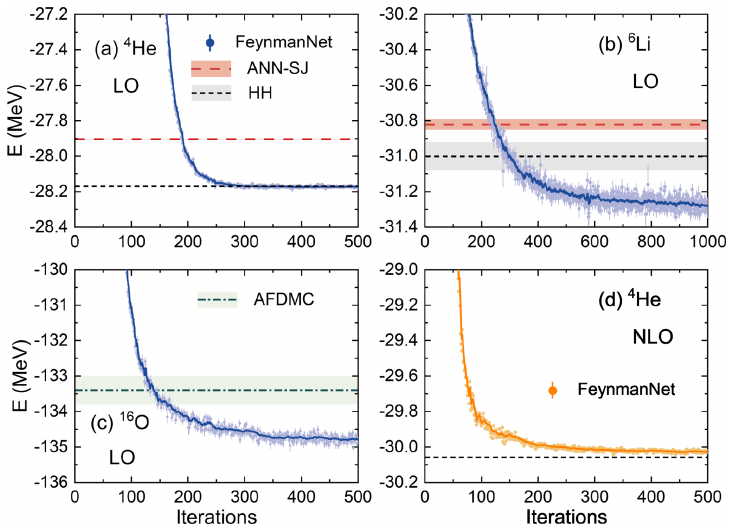} 
\caption{Performance of the FeynmanNet ansatz for the ground states of $^4$He, $^6$Li, and $^{16}$O~\cite{Yang:2023rlw}. Panels (a)–(c) show the convergence of the ground-state energies obtained with the model ``o'' Hamiltonian. Statistical uncertainties from Monte Carlo sampling are indicated by error bars, while the solid curves represent exponential moving averages. Results obtained with the SJ ansatz and the HH method are shown for comparison where available. Panel (d) displays $^4$He results obtained with next-to-leading-order Hamiltonians. }
\label{fig:feynmannet}
\end{figure}

Figure~\ref{fig:feynmannet} illustrates the performance of the FeynmanNet architecture for $^4$He, $^6$Li, and $^{16}$O, using a linear combination of $N_{\rm det}=4$ Slater determinants. Panels (a)–(c) correspond to calculations performed with the model ``o'' Hamiltonian, also used in the HN calculations discussed earlier, which is based on a leading-order pionless effective field theory expansion. Panel (a) shows that the $^4$He energy rapidly converges to a ground-state value consistent with the numerically exact HH result, while improving upon the SJ ansatz. As emphasized in Ref.~\cite{Yang:2023rlw}, this improvement originates from the combined use of multiple determinants and a backflow transformation, which enhances the nodal structure in both spatial and spin–isospin degrees of freedom. For the $p$-shell nucleus $^6$Li, shown in panel (b), where clustering effects increase the complexity of the wave function, FeynmanNet yields lower variational energies than both the SJ ansatz and HH calculations, the latter exhibiting slow convergence for such a halo-like system~\cite{gnech_calculation_2020}. The expressive power of the approach is further demonstrated for $^{16}$O in panel (c), where FeynmanNet attains energies competitive with constrained-path AFDMC results within a purely variational framework. Notably, a comparison with Fig.~\ref{fig:16O_energy} indicates that FeynmanNet achieves lower variational energies than the HN ansatz.
Finally, panel (d) shows results for $^4$He obtained with a next-to-leading-order pionless effective field theory Hamiltonian from Ref.~\cite{schiavilla_two-_2021}, with a three-body range of $R_3 = 2.0$~fm. This Hamiltonian includes tensor and spin–orbit interactions that render the wave function complex valued and reduce the underlying symmetries. Despite this increased complexity, FeynmanNet retains a convergence rate comparable to the leading-order case and reaches energies consistent with HH calculations, highlighting the robustness of the ansatz.

\subsubsection{Essential elements of nuclear binding}
\begin{figure}[!htb]
\centering
\includegraphics[width=0.6\textwidth]{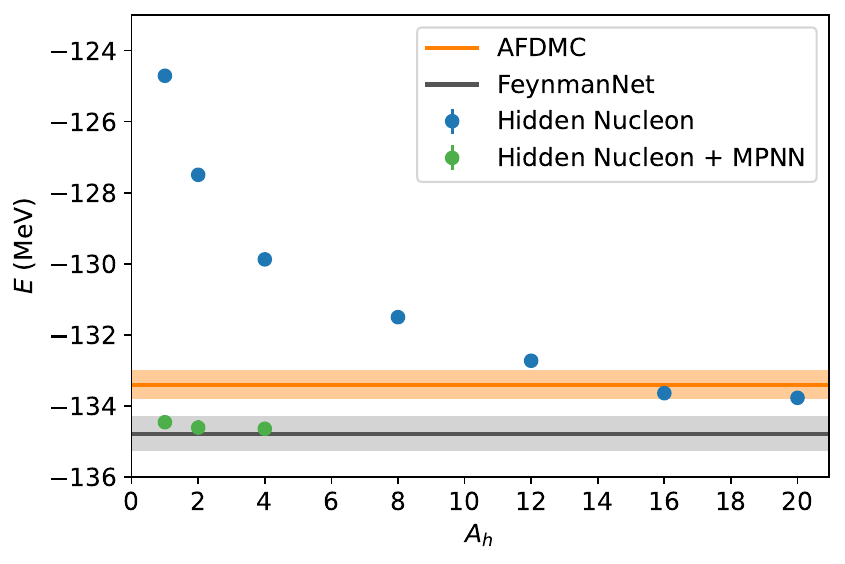} 
\caption{Adapted from Ref.~\cite{gnech_distilling_2024} with permission from the authors. Ground-state energy of $^{16}$O obtained using the Hamiltonian “o” of Ref.~\cite{schiavilla_two-_2021} with $R_3=1.0$ fm. The HN results of Ref.~\cite{lovato_hidden-nucleons_2022} are compared with those obtained by incorporating the simple message-passing neural networks backflow transformations introduced in Ref.~\cite{gnech_distilling_2024}. For reference, the auxiliary-field diffusion Monte Carlo (AFDMC) energy from Ref.~\cite{schiavilla_two-_2021}, together with its uncertainty, is shown by the orange shaded band, while the best FeynmanNet energy reported in Ref.~\cite{Yang:2023rlw} is indicated by the gray band.}
\label{fig:hidden_backflow}
\end{figure}

The authors of Ref.~\cite{gnech_distilling_2024} improved the expressivity of the HN ansatz by introducing backflow transformations acting on the visible coordinates. In particular, they employed the equivariant backflow transformation defined in Eq.~\eqref{eq:mpnn}. The inclusion of backflow substantially increases the flexibility of the HN ansatz, allowing converged energies to be obtained with a HN number $A_h$ that is much smaller than the number of physical nucleons, and in some cases as small as a single HN. Figure~\ref{fig:hidden_backflow}, adapted from the Supplemental Material of Ref.~\cite{gnech_distilling_2024}, shows the convergence of the ground-state energy of $^{16}$O computed using Hamiltonian ``o'' of Ref.~\cite{schiavilla_two-_2021} with $R_3=1.0$, as a function of $A_h$. Even with $A_h=1$, the resulting energy is lower than that obtained with $A_h=16$ in the absence of a backflow transformation. The converged energy is consistent, within uncertainties, with that obtained using the FeynmanNet ansatz, shown in Fig.~\ref{fig:feynmannet}, as well as with the constrained-path AFDMC value.

\begin{figure}[!htb]
\centering
\includegraphics[width=0.55\textwidth]{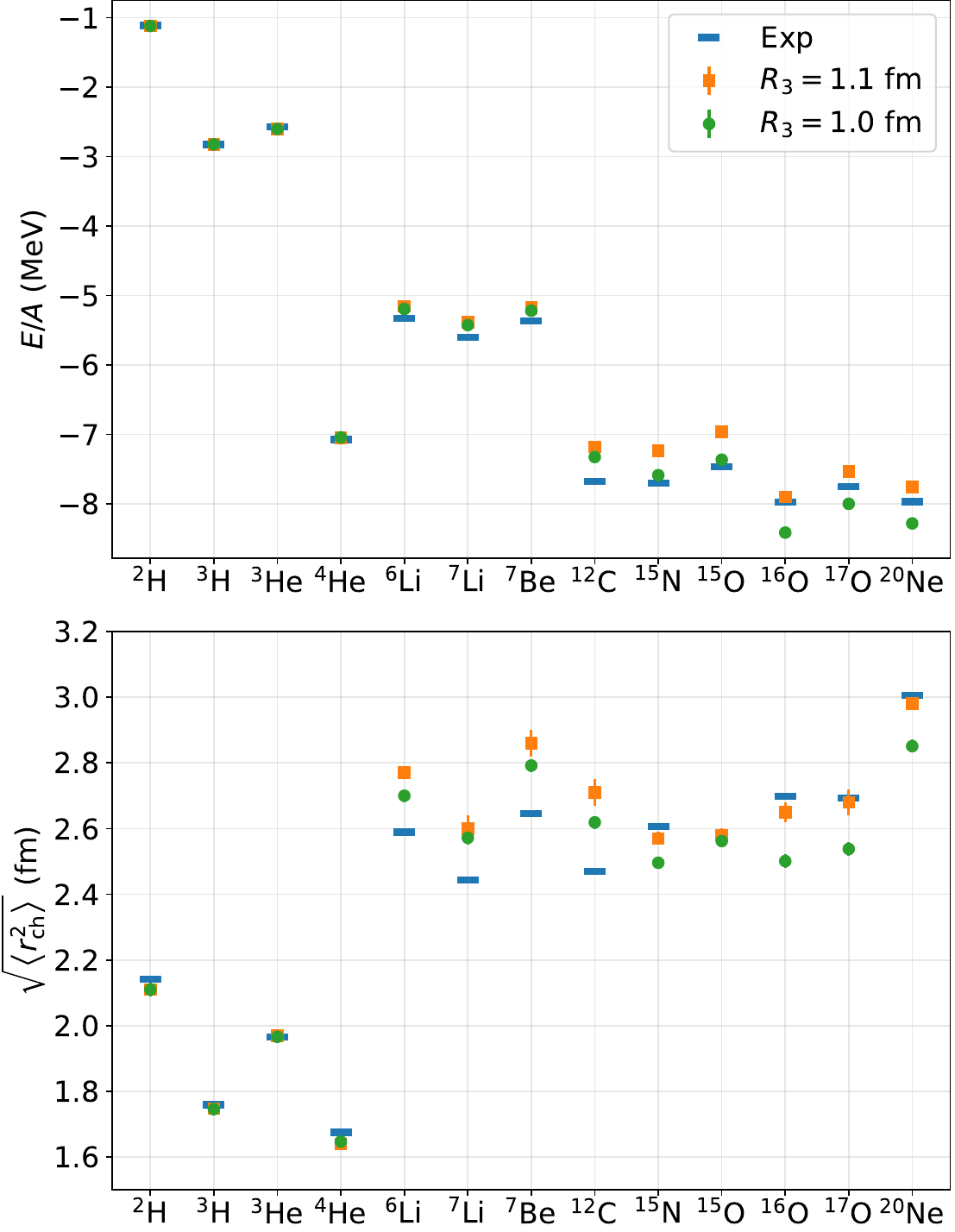} 
\caption{Adapted from Ref.~\cite{gnech_distilling_2024} with permission from the authors. Energies per particle (upper panel) and charge radii (lower panel) of selected nuclei with up to $A = 20$ obtained using the Hamiltonian “o” of Ref.~\cite{schiavilla_two-_2021} with $R_3 = 1.0$ fm and $R_3 = 1.1$ fm, compared with experimental data. }
\label{fig:spectrum_radii}
\end{figure}

This architecture, employing up to $A_h = 4$ HN to ensure convergence, has been used to compute the ground-state energies per nucleon of selected nuclei with $A \leq 20$ for $R_3 = 1.0$ fm and $R_3 = 1.1$ fm. As shown in the top panel of Fig.~\ref{fig:spectrum_radii}, the agreement between the computed and experimental values is remarkably good, given the simplicity of the input Hamiltonian. As noted by the authors of Ref.~\cite{gnech_distilling_2024}, the ground-state energies obtained with this simple Hamiltonian are closer to experiment than those reported in Ref.~\cite{martin_auxiliary_2023} using AFDMC with N$^2$LO chiral EFT interactions. In addition, unlike in NCSM calculations employing consistent N$^2$LO NN and 3N forces~\cite{maris_light_2021}, no increasing overbinding with mass number $A$ is observed.
An important caveat, however, is that neither $R_3 = 1.0$ fm nor $R_3 = 1.1$ fm yield bound ground states for $^6$He, $^8$Li, $^8$B, $^9$C, and $^{17}$F with respect to breakup into smaller clusters. This behavior points to an excessively repulsive character of the Hamiltonian, since increasing $A_h$, and thus the flexibility of the NQS, does not lead to an improvement of the variational energies.

The corresponding charge radii are shown in the lower panel of Fig.~\ref{fig:spectrum_radii}. While the overall trend of the experimental data is well reproduced, consistent with the ground-state energies, the radii of $^6$Li, $^7$Li, $^7$Be, and $^{12}$C are overestimated. By contrast, the radii of $^{15}$N, $^{16}$O, $^{17}$O, and $^{20}$Ne are underestimated with respect to experiment, particularly for $R_3 = 1.0$ fm. Owing to its longer range, the 3N interaction with $R_3 = 1.1$ fm introduces additional repulsion and leads to larger radii in nuclei with $A \geq 15$ compared to the interaction with $R_3 = 1.0$ fm. In contrast to methods based on harmonic-oscillator basis expansions~\cite{caprio_robust_2022}, the radii converge rapidly in VMC-NQS calculations. Consequently, the discrepancies between theoretical predictions and experimental data are most likely attributable to deficiencies in the input Hamiltonian. 

\begin{figure}[!htb]
\centering
\includegraphics[width=0.55\textwidth]{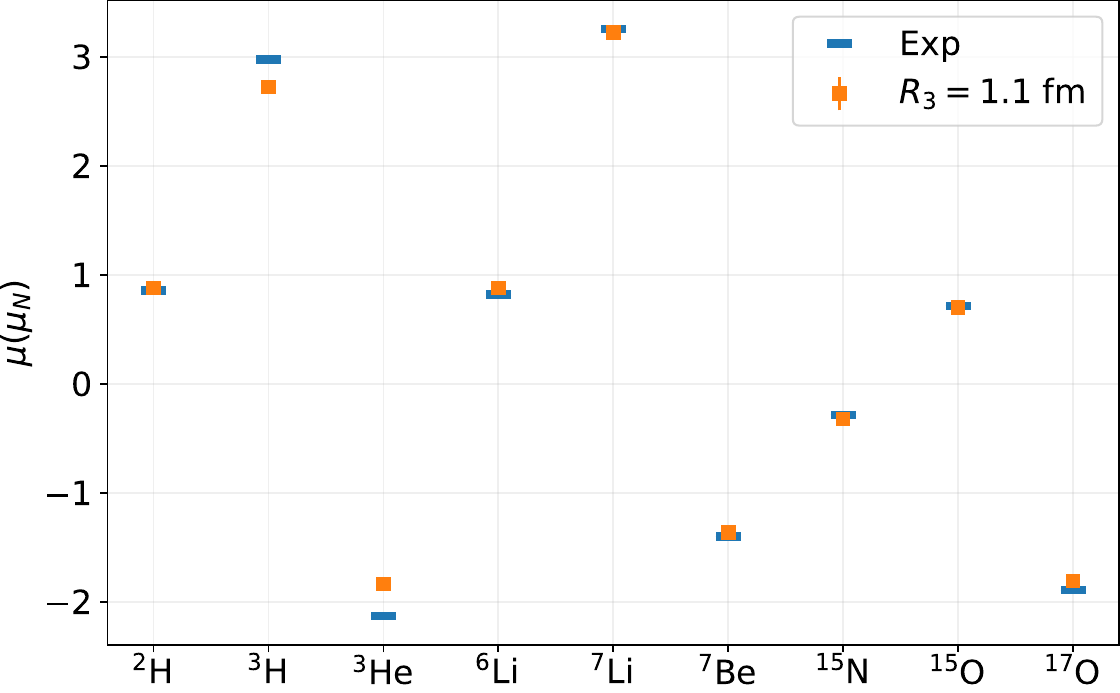} 
\caption{Adapted from Ref.~\cite{gnech_distilling_2024} with permission from the authors. Magnetic moments of selected nuclei with $A \leq 20$ obtained from VMC-NQS calculations using the Hamiltonian “o” of Ref.~\cite{schiavilla_two-_2021} with $R_3 = 1.1$ fm, compared with experimental data.}
\label{fig:moments}
\end{figure}

The magnetic moments of selected nuclei with $A \leq 20$, computed using model “o” with $R_3 = 1.1$ fm, are shown in Fig.~\ref{fig:moments}. The theoretical predictions are in good agreement with experimental data, indicating that the NQS captures the nuclear shell structure, which emerges naturally during the energy minimization. Notably, at the beginning of the training, not only the ground-state energies and charge radii, but also the magnetic moments and angular momenta, differ substantially from their converged and experimentally observed values. The minor discrepancies observed for $^3$H and $^3$He, consistent with GFMC, AFDMC, and HH results~\cite{pastore_quantum_2013,gnech_magnetic_2022,martin_auxiliary_2023}, are likely due to missing two-body current contributions.

\subsubsection[High-resolution potentials]{High-resolution potentials}

Most nuclear-physics applications of VMC methods with NQS take as input interactions based on pionless effective field theory. After the pioneering 2019 work~\cite{keeble_machine_2020}, non-stochastic NQS-based approaches have been employed to solve light nuclear systems with high-resolution interactions. In this context, the authors of Ref.~\cite{wang_neural_2024} introduced a compact neural-network architecture based on a partial-wave expansion of the nuclear wave function, in which the radial components are represented by separate neural networks. This method was shown to accurately reproduce the deuteron ground-state energy and wave function starting from the highly realistic Argonne $v_{18}$ NN potential~\cite{wiringa_accurate_1995}, including its full spin–isospin operator structure. More recently, closely related deterministic neural-network approaches have been extended beyond the two-body sector. In particular, Ref.~\cite{li_solving_2025} introduced an unsupervised deep-learning framework to solve both two- and three-body bound-state problems directly in coordinate space. Building on a discretized representation of the Schrödinger equation, the authors employed deep neural networks to represent the radial wave functions of coupled channels, achieving accurate calculations of $^2$H and $^3$H using the Entem–Machleidt N$^3$LO chiral-EFT potential of Ref.~\cite{Entem2003}.

By contrast, actual VMC calculations based on NQS have more recently been carried out using high-resolution local Hamiltonians that are either phenomenological or derived within chiral EFT~\cite{yang_chiral_2025,wen_neural-network_2025,yang_zemach_2025}. In particular, the authors of Ref.~\cite{yang_chiral_2025} presented QMC calculations of neutron–$\alpha$ D-wave phase shifts using chiral-EFT Hamiltonians at the three lowest orders in the chiral expansion: LO, NLO, and N$^2$LO. Specifically, they employed the local $NN$ and $3N$ potentials constructed in Ref.~\cite{gezerlis_quantum_2013,lynn_quantum_2014}. However, the analysis was limited to the softest cutoff value, $R_0 = 1.2$ fm, which corresponds to a typical momentum cutoff $\Lambda \simeq 400$ MeV.

As in previous QMC studies of neutron--$\alpha$ scattering, the continuum problem is mapped onto a bound-state eigenvalue problem. To this end, an external harmonic-oscillator (HO) confining potential is added to the nuclear Hamiltonian through an additional two-body term of the form
\begin{equation}
V_{\text{HO}} = \sum_{i<j} \frac{1}{2}\,\frac{m_N}{A}\,\omega^2 r_{ij}^2 \, ,
\end{equation}
where $m_N$ denotes the nucleon mass, $\omega$ the HO frequency, and $r_{ij}$ the relative distance between nucleons. The presence of the HO trap discretizes the spectrum, enabling the extraction of scattering information from bound-state energies. The neutron--$\alpha$ D-wave phase shifts are then obtained using the Busch--Englert--Rza\.{z}ewski--Wilkens (BERW) formalism~\cite{busch_two_1998,buttiker_pionnucleon_2000}, which relates the trapped eigenenergies of $^5$He and the ground-state energy of $^4$He to free-space scattering observables.

In Ref.~\cite{yang_chiral_2025}, the energies of both $^4$He and $^5$He are computed using the GFMC method. The lowest $J^\pi = 5/2^+$ eigenstate of $^5$He is projected via imaginary-time propagation, starting from a previously optimized NQS. The trial wave function is taken in the form of Eq.~\eqref{eq:SJ_chiral}, augmented by the transformation defined in Eq.~\eqref{eq:SJ_chiral_2}.
Despite the use of highly sophisticated variational wave functions, the GFMC propagation suffers from a severe fermion-sign problem. This issue is mitigated through a combination of constrained-path propagation followed by a transient estimate~\cite{wiringa_quantum_2000}. Nevertheless, the sign problem remains sufficiently severe that only the softest available cutoff value, $R_0 = 1.2$ fm, could be employed in the chiral nuclear Hamiltonian.

\begin{figure}[!htb]
\centering
\includegraphics[width=0.5\columnwidth]{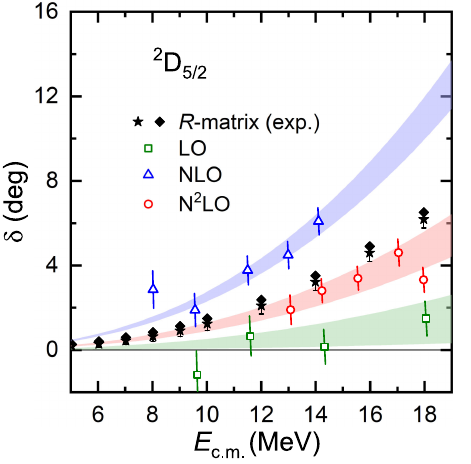}
\caption{From Ref.~\cite{yang_chiral_2025}, reproduced with permission. Neutron–$\alpha$ phase shifts in the ${}^2D_{5/2}$ channel as a function of the center-of-mass energy, predicted at different orders of chiral EFT. Empty symbols denote GFMC results with statistical error bars, while shaded bands indicate the combined statistical and BERW-related systematic uncertainties. Stars correspond to $R$-matrix analyses of experimental neutron–$\alpha$ elastic-scattering data~\cite{bond_determination_1977}, and red diamonds are taken from a private communication between the authors of Ref.~\cite{yang_chiral_2025} and G.~M.~Hale.}
\label{fig:n_alpha_scattering}
\end{figure}

Figure~\ref{fig:n_alpha_scattering} illustrates the neutron–$\alpha$ D-wave phase shifts predicted at different orders of chiral EFT as a function of the center-of-mass energy $E_{\rm c.m.}$. Each point corresponds to an individual GFMC calculation performed in a harmonic-oscillator trap and is extracted using the BERW formalism. The theoretical predictions are compared with phase shifts obtained from $R$-matrix analyses of elastic-scattering data~\cite{bond_determination_1977}.
At leading order, the chiral Hamiltonian yields nearly vanishing phase shifts, in line with general expectations. The dominant contribution emerges at next-to-leading order with the inclusion of the two-pion-exchange $NN$ interaction, but the resulting phase shifts significantly overestimate the empirical values. Corrections at next-to-next-to-leading order reduce the phase shifts and bring the theoretical predictions into closer agreement with the data. At this order, the leading $3N$ force plays a crucial role, while the subleading two-pion-exchange $NN$ interaction has a comparatively minor impact for the soft regulator values considered. Overall, these results identify neutron–$\alpha$ D-wave scattering as a sensitive probe of the long-range structure of the three-nucleon interaction.

In a later work~\cite{yang_zemach_2025}, the same authors improved their variational ansatz by adopting a fully multiplicative form of two-body correlations, rather than the linearized version of Eq.~\eqref{eq:SJ_chiral}. Using this more sophisticated wave function, they computed the binding energies of $^4$He and $^6$Li with accuracy comparable to GFMC and significantly improved relative to earlier VMC and AFDMC calculations. Notably, because VMC does not suffer from a fermion-sign problem, they were able to employ local chiral-EFT interactions from Refs.~\cite{gezerlis_quantum_2013,lynn_quantum_2014} with both $R_0 = 1.2$ fm and the harder cutoff $R_0 = 1.0$ fm. In addition, they considered the high-resolution phenomenological Argonne $v_8^\prime$ plus UIX Hamiltonian.

These calculations resolved the longstanding discrepancy between effective and elastic Zemach radii in $^6$Li and $^7$Li through ab initio nuclear-structure calculations that consistently include nuclear polarizability effects. Enabled by the use of highly sophisticated NQS, the results show that nuclear polarizability effects are negligible in $^7$Li but dominant in $^6$Li, thereby accounting for the observed deviation between the effective and elastic Zemach radii.

The authors of Ref.~\cite{wen_neural-network_2025} employed the wave-function ansatz of Eq.~\eqref{eq:wen_u_corr} to study $A = 3$ nuclei, using as input the local chiral-EFT Hamiltonians of Refs.~\cite{gezerlis_local_2014,lynn_chiral_2016}. Importantly, these interactions include tensor and spin–orbit components in the $NN$ sector, as well as a consistent spin-isospin-dependent three-nucleon force. As discussed in Section~\ref{sec:chiral_ansatz}, a normalizing flow is used to generate the samples required for the Monte Carlo estimation of the energy and its gradient. Notably, the use of normalizing flows avoids the correlation-length problem that typically affects Markov chain Monte Carlo algorithms.

\begin{figure}[!htb]
\centering
\includegraphics[width=0.6\columnwidth]{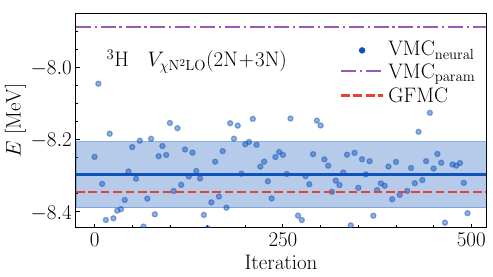}
\caption{From Ref.~\cite{wen_neural-network_2025}, reproduced with permission. Evaluation of the $^3$H ground-state energy obtained with optimized neural-network wave functions using the full N$^2$LO chiral interaction with $(R_0,\Lambda) = (1.2\,\mathrm{fm}, 1\,\mathrm{GeV})$. The average energy (blue horizontal line) and standard deviation (blue shaded band) are estimated over 500 iterations with 40{,}000 samples per iteration. The conventional VMC result is displayed by the purple line}
\label{fig:chiral_A_3}
\end{figure}

Figure~\ref{fig:chiral_A_3} displays the ground-state energy of $^3$H obtained using the N$^2$LO chiral interaction with regulator values $(R_0,\Lambda) = (1.2,\mathrm{fm}, 1,\mathrm{GeV})$. The energy and its uncertainty are estimated over 500 iterations, with 40,000 samples per iteration generated by the normalizing flow. The average NQS energy is $E = -8.30 \pm 0.09$ MeV, which closely matches the GFMC result, $E = -8.34$ MeV. The two values agree within one-half of a standard deviation and differ by only $0.45$\%.

These results indicate that the neural-network-generated wave function can effectively approximate the $^3$H ground state with the soft-core interaction, without any imaginary-time evolution, even in the presence of complicated pion-exchange terms as well as tensor, spin-orbit, charge-independence-breaking, and charge-symmetry-breaking interactions.

\subsection{Nuclear matter}
\label{sub:matter}

Multi-messenger astronomy is creating new observations of matter at densities and isospin asymmetries which are not directly accessible by terrestrial experiments, expanded most recently with the observation of GW170817~\cite{LIGOScientific:2017vwq,Abbott:2017,Sabatucci:2020xwt,senger2021}. Theoretical modeling of this matter, in particular material similar to the inner crust of neutron stars, poses significant challenges due to the inherent complexity of clustering phenomena, the emergence of superfluidity, and the existence of both free neutrons and neutron-rich nuclei. Various phenomenological approaches have been developed to address some of these challenges, including the compressible liquid-drop model~\cite{ Baym:1971ax,Baym:1971pw,Ravenhall:1983uh,douchin2001,Newton:2011dw,Gulminelli:2015csa,Lim:2017luh,Carreau:2019zdy,Grams:2021lzx} and self-consistent mean-field models~\cite{negele1973a,Baldo:2006jr,Grill:2011dr}. In the former, the energy is parameterized as a function of global properties, and the nucleons within clusters are treated separately from the free neutrons, all of which are assumed to be uniformly distributed. While this treatment has a low computational cost, it neglects quantum mechanical shell effects, which are critical for determining the equilibrium composition of the crust. In contrast, self-consistent mean-field models are fully quantum mechanical, constructing the many-body state in terms of single-particle or quasi-particle wave functions. However, since they neglect correlations beyond the mean-field approximation, they have limited applicability for low-density nuclear matter, where strong correlations may lead to significant deviations from the mean-field behavior.

Recently, VMC methods utilizing NQS have been used to model nuclear matter without making similar, limiting assumptions within both pure neutron matter (PNM) and symmetric nuclear matter (SNM) using a pionless EFT potential~\cite{fore2023,fore_investigating_2024}. The initial results show promise through capturing signs of density dependent neutron superfluidity in PNM as well as successfully modeling clustering in nuclear matter, without any assumptions on the structure of the material.

\subsubsection{Pure neutron matter}

\begin{figure}[!htb]
\centering
    \includegraphics[width=0.6\columnwidth]{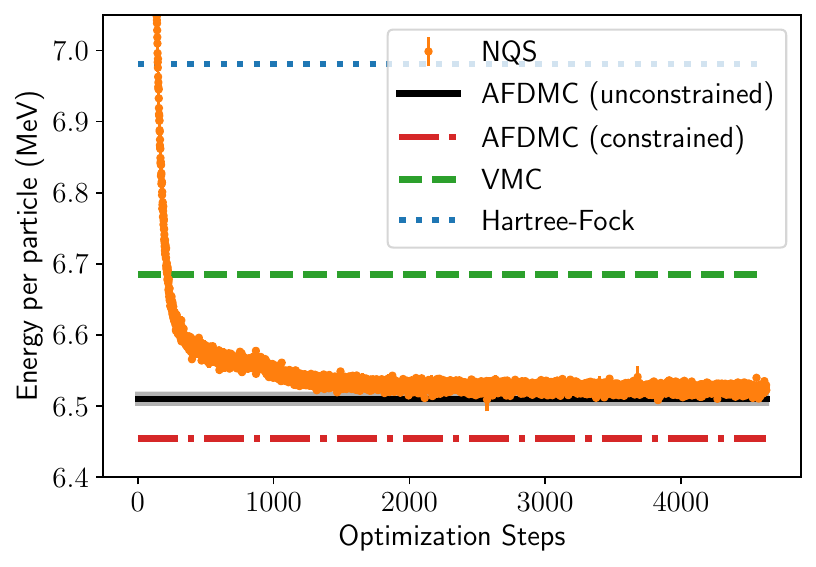}
\caption{NQS training data in neutron matter at $\rho = 0.04$ fm$^{-3}$ (data points) compared with Hartree-Fock (dotted line), conventional VMC (dashed line), constrained-path AFDMC (dash-dotted line) and unconstrained-path AFDMC results (solid line).}
\label{fig:PNM_training}
\end{figure}

Low-density neutron matter is characterized by fascinating emergent quantum phenomena, such as the formation of Cooper pairs and the onset of superfluidity.
Quantum Monte Carlo approaches~\cite{carlson_quantum_2015}, in particular the auxiliary-field diffusion Monte Carlo (AFDMC) method~\cite{Schmidt:1999lik}, have been extensively applied to accurately compute neutron-matter properties~\cite{lonardoni_nuclear_2020,piarulli_benchmark_2020,Lovato:2022apd}. In the low-density regime, AFDMC calculations have convincingly shown a depletion of the superfluid gap with respect to BCS theory~\cite{Gandolfi:2008id,Gandolfi:2022dlx}. However, because of the fermion sign problem, AFDMC predictions depend upon the starting variational wave function. For instance, the superfluid phase must be assumed {\it a priori}. No assumption of the phase is necessary when NQS are used to compute the initial variational state. It can be seen in Fig.~\ref{fig:PNM_training} that the final trained state for a simulation of 14 neutrons with periodic boundary conditions that the variational NQS is able to recover almost an identical ground state energy to the unconstrained AFDMC calculation.

\begin{figure}
\includegraphics[width=\columnwidth]{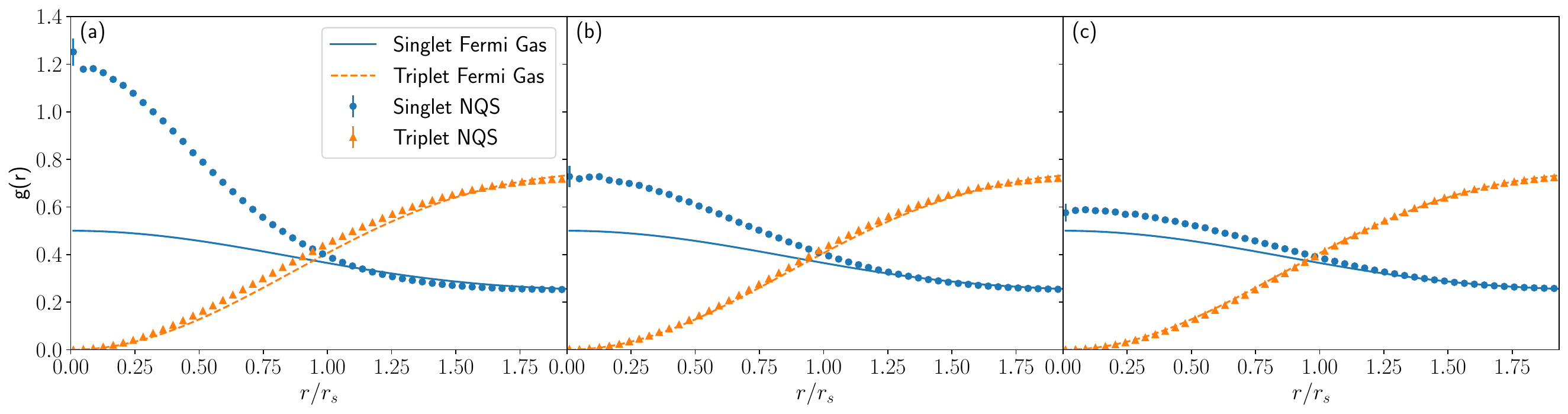}
\caption{Spin-singlet and triplet two-body distribution functions at $\rho=0.01$ fm$^{-3}$ (panel a), $\rho = 0.04$ fm$^{-3}$ (panel b), and $\rho = 0.08$ fm$^{-3}$ (panel c) vs pair distance in units of the Wigner-Seitz radius. The NQS calculations (solid symbols) are compared with non-interacting Fermi Gas results (solid lines).}
\label{fig:pair_dist}
\end{figure}

The trained NQS wave functions from these simulations are also used to evaluate two-body pair distributions as defined in Ref.~\cite{Gandolfi2009}. Fig.~\ref{fig:pair_dist} shows these distributions at $\rho=0.01$ fm$^{-3}$ (panel a), $\rho=0.04$ fm$^{-3}$ (panel b), and $\rho=0.08$ fm$^{-3}$ (panel c). The significant increase in the spin-singlet channel compared to the non-interacting Fermi Gas indicates that the NQS wave function can capture the emergence of the $^1S_0$ neutron pairing, despite not being explicitly encoded in the ansatz. Consistent with the behavior of the pairing gap~\cite{Gandolfi:2008id,Benhar:2017mof}, the enhancement is more prominent at low densities and vanishes at higher densities. On the other hand, at these densities, no pairing correlations are present in the spin-triplet channel.

\subsubsection{Clustering}

Simulations of matter with protons as well as neutrons are complicated by the nuclear interaction acting simultaneously in distinct spin-isospin channels, most notably the $^1S_0$ and $^3S_1$ channels, rather than collapsing to a single dominant channel as in pure neutron matter (see Section \ref{sub:hamiltonian}). This richer structure leads to pronounced clustering that is difficult for many methods to capture without simplifying assumptions. NQS, however, excel at learning these complex relationships. 

\begin{figure}[!t]
\centering
\includegraphics[width=0.6\columnwidth]{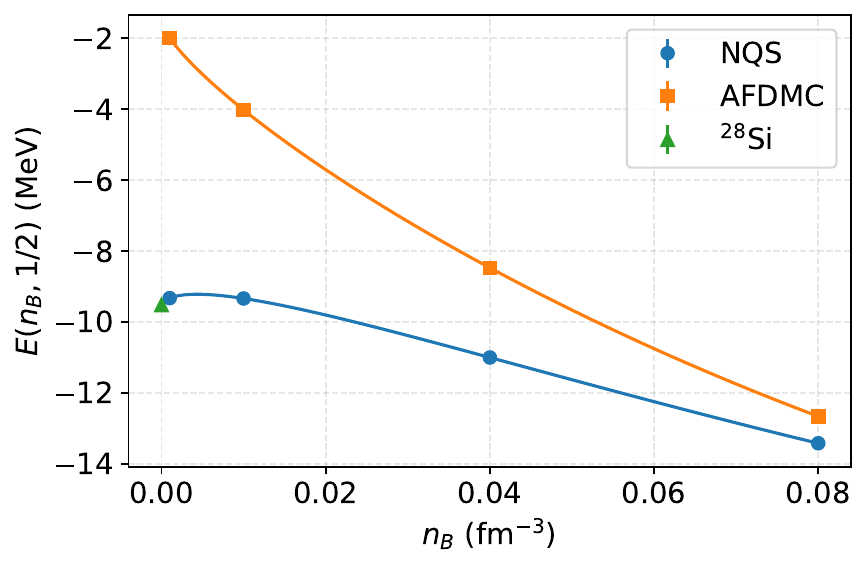}
\caption{Energy per particle of SNM from NQS and AFDMC simulations. The NQS results (blue circles) indicate stronger binding compared to AFDMC (orange squares), attributed to clustering at low densities. Our low-density results are compared to $^{28}$Si with no electromagnetic contribution, which serves as the expected zero-density ground state for a simulation of $N=28$ nucleons with $x=1/2$. The error bars represent 95\% confidence intervals for the mean.}
\label{fig:SNM_energy}
\end{figure}

Reference~\cite{fore_investigating_2024} utilizes the MPNN backflow method with the Pfaffian--Jastrow architecture to efficiently learn pairing correlations within isospin symmetric and asymmmetric matter. With this approach, NQS are able to correctly learn the structure of matter through a range of densities corresponding to the inner crust of neutron stars. Figure~\ref{fig:SNM_energy} shows ground state energies learned by the NQS are significantly improved compared to state-of-the-art AFDMC calculations given identical assumptions. For a simulation containing 28 particles, the expected low-density limit of periodic symmetric matter corresponds to the ground-state energy of $^{28}$Si in the absence of electromagnetic interactions. The NQS results approach this limit, providing clear evidence that clustering effects are being accurately captured. Further improvements may be achieved by initializing AFDMC calculations from a NQS wave function. 

\subsubsection{Beta-equilibrated matter}

Cold nuclear matter at densities found in the inner crust of neutron stars is mostly composed of neutrons, protons, and electrons. Assuming the matter is neutrino transparent, the relative abundance of these particles at a specific density can be determined using a model for the energy as a function of proton fraction, under the well-supported assumptions of charge neutrality, $n_p = n_e$, and beta equilibrium, $\mu_e=\mu_n-\mu_p$~\cite{benhar2023}. This model, along with simulation results for SNM and PNM, can be used to construct an equation of state for beta-equilibrated nuclear matter given a particular Hamiltonian. 

At zero temperature, the beta-equilibrium condition directly relates the proton fraction to the density dependence of the symmetry energy, defined as the difference between the energies of PNM and SNM. Working out these relations yields
\begin{equation}
\mu_n - \mu_p = -\frac{\partial E(n_B,x)}{\partial x} = 4(1-2x) S(n_B),
\label{eq:mu_hat}
\end{equation}
together with the expression for the electron density,
\begin{equation}
x \equiv \frac{n_p}{n_B} = \frac{n_e}{n_B} = \frac{(\mu_e^2 - m_e^2)^{3/2}}{3\pi^2 n_B},
\label{eq:x_from_ne}
\end{equation}
which jointly determine the proton fraction in beta-equilibrated matter.

Due to the low proton fractions in the crust, directly modeling nuclear matter at these isospin asymmetries may require hundreds of nucleons to suppress finite-size effects associated with clustering. Such calculations are currently infeasible with \emph{ab initio} variational Monte Carlo methods. Instead, in Ref.~\cite{fore_investigating_2024} simulations of PNM and SNM are performed, and ground-state energies at intermediate isospin asymmetries are inferred using well-known expansions in the neutron richness, given by $E(n_B,x)=E(n_B, 1/2)+(1-2x)^2 S(n_B)$.

\begin{figure}[!h]
\centering
\includegraphics[width=0.7\columnwidth]{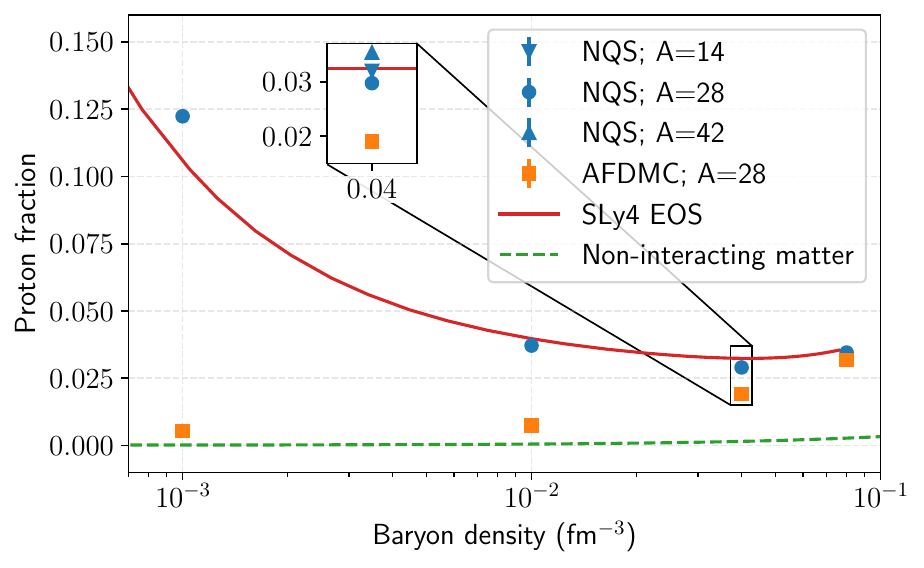}
\caption{Proton fraction in beta-equilibrated matter as a function of baryon density. Blue circles and orange squares represent NQS and AFDMC calculations for $A=28$ nucleons, respectively. Blue up-triangles and down-triangles indicate NQS results for $A=42$ and $A=14$, respectively, highlighting finite-size effects. The green dashed line shows non-interacting matter, and the red solid curve represents a Skyrme parametrization of the equation of state (SLy4 EOS) for the inner crust of neutron stars~\cite{chabanat1998, douchin2001}. The inset displays NQS results for varying particle numbers at a density of 0.04 fm$^{-3}$ to illustrate finite-size effects.}
\label{fig:x_vs_rho}
\end{figure}

The resulting proton fraction in beta-equilibrated matter as a function of baryon density is displayed in Fig.~\ref{fig:x_vs_rho}. The NQS predictions align closely with the phenomenological Skryme parameterizations of the equation of state (SLy4 EOS) for the inner crust of neutron stars~\cite{douchin2001,chabanat1998}. Figure~\ref{fig:x_vs_rho} underscores the flexibility of the NQS ansatz: by leveraging identical neural network architectures across all densities, they successfully capture the complexity of nuclear structures relevant to the inner crust of neutron stars. In contrast, the AFDMC yields proton fractions that tend towards those of non-interacting matter. This difference arises because AFDMC does not account for the onset of nuclear clusters at low densities, which significantly lowers the energy per particle in isospin-symmetric nuclear matter, as captured by NQS. For instance, at $n_B=0.001$ fm$^{-3}$, the NQS predicts an energy per particle that is 7.334(46) MeV lower than AFDMC, while at $n_B=0.08$ fm$^{-3}$, this difference reduces to 0.758(35) MeV.
In contrast, both methods yield similar energies for pure neutron matter, where clustering effects are absent, as shown in Ref.~\cite{fore2023}. As a result, when solving the beta-equilibrium equations, the NQS predicts a higher proton fraction compared to AFDMC, reflecting its ability to capture nuclear clustering effects at low densities. All in all, these initial results showcase the potential of NQS simulations for dense matter physics of relevance for astrophysical simulations.


\subsection{Electroweak interactions}
\label{sub:electroweak}
Studying the response of a many-body system to perturbative probes is of great relevance for extracting information about the system’s dynamical structure at both the nuclear and nucleon levels~\cite{bacca_electromagnetic_2014}. These calculations are crucial for interpreting electron-scattering experiments, including assessing whether explicit QCD effects are required to explain the measured response functions~\cite{lovato_charge_2013,cloet_relativistic_2016,lovato_electromagnetic_2016}. Moreover, they are essential for fully exploiting current and next-generation accelerator-based neutrino oscillation experiments~\cite{lovato_ab_2020,acharya_16o_2024}, which use atomic nuclei in their detectors to enhance event rates.

In the linear-response regime, the interaction of an atomic nucleus with electroweak probes is described by the nuclear response function
\begin{equation}\label{eq:resp}
  {\cal R}(\omega)=\sumint_f |\langle\Psi_f | \hat O | \Psi_0 \rangle |^2 \delta{(E_f-E_0-\omega)} \,,
\end{equation}
which encapsulates the dynamical part of the reaction.
In the above equation, \( \hat O \) denotes the excitation operator, while
\( \lvert \Psi_0 \rangle \) and \( \lvert \Psi_f \rangle \) represent the initial and final
states of the system, with energies \( E_0 \) and \( E_f \), respectively.
The final state may be either bound (belonging to the discrete part of the spectrum)
or unbound (in the continuum).
A direct evaluation of the response function therefore, in principle,
requires calculating each individual transition amplitude.

Microscopically computing continuum wave functions for medium-mass nuclei remains an outstanding theoretical challenge. At each continuum energy, the many-body wave function fragments into numerous channels, each representing a distinct breakup configuration, leading to substantial complexity in both momentum and coordinate representations. Integral transform techniques, such as the Laplace transform~\cite{carlson_euclidean_1992} and the Lorentz integral transform (LIT)~\cite{efros_lorentz_2007}, circumvent the difficulties associated with the direct calculation of continuum final states, reformulating the problem as a bound-state calculation. 

The LIT is defined as
\begin{equation} \label{LIT}
  {\cal L}(\omega_0,\Gamma)=\int_{\omega_{th}}^\infty d\omega
  \frac{{\cal R}(\omega)}{(\omega-\omega_0)^2+\Gamma^2}\,,
\end{equation}
where \( \omega_{th} \) is the threshold energy and \( \Gamma>0 \) is the Lorentzian width, serving as a resolution parameter. The LIT method proceeds in two steps. First, \( {\cal L}(\omega_0,\Gamma) \) is computed directly, without requiring explicit knowledge of \( {\cal R}(\omega) \). In a second step, the dynamical response function is obtained through an inversion of the LIT.

The function ${\cal L}(\omega_0,\Gamma)$ can be computed starting from its definition and substituting the expression in Eq.~\eqref{eq:resp} for ${\cal R}(\omega)$. Using the completeness relation of the Hamiltonian eigenstates 
\begin{equation}
\sumint_f | \Psi_f \rangle \langle \Psi_f | =1 
\end{equation}
yields 
\begin{align}
{\cal L}(\omega_0,\Gamma)=\frac{\Gamma}{\pi} \langle \Psi_0|{ \hat O}^\dagger \frac{1}{( \hat H - z) } \frac{1}{(\hat H - z^\ast)} \hat O |\Psi_0 \rangle\, .
\end{align}
We denote with $\Psi_L$ the solution of the inhomogeneous  Schr\"odinger equation
\begin{equation}\label{psiL}
( \hat H - z) |\Psi_L\rangle = \hat O | \Psi_0 \rangle \;,
\end{equation}
where \(z = E_0 + \omega_0 + i\Gamma\). Finding the solution for different values of $\omega_0$ and $\Gamma$ lead directly to the transform
\begin{equation}
{\cal L}(z)= \frac{\Gamma}{\pi}\langle \Psi_L |\Psi_L\rangle\, .
\end{equation}
Because ${\cal L}(z)$ is finite, the solution of Eq.~\eqref{psiL} has the same asymptotic boundary conditions as a bound state.

In Ref.~\cite{parnes_nuclear_2026}, the LIT of the dipole operator was computed for the first time using NQS, specifically the Pfaffian--Jastrow ansatz of Eq.~\eqref{eq:pf_even}, to represent both \( \Psi_0 \) and the LIT state \( \Psi_L \). In that work, the LIT equation is solved by maximizing the quantum fidelity between the auxiliary state
\( \lvert \Psi \rangle \equiv (\hat H - z)\lvert \Psi_L \rangle \)
and the source state
\( \lvert \Phi \rangle \equiv \hat O \lvert \Psi_0 \rangle \),
where the fidelity is defined as~\cite{sinibaldi_unbiasing_2023}
\begin{equation}
\mathcal{F}(\Psi, \Phi) =
\frac{\langle \Psi | \Phi \rangle \langle \Phi | \Psi \rangle}
{\langle \Psi | \Psi \rangle \langle \Phi | \Phi \rangle}\,.
\end{equation}
The optimal parameters $\params$ of the NQS are obtained by maximizing a linear combination of the fidelity defined above and a reverse Kullback--Leibler divergence between the target and proposal distributions $\mathrm{KL}\!\left(\pi_{\Psi}\,\|\,\pi_\Phi\right)$,
where 
\begin{equation}
\pi_{\Psi}(X)=\frac{|\Psi(X)|^2}{\langle\Psi|\Psi\rangle}\quad , \quad \pi_\Phi(X)=\frac{|\Phi(X)|^2}{\langle\Phi|\Phi\rangle}\, .
\end{equation}
The gradient of the loss function reads
\begin{equation}
g_{\params}=\nabla_{\params}\mathcal{F}
-\lambda\,\nabla_{\params}\mathrm{KL}\!\left(\pi_{\Psi}\,\|\,\pi_\Phi\right)\,,
\end{equation}
with \( \lambda=1 \) in typical applications.
Stable parameter updates are achieved by employing the \textsc{spring} algorithm (see Section~\ref{subsec:opt})~\cite{goldshlager_kaczmarz-inspired_2024}, which leads to the damped linear system
\(
\bigl(S+\epsilon I\bigr)\delta{\params}
=g_{\params}+\epsilon\,\mu\,\delta{\params}_{\text{prev}}\,.
\)
To ensure scale invariance, the authors of Ref.~\cite{parnes_nuclear_2026} set \( \epsilon=\varepsilon\langle\mathrm{diag}(S)\rangle \)
with a small \( \varepsilon=10^{-3} \).
Note that maximizing the fidelity only fixes the auxiliary state $|\Psi\rangle$ up to an overall complex normalization constant~\cite{hendry_machine_2019}, given by $\mathcal{N} = \langle \Phi | \Psi\rangle / \langle \Phi | \Phi \rangle$, which is estimated stochastically by sampling from $\pi_{\Phi}(X)$. 


Computing the LIT as the norm of \(\Psi_L\) is computationally challenging due to the slowly decaying and oscillatory tails of this wave function. To address this difficulty, one can rewrite the norm as  
\begin{equation}
\langle\Psi_{L}|\Psi_{L}\rangle=\langle\Phi|\frac{1}{\hat{H}-z^{*}}\frac{1}{\hat{H}-z}|\Phi\rangle\nonumber=\frac{1}{\Gamma}\operatorname{Im}\langle\Phi | \Psi_{L}\rangle
\label{eq:lit_xilin}
\end{equation}
which can instead be estimated in a numerically stable manner by sampling from \( \pi_\Phi(X) = |\Phi(X)|^2 / \langle \Phi | \Phi \rangle \). 
The authors of Ref.~\cite{parnes_nuclear_2026} also provide the following upper bound on the LIT uncertainty
\begin{align}
\Delta\mathcal{L}(\omega_{0},\Gamma) &\leq 
\mathcal{D} \; \frac{\mathcal{N}^{-1}||\Phi\rangle|}{\Gamma} \sqrt{\frac{1 - \mathcal{F}  }{ \mathcal{F} }},
\label{eq:error}
\end{align}
where
\begin{equation}
    \mathcal{D} = \min\Big(
\left|(1 - P_{\Phi})|\Psi_L\rangle\right|,\Big|(1 - P_{\Phi})  \frac{H}{\left|z\right|}|\Psi_L\rangle\Big|
\Big),
\end{equation}
and $P_\Phi=|\Phi\rangle\langle \Phi|/\langle \Phi|\Phi\rangle$. This relation implies that in the limit $\Gamma\to 0$, $\omega_0\approx\omega$, and
${\cal L} \to (\pi/\Gamma) {\cal R}$ one finds $\Delta {\cal L}/{\cal L} \propto \sqrt{(1-{\cal F})/\Gamma}$.
Hence, choosing a smaller $\Gamma$ not only makes the required wave function more complex, but also demands correspondingly higher fidelity in the solution of the LIT equation.

Once the central values and the corresponding uncertainties of the LIT are determined, one needs to invert the 
integral transform to retrieve the response function and obtain the corresponding cross sections to be compared with experimental data. In Ref.~\cite{parnes_nuclear_2026} two different techniques have been investigated. The first is a regularized version of the standard inversion procedure introduced in Ref.~\cite{efros_lorentz_2007}, which relies on a suitable basis expansion of the response function. The second is an improved version of the so-called Bryan's version of the Maximum Entropy method~\cite{jarrell_bayesian_1996} which enables the propagation of the uncertainties in $\mathcal{L}(\omega_0, \Gamma)$ into the reconstructed ${\cal R}(\omega)$ using Bayes' theorem.  

\begin{figure}[!t]
    \centering
    \includegraphics[width=0.5\columnwidth]{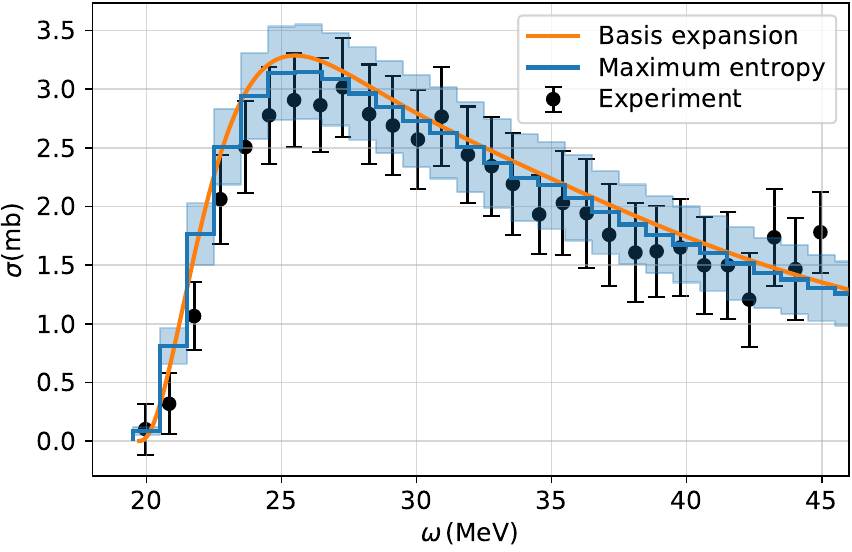}
    \caption{Adapted from Ref.~\cite{parnes_nuclear_2026} with permission from the Authors. Photo-disintegration cross section of $^4$He as a function of photon energy, obtained from NQS calculations of the LIT. The basis function results (solid orange lines) and the Maximum Entropy reconstruction (blue histogram with error bars) show good agreement with experimental data from Ref.~\cite{bacca_electromagnetic_2014}.}
    \label{fig:He4_sigma}
\end{figure}

Figure \ref{fig:He4_sigma}, adapted from  Ref.~\cite{parnes_nuclear_2026} displays the photodisintegration cross section of $^4$He, obtained by inverting the LIT computed for $\Gamma = 10$ MeV within the NQS framework, and compare it with experimental data from Ref.~\cite{bacca_electromagnetic_2014}. The basis-expansion inversion method in orange and the maximum entropy approach in blue yield results that agree remarkably well with experiment, despite the simplicity of the input Hamiltonian.

\subsection{Pairing interactions and the occupation-number formalism}
\label{sub:pairing}

All $A>2$ nuclear physics applications of NQS discussed so far entail solving the quantum many-body problem in coordinate space, with the antisymmetry of the wave function enforced explicitly through tailored architectures, such as Slater determinants or Pfaffians.
In this setting, fermionic statistics are hard-coded at the level of the wave function ansatz.
A complementary and conceptually distinct approach consists in working in the occupation-number (second-quantized) formalism~\cite{choo_fermionic_2020}, where the many-body basis is built from antisymmetrized Fock states.
In this representation, fermionic antisymmetry is not imposed on the neural-network architecture itself, but is instead automatically taken into account when evaluating expectation values of quantum-mechanical operators through the underlying algebra of creation and annihilation operators~\cite{fetter_quantum_2012}.
This shifts the burden of enforcing antisymmetry from the variational ansatz to the operator formalism.

In the second quantization setting within the occupation number formalism, a NQS representation can be chosen to be
\begin{equation}
\Psi(N) \equiv \langle N | \Psi \rangle\,,
\end{equation}
where $|N\rangle = |n_1,\dots,n_P\rangle$ denotes a basis state in Fock space. The state is specified by occupation numbers $n_i$ indicating whether the fermionic mode $i$ is occupied ($n_i=1$) or empty ($n_i=0$).
Within this general framework, a particularly important class of applications involves pairing correlations, which provide a natural setting for occupation-number--based NQS approaches.
In many physically relevant applications, the full fermionic Fock space is further restricted to a subspace adapted to the dominant correlations of the problem.
For pairing interactions, this corresponds to limiting the Hilbert space to seniority-zero configurations, in which fermions occupy time-reversed states in correlated pairs.
Within this restricted subspace, the occupation-number representation naturally encodes the presence or absence of fermion pairs in a set of active modes.

From a physical perspective, pairing correlations underlie a wide range of phenomena in interacting fermionic systems~\cite{bohr_possible_1958,ring_nuclear_1980,dean_pairing_2003}.
BCS theory provides the canonical framework for understanding pairing in electronic superconductors, wherein fermions near the Fermi surface form correlated Cooper pairs, giving rise to superconductivity.
Soon after the formulation of BCS theory, Bohr, Mottelson, and Pines identified an analogous pairing mechanism in atomic nuclei, in which nucleons near the Fermi surface form spin-singlet pairs~\cite{bohr_possible_1958}.

In finite nuclei, pairing manifests itself through characteristic signatures such as enhanced binding of even--even systems, energy gaps to the first excited states, and systematic trends in excitation spectra and moments of inertia.
These effects are commonly described within mean-field frameworks such as BCS and its particle-number conserving extensions, most notably the Hartree--Fock--Bogoliubov formalism~\cite{dobaczewski_pairing_2001}.
Beyond the mean-field level, symmetry restoration and configuration mixing techniques further refine the treatment of pairing correlations, particularly in finite systems.

Pairing Hamiltonians also play an important methodological role.
Despite their apparent simplicity, they capture essential nonperturbative physics and include integrable subclasses, such as the Richardson--Gaudin models~\cite{richardson_restricted_1963,richardson_application_1963,gaudin_diagonalisation_1976}, which admit essentially exact solutions.
As a result, pairing models provide valuable benchmarks for many-body methods, allowing controlled assessments of accuracy across weak- and strong-coupling regimes.

In its most compact form, the pairing Hamiltonian proposed by Richardson is given by
\begin{equation}\label{eq:pairing_ham}
H = \sum_{p=1}^P d_p N_p - \sum_{p, q = 1}^P g_{pq} A_p^\dagger A_q,
\end{equation}
where $N_p = \sum_{\sigma \in \{\minus, \plus\}} a_{p \sigma}^\dagger a_{p \sigma}$ is the pair number operator.
The operators $A_p^\dagger = a_{p\plus}^\dagger a_{p\minus}^\dagger$ and $A_p = a_{p\plus} a_{p\minus}$ create and annihilate fermion pairs, respectively.
Exact solutions for this model can be obtained both for uniform pairing strengths, $g_{pq} = g$ for all $p,q = 1,\ldots,P$, and for separable hyperbolic couplings of the form
$g_{pq} = 2g\,\sqrt{(\alpha - d_p)(\alpha - d_q)}$.

Once a neural-network quantum-state ansatz $\Psi_V(N)$ is adopted to represent the many-body wave function, quantum-mechanical expectation values can be estimated stochastically using the Metropolis--Hastings algorithm.
Since the number of fermion pairs is conserved, admissible Monte Carlo updates consist of exchanging occupations between pair modes, ensuring ergodicity of the sampling.
This procedure is directly analogous to the isospin-exchange moves discussed in Section~\ref{sec:nqs} within the first-quantized formulation.

As in coordinate-space approaches, the optimal values of the variational parameters defining the NQS are determined by invoking the variational principle.
To this end, all optimization strategies introduced in Section~\ref{sec:nqs} remain applicable in the occupation-number representation, including the stochastic reconfiguration method supplemented with the RMSProp-inspired shift of Eq.~\eqref{eq:sr_rmsprop}.

The main advantage of this formalism is that the NQS itself does not need to be antisymmetric.
As a result, relatively simple architectures, such as restricted Boltzmann machines and feed-forward neural networks, provide valid representations of the many-body wave function (see however Refs.~\cite{passetti_can_2023,denis_comment_2025} for a critical assessment of the representational power of NQS in second quantization).
To demonstrate the power of this approach, the authors of Ref.~\cite{rigo_solving_2023} consider the pairing Hamiltonian in Eq.~\eqref{eq:pairing_ham} for the exactly solvable uniform-coupling and separable-coupling cases. The NQS is parameterized as
\begin{equation}
    \Psi_V(N) = e^{\mathcal{U}(N)} \tanh \big[ \mathcal{V}(N) \big],
\end{equation}
where both $\mathcal{U}(N)$ and $\mathcal{V}(N)$ are MLPs. Although this functional form resembles those employed in Refs.~\cite{adams_variational_2021,gnech_calculation_2020,yang_consistent_2022}, the resulting architectures are significantly simpler, as they do not rely on permutation-invariant Deep Sets constructions~\cite{fuchs_deepsets_2019}.

\begin{figure}[t]
  \centering
  \includegraphics[width=0.48\textwidth]{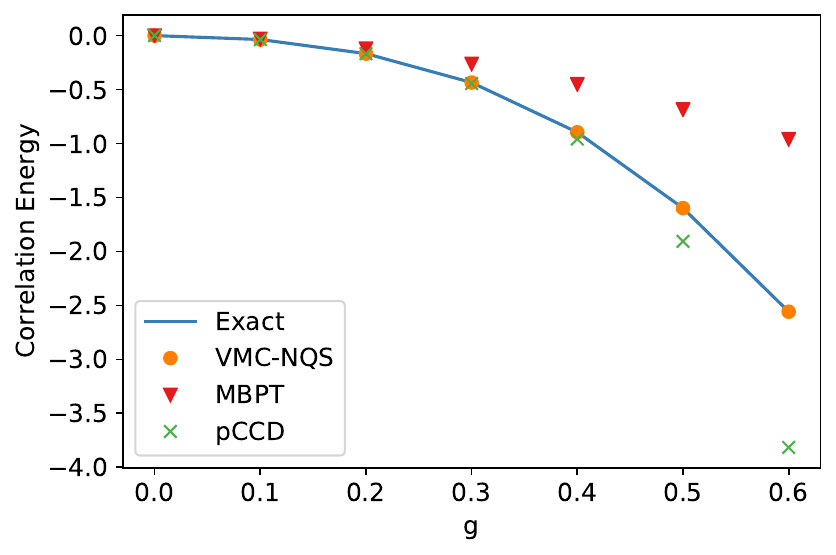}\hfill
  \includegraphics[width=0.48\textwidth]{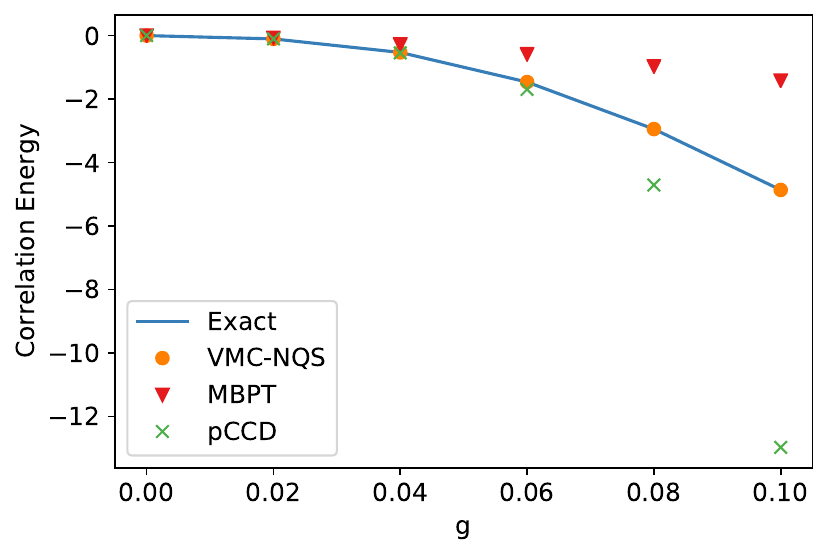}
  \caption{Correlation energies as a function of the interaction strength $g$ for the constant-coupling pairing model (left) and the separable-coupling model (right), both with $P=10$ energy levels and five fermion pairs.
  Results from VMC-NQS (orange circles), MBPT (red triangles), and pCCD (green crosses) are compared with the exact solution (solid blue line).}
  \label{fig:pairing_side_by_side}
\end{figure}

In Fig.~\ref{fig:pairing_side_by_side}, Ref.~\cite{rigo_solving_2023} compares correlation energies obtained with VMC--NQS, many-body perturbation theory (MBPT), and pair coupled-cluster doubles (pCCD) against the exact solution for both the constant-coupling and separable-coupling pairing models.
While all approaches yield similar results in the weak-coupling regime, MBPT rapidly deviates from the exact solution as the interaction strength increases, reflecting the breakdown of perturbation theory in the presence of strong pairing correlations.
The pCCD method exhibits a different failure mode, significantly overbinding at intermediate and strong coupling and thereby violating the variational principle, with particularly pronounced deviations for the separable interaction.
In contrast, the VMC--NQS results closely track the exact correlation energies across the entire range of coupling strengths shown, remaining accurate deep into the non-perturbative regime.
Taken together, these results demonstrate that occupation-number--based NQS provide a highly accurate and variationally controlled description of pairing Hamiltonians, successfully capturing collective correlations across weak- and strong-coupling regimes where conventional many-body methods fail.

At the same time, this success must be understood within the scope of the models considered.
When pairing Hamiltonians are formulated in terms of an underlying single-particle basis, realistic nuclear interactions are strongly non-perturbative.
Converged calculations therefore require a large number of single-particle orbitals.
Consequently, the dimension of the single-particle basis typically satisfies $P \gg A$, where $A$ is the number of nucleons.
In state-of-the-art no-core shell-model calculations~\cite{barrett_ab_2013}, $P$ can reach several thousand.
This, in turn, leads to neural-network representations of prohibitive size.
As a result, the applicability of occupation-number--based NQS approaches to fully realistic nuclear Hamiltonians remains limited at present, motivating the exploration of complementary representations and hybrid strategies.

\label{sec:applications}

\newpage
\section{Connections to condensed matter physics}
\label{sub:condensed_matter} 
The development of neural quantum states (NQS) has been deeply influenced by challenges and innovations in condensed matter physics, where strongly correlated quantum many-body systems demand accurate and scalable computational approaches. Many of the foundational ideas behind NQS, including pioneering work using restricted Boltzmann machines~\cite{carleo_solving_2017}, emerged from efforts to model lattice spin systems. These were subsequently extended to continuous-space systems such as the homogeneous electron gas~\cite{pescia_message-passing_2024} and ultracold Fermi gases~\cite{kim_neural-network_2024}. Such advances offer a natural conceptual bridge to nuclear systems, which pose many of the same computational challenges. As a general-purpose variational ansatz, NQS provides greater flexibility than traditional techniques and holds promise for systems governed by complex interactions and symmetry constraints. We now review two condensed matter NQS applications which are closely tied to nuclear approaches. These approaches employ a continuous formulation and tackle either low dimensionality or extended, strongly-correlated systems.

\subsection{Polarized fermions}
\label{sub:polarized_fermions}

One-dimensional fermionic systems offer a particularly useful and conceptually rich setting for studying quantum many-body physics, combining analytical insight with computational accessibility.
Due to their reduced dimensionality, such systems admit simplified theoretical descriptions and, in some cases, exact or highly controlled solutions~\cite{busch_two_1998, bethe_zur_1931}, while still exhibiting nontrivial correlation effects that challenge mean-field approaches.
From a computational standpoint, one-dimensional models provide a natural proving ground for new variational methods, as they allow detailed benchmarking against established many-body techniques at modest numerical cost.
In addition, one-dimensional fermions display distinctive phenomena, such as Fermi--Bose dualities and enhanced interaction effects, that have no direct analogue in higher dimensions and are of intrinsic theoretical interest~\cite{girardeau_relationship_1960,valiente_bose-fermi_2020,girardeau_effective_2004}.
These features make one-dimensional fermionic systems an attractive platform for exploratory studies of neural-network quantum states, serving as a stepping stone toward more realistic three-dimensional and spinful systems of relevance to nuclear and condensed-matter physics~\cite{keeble_machine_2020,adams_variational_2021,gnech_nuclei_2022,lovato_hidden-nucleons_2022}.


This strategy has been adopted in several early applications of neural-network quantum states to continuous-space fermionic systems~\cite{keeble_machine_2023,bedaque_machine_2024}. 
In particular, Ref.~\cite{keeble_machine_2023} investigated fully polarized, or equivalently spinless, fermions confined to one dimension. 
In this setting, the absence of internal spin degrees of freedom allows the fermionic antisymmetry to be enforced entirely through the spatial structure of the wave function, making it a natural test case for NQS architectures without explicit spin inputs. 
To ensure nontrivial interactions in the polarized system, finite-range pairwise forces must be employed, since zero-range interactions do not contribute to the energy of identical fermions in one dimension~\cite{keeble_machine_2023}.

The system considered in Ref.~\cite{keeble_machine_2023} consists of $A$ fermions in a harmonic trap interacting through a finite-range Gaussian potential, described by the Hamiltonian
\begin{align}
\hat H = - \frac{1}{2} \sum_{i=1}^A \nabla_i^2 
+ \frac{1}{2} \sum_{i=1}^A x_i^2
+ \frac{V_0}{\sqrt{2\pi}\sigma_0}
\sum_{i<j} \exp\!\left[-\frac{(x_i-x_j)^2}{2\sigma_0^2}\right].
\end{align}
Here, $V_0$ and $\sigma_0$ denote the interaction strength and range, respectively. 
The Hamiltonian is expressed in harmonic-oscillator units corresponding to a trap frequency $\omega$, such that lengths are measured in units of $a_{\mathrm{ho}} = \sqrt{\hbar/m\omega}$ and energies in units of $\hbar\omega$.

The NQS employed for this system was inspired by the FermiNet architecture~\cite{pfau_ab_2020}. 
In one dimension, the many-body wave function depends only on the particle coordinates $\Psi(x_1,\ldots,x_A)$. 
The neural ansatz takes as input the set of particle positions and augments them with a permutation-equivariant feature, the mean position
$\mu = \frac{1}{A}\sum_{i=1}^A x_i$. 
These features are propagated through two equivariant hidden layers with element-wise $\tanh$ nonlinearities, followed by a linear layer that constructs a symmetric generalized Slater matrix. 
The determinant of this matrix yields the full antisymmetric many-body wave function~\cite{keeble_neural_2022,keeble_machine_2023}. 
For particle numbers in the range $A=2$--$6$, this architecture contains approximately $8{,}500$--$9{,}000$ variational parameters, with only mild scaling as $A$ increases.

\begin{figure}[!t]
\centering
\includegraphics[width=0.5\columnwidth]{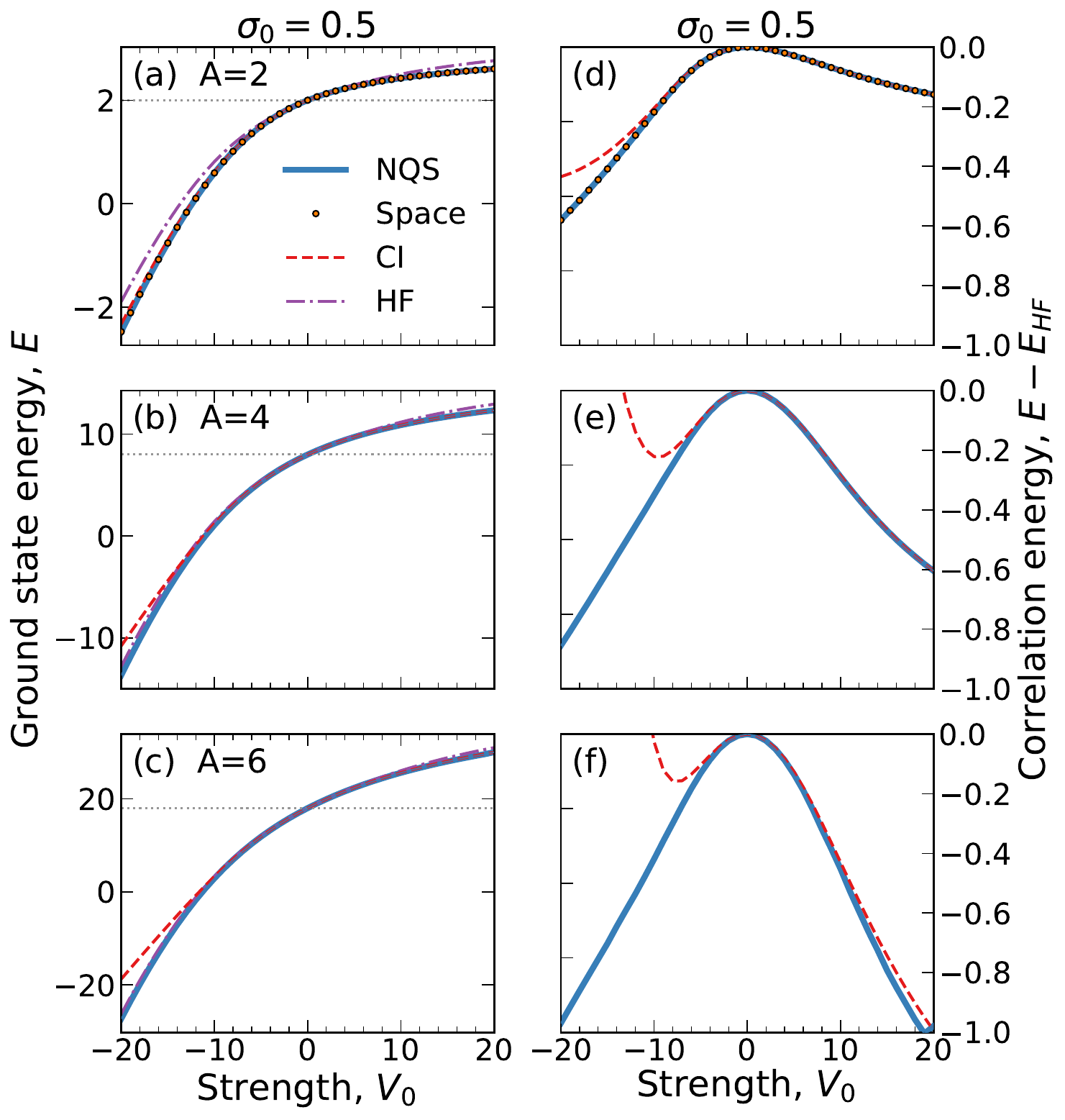}
\caption{Ground-state energies for few-body systems as a function of the interaction
strength \(V_0\) at fixed range \(\sigma_0 = 0.5\).
Panels (a)–(c) show the total energy \(E\) for systems with
\(A = 2, 4,\) and \(6\), while panels (d)–(f) show the corresponding
correlation energies \(E - E_{\mathrm{HF}}\).
Neural-network quantum state (NQS) results are shown as solid blue lines.
The real-space solution for \(A = 2\) is indicated by filled circles.
Configuration-interaction (CI) results are shown as red dashed lines, and the
Hartree--Fock (HF) approximation as purple dash-dotted lines, following
Ref.~\cite{keeble_machine_2023}.}
\label{fig:1D_energy}
\end{figure}

In Fig.~\ref{fig:1D_energy}, NQS results are compared with configuration interaction (CI), Hartree–Fock (HF), and real-space solutions (Space), where available, for a fixed width $\sigma_0 = 0.5$. For all particle numbers $A = 2, 4,$ and $6$, the NQS yields energies below the HF results, with the discrepancy increasing as the interaction becomes either strongly attractive or strongly repulsive. This architecture only incorporates a single generalised Slater determinant, so the difference between the HF and NQS results can directly be linked to backflow correlations included by construction through the equivariant features. 

While the NQS performs consistently across all interaction regimes and reproduces the exact results for $A = 2$, the fixed-dimension CI calculations exhibit noticeable basis-truncation effects as the interaction strength increases, particularly in the attractive regime. This behavior can be traced to the emergence of short-range structure in the wave function under strong attraction. Resolving accurately these structure requires progressively larger single-particle bases. 

\begin{figure}[!t]
\centering
\includegraphics[width=0.8\columnwidth]{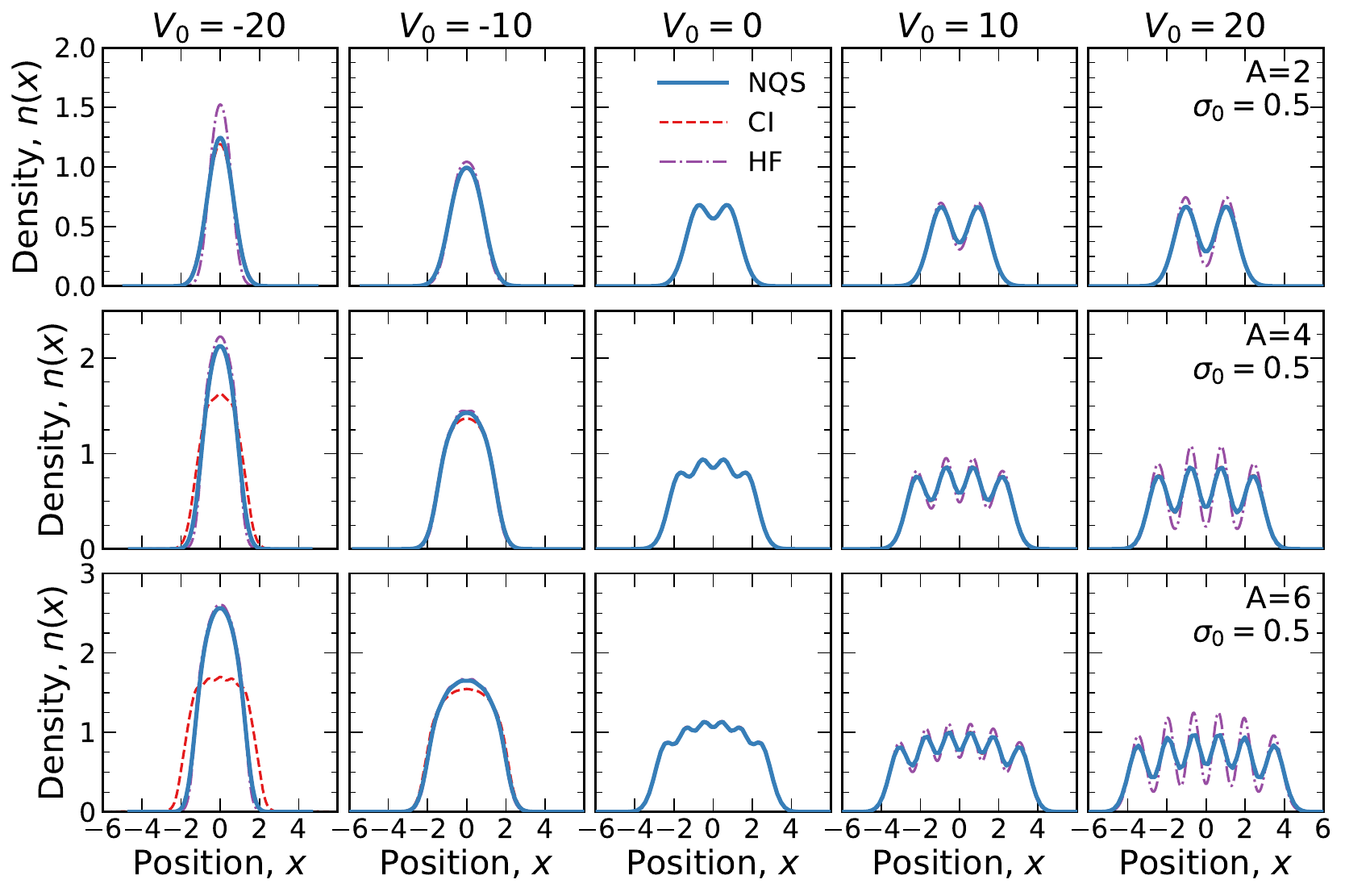}
\caption{Density profiles \(n(x)\) for one-dimensional spinless trapped fermions as a function of position \(x\) for different values
of \(V_0\) at fixed range \(\sigma_0 = 0.5\). From left to right, \(V_0\) varies
from \(-20\) to \(20\) in increments of 10. From top to bottom, results are
shown for \(A = 3, 4, 5,\) and \(6\) particles. Figure sourced from Ref.~\cite{keeble_machine_2023}.}
\label{fig:1D_densities}
\end{figure}

The density distributions of one-dimensional systems directly pinpoint the effect of complex many-body correlations. We show in Fig.~\ref{fig:1D_densities} the density distributions across a wide range of strengths, from very attractive (left panels) to very repulsive (right panels). The repulsive regime can be easily understood in terms of Wigner crystallization~\cite{keeble_machine_2023}. In contrast, the density distribution in the strongly attractive regime
displays a distinct bell shape that is reminiscent of bosonic systems. An associated pairing-like effect in the occupation numbers of the system further points towards the emergence of bosonic features in the many-body wavefunction~\cite{keeble_machine_2023}. CI approaches have a difficulty capturing such results. These features, however, can be understood in terms of a duality between strongly attractive fermions and weakly interacting bosons in 1D systems~\cite{girardeau_relationship_1960,valiente_bose-fermi_2020,girardeau_effective_2004}. 
Overall, these results demonstrate that relatively compact, permutation-equivariant NQS architectures can capture strong-correlation effects in continuous-space fermionic systems. 
The one-dimensional exactly solvable case thus provides a clean benchmark and a foundation for extensions to higher-dimensional and spinful systems.

\subsection{The homogeneous electron gas}
\label{sub:heg}

The three-dimensional homogeneous electron gas (HEG) has long served as a prototypical model for studying strongly correlated fermions with long-range Coulomb interactions. Despite its apparent simplicity, with electrons interacting via the Coulomb potential in a uniform, neutralizing background, the HEG exhibits rich many-body behavior. At low densities, where Coulomb repulsion dominates over kinetic energy, the system is expected to undergo Wigner crystallization into a body-centered cubic (BCC) lattice. Due to translational invariance, this crystal may appear as a ``floating'' phase, characterized by crystalline order in correlations rather than in the density itself.

The HEG has been investigated with a variety of NQS architectures in recent years, including
LiNet~\cite{li_ab_2022}, FermiNet~\cite{cassella_discovering_2023}, and WAPNet~\cite{wilson_neural_2023}, as well as with conventional many-body methods such as diffusion Monte Carlo~\cite{lopez_rios_inhomogeneous_2006,azadi_low-density_2022} and coupled cluster approaches~\cite{liao_towards_2021}.
Rather than surveying these applications exhaustively, we focus here on the approach of Ref.~\cite{pescia_message-passing_2024}, for which the HEG served as the primary development and validation platform for the permutation-equivariant message-passing neural network (MPNN) architecture that appears throughout this review.
The insights gained in this setting subsequently informed applications to other continuous-space fermionic systems, including ultracold Fermi gases~\cite{kim_neural-network_2024}, nuclear matter~\cite{fore_investigating_2024}, and finite nuclei~\cite{gnech_distilling_2024,donna_hypernuclei_2025}.

\begin{figure}[!t]
\centering
\includegraphics[width=0.5\columnwidth]{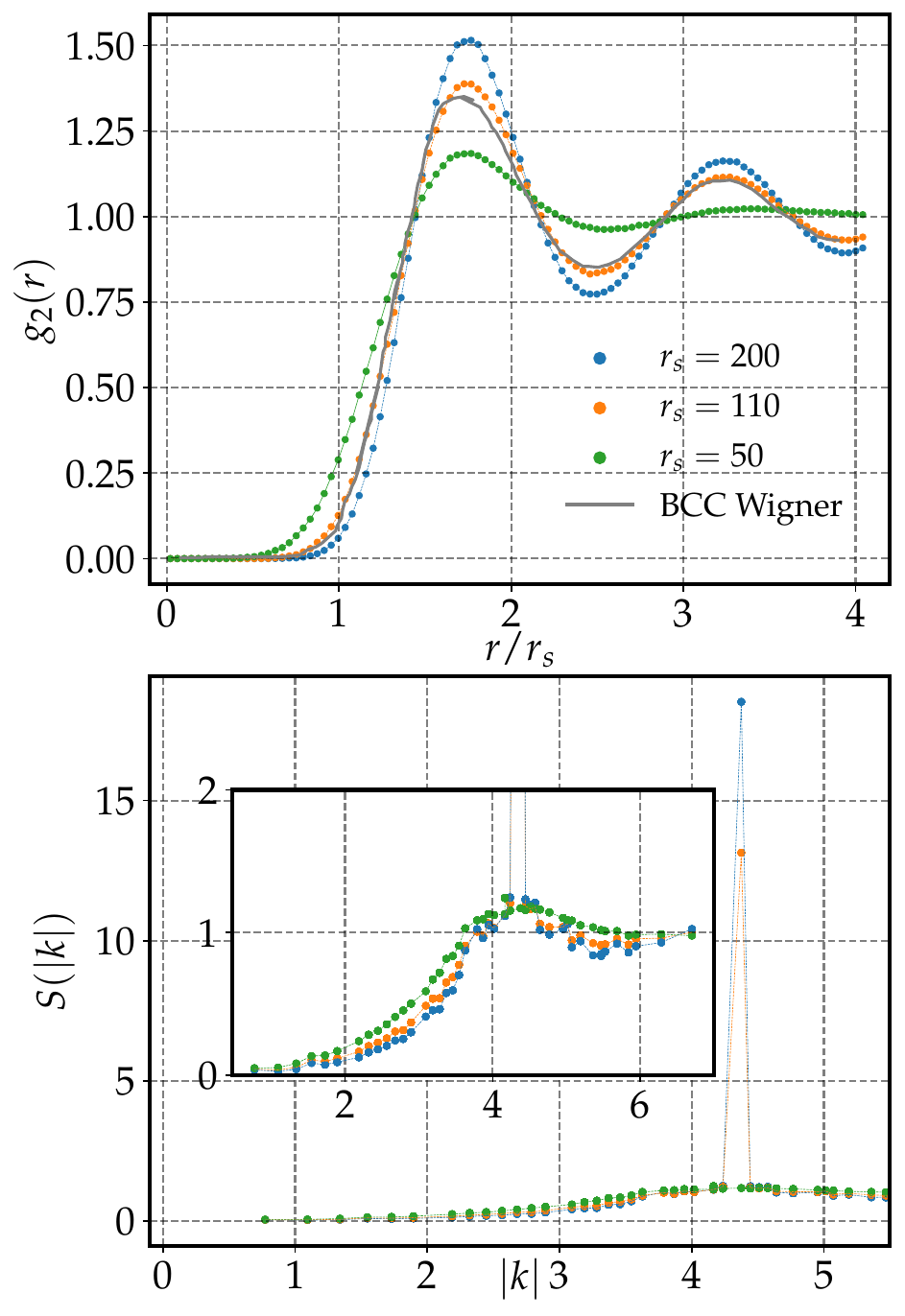}
\caption{Spin-averaged radial distribution function (top panel) and the corresponding
structure factor (bottom panel) for the homogeneous electron gas with
\(N = 128\) electrons, evaluated at \(r_s = 50, 110,\) and \(200\).
A permutation-equivariant message-passing neural network with an attention
mechanism~\cite{pescia_message-passing_2024} was used to construct backflow
correlations within a Slater--Jastrow NQS. Plane-wave orbitals were used at
\(r_s = 50\), while Gaussian orbitals centered on body-centered cubic (BCC)
lattice sites were employed at \(r_s = 110\) and \(200\). }
\label{fig:heg}
\end{figure}

To resolve the emergence of the floating Wigner crystal, NQS must scale to large particle numbers while accurately capturing subtle many-body correlations.
This requirement directly motivated the development of a permutation-equivariant MPNN architecture augmented with an attention mechanism~\cite{pescia_message-passing_2024}, which encodes how the state of a single electron is influenced by the instantaneous positions and spins of all other electrons.
This approach enabled simulations of the HEG with up to 128 electrons, pushing far beyond the system sizes accessible in previous NQS studies~\cite{li_ab_2022}, and made it possible to detect Wigner-like order in the structure factor at low densities (see Fig.~\ref{fig:heg}).

In this HEG application, the construction of the MPNN begins by representing each electron as a node in a fully connected graph, with edges encoding the pairwise relationships between electrons. The goal of the MPNN is to produce spatial displacements $\delta \mathbf{r}_i(\mathbf{X})$ for each particle $i = 1, \ldots, N$, thereby incorporating backflow correlations into the single-particle orbitals of a neural Slater determinant. During each message-passing iteration, both the original one- and two-body features from the simulation and a set of auxiliary hidden one- and two-body vectors are propagated and updated, enabling the network to progressively refine its representation of correlations. Specifically, the visible two-body features for the HEG are taken to be
\begin{equation}
    \mathbf{v}_{ij} = \big[ \mathbf{r}_{ij}, \ \| \mathbf{r}_{ij} \|, \ s_i^z \cdot s_j^z \big],
\end{equation}
where the pair displacements $\mathbf{r}_{ij} = \mathbf{r}_i - \mathbf{r}_j$ and distances $\|\mathbf{r}_{ij}\|$ are mapped to their $L$-periodic versions:
\begin{align}
    \mathbf{r}_{ij} \ \mapsto \ 
    \left[ 
    \sin \left( \frac{2 \pi}{L} \mathbf{r}_{ij} \right),
    \cos \left( \frac{2 \pi}{L} \mathbf{r}_{ij} \right) \right], \quad
    \|\mathbf{r}_{ij}\| \ \mapsto \
    \Bigg\| \sin\left(\frac{\pi}{L} \mathbf{r}_{ij} \right) \Bigg\|.
\end{align}
The initial one-body features, on the other hand, are taken as a trainable embedding vector that is independent of the particle index,
\begin{equation}
    \mathbf{v}_i = \mathbf{e}.
\end{equation}
Note that this choice of initial one-body features excludes dependence on absolute positions $\mathbf{r}_i$ to ensure translational invariance.

Beyond the initial one- and two-body features and their processing into visible features, the main difference between this MPNN and the one in Section~\ref{subsub:backflow_correlations} is the use of an attention mechanism~\cite{vaswani_attention_2017} in the message construction. Originally developed for natural language processing, attention mechanisms learn to assign weights to different inputs, enabling the model to focus on the most relevant information when updating each particle's features (see Section~\ref{sec:deep_learning} for more details).
Unlike Eq.~\eqref{eq:mpnn-message}, where a feedforward neural network is applied directly to the pairwise features to produce the message, Ref.~\cite{pescia_message-passing_2024} first computes an attention score
\begin{equation}
    \boldsymbol{\omega}_{ij}^{(t)} = \mathrm{GELU}
    \left(\sum_{k=1}^N \mathbf{Q}_{ik}^{(t)} \mathbf{K}_{kj}^{(t)} \right),
\end{equation}
which modulates the signal element-wise based on similarities between the query tensor $\mathbf{Q}_{ij}^{(t)} = W_Q \mathbf{g}_{ij}^{(t)}$ and the key tensor $\mathbf{K}_{ij}^{(t)} = W_K \mathbf{g}_{ij}^{(t)}$.
Note that the weight matrices used to produce the query and key are the same for each pair to ensure permutation equivariance.
The message is then constructed as
\begin{equation}
    \mathbf{m}_{ij}^{(t)} = \boldsymbol{\omega}_{ij}^{(t)} \odot f_V^{(t)}(\mathbf{h}_{ij}^{(t-1)}),
\end{equation}
where $f_V^{(t)}$ serves as the value network, providing the information to be passed through a nonlinear transformation, and $\odot$ denotes element-wise multiplication rather than the usual dot product. This construction provides flexibility in defining the message content while simultaneously weighting its relevance.

The MPNN approach achieves ground-state energies for the HEG that are on par with or better than those obtained with state-of-the-art NQS, such as FermiNet~\cite{cassella_discovering_2023} and WAPNet~\cite{wilson_neural_2023}, as well as fixed-node and backflow diffusion Monte Carlo methods~\cite{lopez_rios_inhomogeneous_2006, azadi_low-density_2022}. For small systems, it reaches chemical accuracy relative to exact methods, while for larger systems it outperforms comparable variational and diffusion Monte Carlo approaches, particularly at high densities. The method systematically improves with additional message-passing iterations and maintains accuracy across different densities and polarizations. Most notably, the MPNN uses orders of magnitude fewer parameters ($\sim 19{,}000$) than other NQS, and can simulate continuous-space systems with up to 128 electrons, more than double the size previously accessible.

The method also reproduces the expected liquid-to-Wigner-crystal phase transition around $r_s = 100$, in agreement with previous studies~\cite{azadi_correlation_2023,drummond_diffusion_2004, ceperley_ground_1980}. Gaussian orbitals were found to more easily capture the crystalline phase for large $r_s$ compared to plane-wave orbitals, as shown in Fig.~\ref{fig:heg}. The prominent peak in the structure factor that appears between $r_s = 50$ and $r_s = 110$ indicates the emergence of a crystalline state from a fluid state. These results demonstrate that MPNN-based NQS can efficiently and accurately model both delocalized and localized phases in extended systems, providing a framework applicable to other systems where backflow correlations play a central role.

\subsection{Ultra-cold Fermi gases}
\label{sub:ufg}

The unitary Fermi gas (UFG) is a paradigmatic strongly correlated quantum system of two-component fermions interacting via a short-range potential tuned to infinite scattering length and negligible effective range. In this regime, the system exhibits universal behavior, with properties determined solely by the particle density rather than the microscopic details of the interaction. The UFG lies at the crossover between the BCS superfluid and the Bose--Einstein condensate (BEC) limits, and displays strong pairing correlations and superfluidity in the absence of a small expansion parameter. These features make the UFG both a challenging nonperturbative many-body problem and an ideal benchmark for testing wave-function ans\"atze and many-body methods. Moreover, low-density neutron matter in the inner crust of neutron stars shares similar universal characteristics, making the UFG a useful proxy for studying pairing and superfluidity in astrophysical systems.

In this review, we focus on the work of Ref.~\cite{kim_neural-network_2024}, which introduces a Pfaffian--Jastrow NQS augmented by neural backflow transformations generated through a permutation-equivariant message-passing architecture~\cite{pescia_message-passing_2024}. This construction combines a fully trainable Pfaffian pairing structure with many-body backflow effects, enabling the wave function to capture fermionic antisymmetry and strong pairing correlations within a unified framework, without imposing a predefined decomposition into singlet and triplet pairing channels, fixed functional forms for the pairing orbitals, or a prescribed spin ordering. As a result, the ansatz can be naturally extended to nuclear systems with explicit spin- and isospin-exchange interactions and nontrivial coupling to the spatial degrees of freedom. In parallel with  Ref.~\cite{kim_neural-network_2024}, an NQS approach to the unitary Fermi gas based on the FermiNet architecture was developed around the same time~\cite{lou_neural_2024}. However, its formulation targets systems with fixed internal spin degrees of freedom and can not directly address the explicit spin--isospin exchange structure of nuclear interactions.

\begin{figure}[!t]
\centering
\includegraphics[width=0.75\columnwidth]{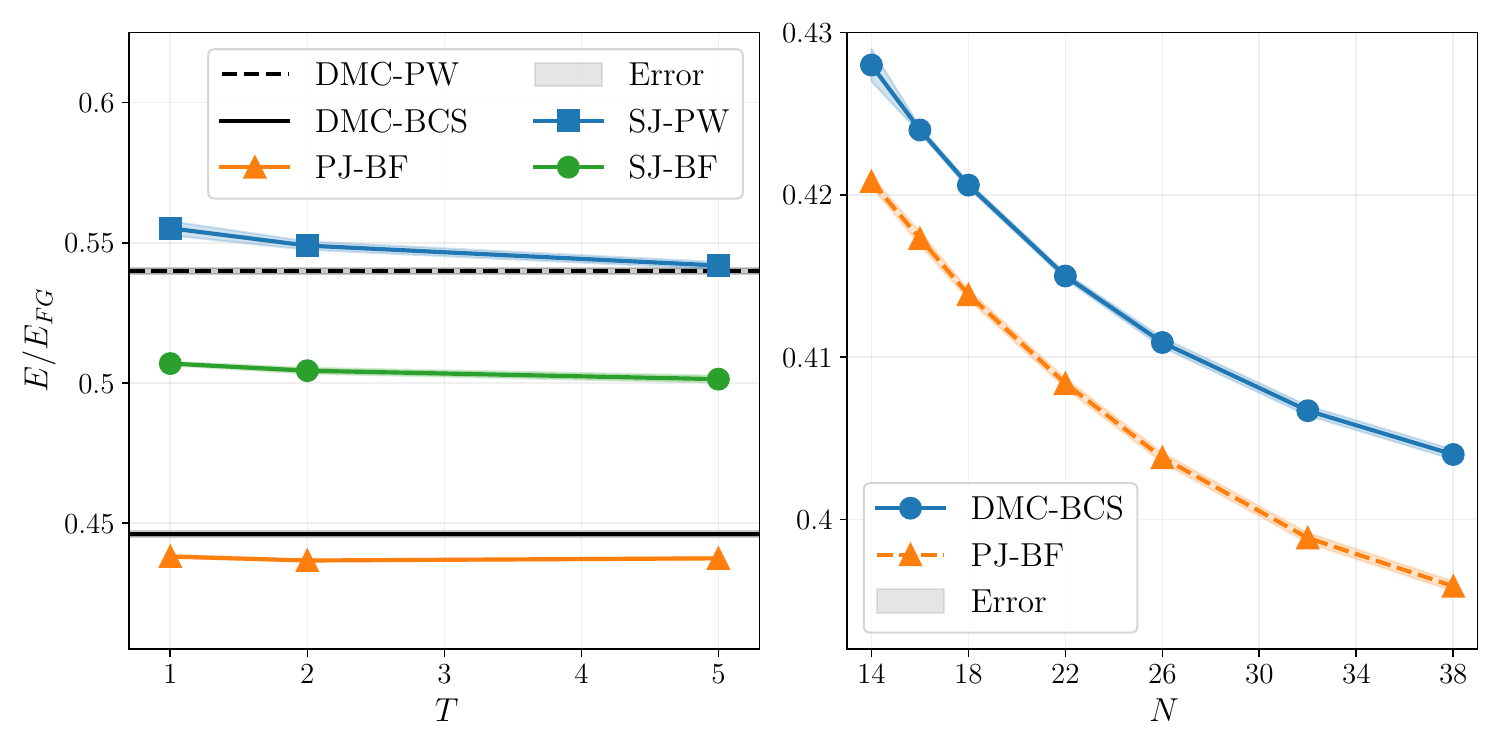}
\caption{
Ground-state energy per particle of the UFG, normalized by the noninteracting Fermi gas energy in the thermodynamic limit, $E_{FG}$, as a function of message-passing depth \(T\) (left panel) and particle number \(N\) (right panel).
Details of the models, interaction parameters, and uncertainty estimates are discussed in the main text.
}
\label{fig:ufg-energies-T-N}
\end{figure}

\begin{figure}[!t]
\centering
\includegraphics[width=0.75\columnwidth]{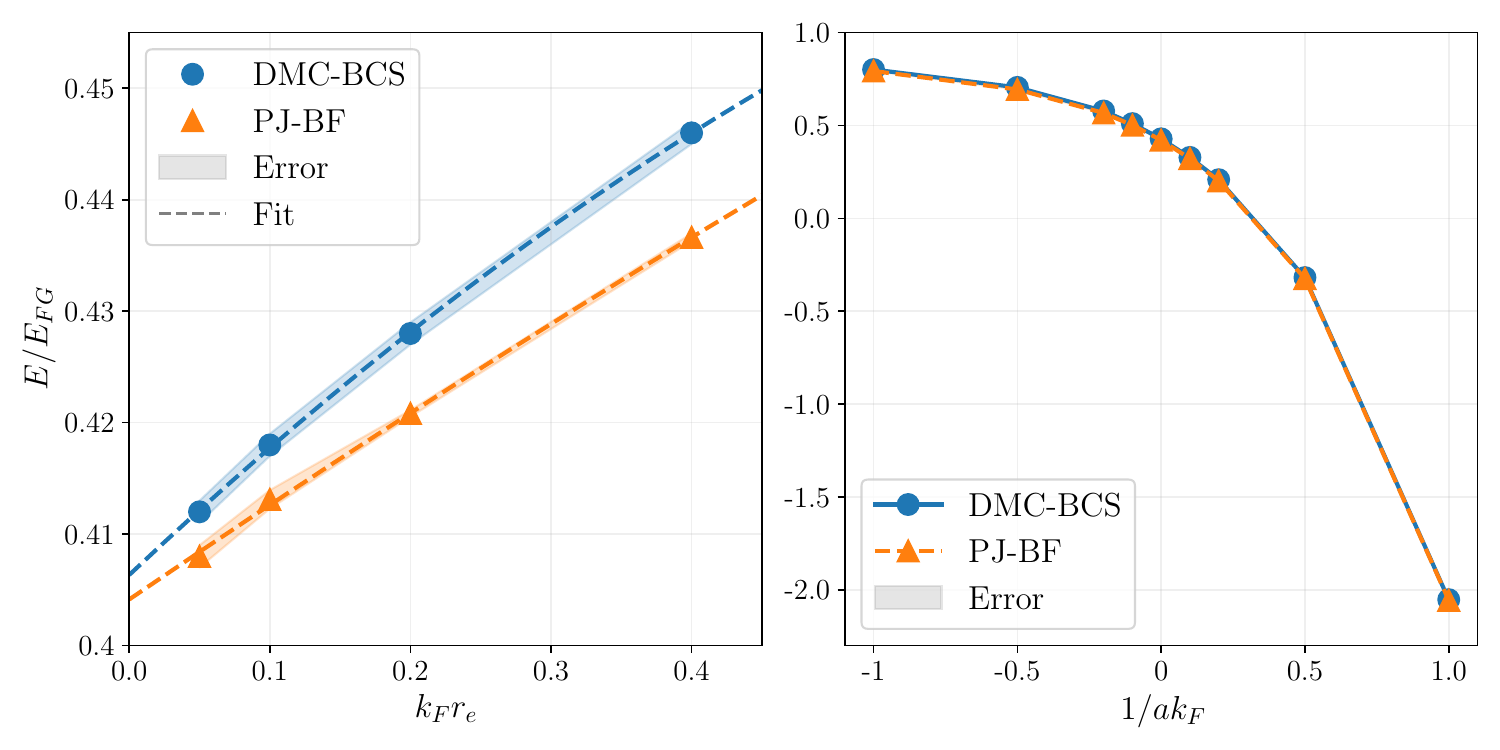}
\caption{Ground-state energy per particle of the UFG, normalized by the noninteracting Fermi gas energy in the thermodynamic limit, $E_{FG}$, as a function of effective range \(k_F r_e\) (left panel) and inverse scattering length \(1/(a k_F)\) (right panel).
Details of the calculations and uncertainty estimates are discussed in the main text.
}
\label{fig:ufg-energies-reff-a}
\end{figure}

As discussed in Sections~\ref{subsub:pfaffian_jastrow} and~\ref{sub:heg}, Ref.~\cite{kim_neural-network_2024} employs a Pfaffian--Jastrow NQS combined with a permutation-equivariant message-passing neural network. The ground-state energy per particle, normalized by the noninteracting Fermi gas energy in the thermodynamic limit, is shown in Figs.~\ref{fig:ufg-energies-T-N} and~\ref{fig:ufg-energies-reff-a}. For the NQS results, converged energies are obtained by averaging over the final 100 optimization iterations, with shaded bands indicating the corresponding standard deviations. DMC uncertainty bands correspond to standard errors extracted from block-averaged energies.

Figure~\ref{fig:ufg-energies-T-N} presents an initial comparison of three NQS architectures and two DMC benchmarks as a function of MPNN depth \(T\). The NQS architectures include Slater--Jastrow ans\"atze with plane-wave orbitals (SJ--PW) and with backflow correlations (SJ--BF), together with the Pfaffian--Jastrow (PJ--BF) ansatz. In the left panel, the interaction parameters are fixed such that \(k_F r_e = 0.4\), and DMC reference energies with (DMC--BCS) and without (DMC--PW) pairing are shown for comparison.
As the MPNN depth increases, the SJ--PW energies approach the DMC--PW benchmark, indicating that the MPNN captures correlations at a level comparable to DMC with a restricted nodal surface. 
However, incorporating backflow correlations generated by the MPNN into the single-particle orbitals of the Slater determinant, as in the SJ--BF ansatz, is still insufficient to capture the strong pairing correlations.
By contrast, the PJ--BF ansatz, which explicitly encodes these pairing correlations, outperforms DMC--BCS at all message-passing depths shown.
The right panel shows the system-size dependence of the energy at fixed effective range \(k_F r_e = 0.2\). Since all the MLPs used to construct the PJ--BF ansatz act only on pairs or single particles, the number of variational parameters does not explicitly depend on the particle number. For all particle numbers shown, the Pfaffian--Jastrow ansatz systematically yields lower energies than the DMC--BCS reference, with the difference between the two energies growing as the number of particles increases.

Figure~\ref{fig:ufg-energies-reff-a} shows energies at unitarity as a function of effective range, including extrapolations to zero effective range using quadratic fits, as well as results across the BCS--BEC crossover at fixed \(k_F r_e = 0.2\). These results demonstrate that the Pfaffian--Jastrow ansatz is robust across all parameter regions explored, indicating that the Pfaffian pairing structure provides a consistently more accurate variational description of strongly correlated pairing physics for ultracold Fermi gases near the unitary limit.

A central practical outcome of this work is the demonstration that transfer learning can stabilize and accelerate optimization in regimes characterized by strong, short-range interactions and small effective ranges. Because the ansatz does not encode explicit assumptions about short-distance structure, it remains broadly applicable. By progressively pretraining on softer interactions and smaller systems, a single unified ansatz can be extended to harder regimes and larger particle numbers, enabling a controlled exploration of the BCS--BEC crossover and finite-size effects.

Beyond ground-state energies, the Pfaffian--Jastrow NQS accurately describes observables sensitive to the quality of the variational wave function, including pair distribution functions and pairing gaps. The predicted pairing gaps are closer to experimental values than those obtained from DMC calculations based on BCS trial states, and are consistent within uncertainties with DMC results for substantially larger systems. This highlights the potential of neural-network-based Pfaffian ans\"atze for providing a faithful description of both energetic and pairing properties in strongly correlated fermionic systems.

\newpage
\section{Conclusions and perspectives}
\label{sec:conclusions}
In this review, we report recent progress in the application of neural-network quantum states to nuclear many-body problems. Starting from the pioneering studies of the deuteron~\cite{keeble_machine_2020} and very light nuclei~\cite{adams_variational_2021}, the field has advanced enormously over the past five years. Thanks to algorithmic developments and the increased availability of GPU resources, it is now possible to compute ground-state properties of medium-mass nuclei~\cite{gnech_distilling_2024} and infinite nuclear matter, simulating up to $42$ particles in periodic boundary conditions~\cite{fore_investigating_2024}. While simulations of these relatively large systems have so far been carried out using ``essential'' nuclear interactions~\cite{schiavilla_two-_2021}, NQS capable of tackling high-resolution potentials derived within chiral EFT have recently been successfully developed~\cite{wen_neural-network_2025,yang_zemach_2025}. Notably, these applications are not limited to ground-state properties, but also extend to neutron–nucleus scattering~\cite{yang_chiral_2025}.

In addition to extending the reach of conventional continuum QMC methods to medium-mass nuclei, the use of NQS has enabled the study of properties of nuclear systems that are inaccessible to conventional QMC approaches. A notable example is the self-emergence of nuclear clusters in the crust of neutron stars~\cite{fore_investigating_2024}, which lowers the energy per particle and increases the proton fraction. Previous continuum QMC studies in this density regime were unable to capture this phenomenon due to the fermion-sign problem and the reliance on variational ans\"atze that are only suitable for describing the uniform liquid phase. As a second example, the combination of the Lorentz Integral Transform with the VMC--NQS method enables QMC calculations of electroweak response functions~\cite{parnes_nuclear_2026} in the low-energy regime. This region is particularly challenging for conventional GFMC approaches based on imaginary-time current--current correlators, since the inversion of the corresponding Euclidean responses is an ill-posed problem and, in practice, becomes increasingly unreliable at low energy transfer, even when using Maximum Entropy techniques~\cite{lovato_electromagnetic_2016}. Importantly, the low-energy response is of central relevance for neutrino experiments, especially those involving astrophysical neutrinos, such as supernova and diffuse supernova neutrinos~\cite{scholberg_supernova_2012}.

In this review, we have also highlighted the fertile connections with condensed-matter applications. One-dimensional fermionic systems provide an excellent, yet challenging, testing ground for NQS architectures~\cite{keeble_machine_2023} and for the development of new ones, including alternatives to standard MLPs, such as Kolmogorov--Arnold wave functions~\cite{bedaque_machine_2024}. In addition, the first calculations of the dilute neutron-matter equation of state using NQS were enabled by the introduction of periodic coordinates as network inputs, an approach originally developed for periodic bosonic systems~\cite{pescia_neural-network_2022}. These include one- and two-dimensional interacting quantum gases with Gaussian interactions, as well as $^4$He confined in a one-dimensional geometry. Similarly, the highly expressive MPNN backflow transformation, now commonly employed to describe confined and periodic nuclear systems, was originally proposed in the context of the homogeneous electron gas~\cite{pescia_message-passing_2024}. Analogously, the Pfaffian--Jastrow ansatz, before being applied to elucidate the onset of nuclear clustering in the neutron-star crust~\cite{fore_investigating_2024}, proved to be of critical importance for an accurate description of the unitary Fermi gas~\cite{kim_neural-network_2024}.

While not explicitly discussed in this review, recent advances in NQS applications to the nuclear quantum many-body problem include systematic studies of hypernuclei~\cite{donna_hypernuclei_2025,zhang_machine_2025}. In addition to their intrinsic interest, as evidenced by intense experimental campaigns~\cite{hashimoto_spectroscopy_2006}, an accurate determination of nucleon--hyperon and nucleon--nucleon--hyperon interactions is of critical importance for the description of matter in the inner core of neutron stars~\cite{lonardoni_hyperon_2015}. In particular, a quantitative understanding of the repulsive components of the nucleon--nucleon--hyperon interaction is widely regarded as a key ingredient for resolving the so-called hyperon puzzle, namely the apparent tension between the onset of hyperons in dense matter and the existence of neutron stars with masses of about two solar masses~\cite{bombaci_hyperon_2017}.

Based on the rapid progress achieved over the past few years, VMC--NQS approaches are expected to play an increasingly important role in the nuclear many-body community in the coming years.
Potential future applications include the combination of NQS with eigenvector continuation techniques~\cite{frame_eigenvector_2018,duguet_colloquium_2024} to enable systematic studies of nuclear interactions and the reliable quantification of theoretical uncertainties. In this context, foundation models recently developed in condensed-matter physics are also expected to play an important role~\cite{rende_foundation_2025}. Both eigenvector continuation and foundation models offer efficient avenues for exploring families of nuclear Hamiltonians characterized by varying couplings and regulator choices. These approaches are complementary to, and potentially more accurate than, Gaussian-process emulators, which have already been employed in conjunction with NQS to study the sensitivity of hypernuclear ground-state energies to the strength and range of the $\Lambda$--nucleon--nucleon interaction~\cite{donna_hypernuclei_2025}.

Another promising future direction is the calculation of real-time quantum dynamics, which is of critical importance for an \emph{ab initio} description of scattering, fission, and fusion processes. Compared with time-dependent Hartree--Fock approaches~\cite{simenel_heavy-ion_2018}, time-dependent variational Monte Carlo (tVMC)~\cite{carleo_localization_2012} fully incorporates dynamical correlations at both short and long distances. Applications of tVMC to NQS in the continuum are still in their infancy, even in condensed-matter systems~\cite{nys_ab-initio_2024}. Consequently, beyond their direct relevance for the experimental program, employing these techniques to solve the nuclear time-dependent Schr\"odinger equation would also provide a stringent testbed, as nuclear dynamics requires the simultaneous treatment of discrete bound states and continuum degrees of freedom. Achieving this goal would however widen the reach of this promising \textit{ab initio} approach to the realm of nuclear dynamics, potentially addressing open questions in fusion, low-energy transfer and, eventually, fission reactions.

\newpage
\section*{Acknowledgments}
We acknowledge useful discussions with James Keeble, Mehdi Drissi, and Javier Rozalén-Sarmiento.

The present research is supported by the U.S. Department of Energy, Office of Science, Office of Nuclear Physics, under contracts DE-AC02-06CH11357 (A.~L., B.~F.) and DE-FG02-93ER40756 (J.~K.), by the DOE Early Career Research Program (A.~L., B.~F., J.~K.), by the Fermi Research Alliance, LLC under Contract No. DE-AC02-07CH11359 with the U.S. Department of Energy, Office of Science, Office of High Energy Physics (N.~R.), by the SciDAC-5 NeuCol and NUCLEI programs (A.~L., N.~R.). (A.~L., N.~R.) also acknowledge financial support by grant PID2023-147458NB-C21 funded by MCIN/AEI/10.13039/501100011033 and by the European Union. 

A. R. acknowledges financial support from MCIN/AEI/10.13039/501100011033 through grants PID2021-127890NB-I00,; PID2023-147112NB-C22; CNS2022-135529 funded by the “European Union NextGenerationEU/PRTR”, 
and CEX2024-001451-M  to the “Unit of Excellence Mar\'ia de Maeztu 2025-2031” award to the Institute of Cosmos Sciences; and by the Generalitat de Catalunya, through grant 2021SGR01095. 

M. H. J. has partly been partially supported by the U.S. Department of Energy through grant DE-SC0026198 and the U.S. National Science Foundation through grant PHY-2310020.  
	

\bibliography{zotero_references}

\newpage
\appendix
\renewcommand*{\thesection}{\Alph{section}}
	
		
\end{document}